\documentclass[twoside,openright,
               a4paper,11pt,BCOR 7mm,DIV=11,
               liststotoc,idxtotoc,bibtotoc,
               chapterprefix,pointlessnumbers,headsepline]
              {scrreprt}

\usepackage[utf8]{inputenc}
\usepackage[T1,T2A]{fontenc}
\usepackage[colorlinks=true,citecolor=blue,linkcolor=blue]{hyperref}
\usepackage[utopia]{quotchap}
\usepackage{amsmath,amssymb,slashed}
\usepackage{epsfig}  
\usepackage{graphicx}               
\usepackage{url}
\usepackage{hyperref}
\usepackage{siunitx}
\usepackage{color}
\usepackage{multirow,hhline}
\usepackage{placeins}
\usepackage[dvipsnames]{xcolor}
\usepackage{dcolumn}
\usepackage[numbers,sort&compress]{natbib}

\setkomafont{pagehead}{\scshape}         
\addtokomafont{captionlabel}{\bfseries}  
\setcapindent{0cm}                       
\usepackage{tikz}
\usetikzlibrary{trees}
\usetikzlibrary{decorations.pathmorphing}
\usetikzlibrary{decorations.markings}

\makeatletter
\renewcommand{\@makechapterhead}[1]{\chapterheadstartvskip%
  \vspace*{-1cm}
  {\size@chapter{\sectfont\raggedleft
      {\chapnumfont
        \ifnum \c@secnumdepth >\m@ne%
        \if@mainmatter\thechapter%
        \fi\fi
        \par\nobreak}%
      {\raggedleft\advance\leftmargin10em\interlinepenalty\@M\Huge\mdseries #1\par}}
    \nobreak\chapterheadendvskip}}
\makeatother

\definecolor{jblue}  {RGB}{20,50,100}
\definecolor{npurple}  {RGB} {153, 51, 204}  
\definecolor{wred}   {RGB}{217,0,56}
\definecolor{white}   {RGB}{255,255,255}

\definecolor{korange}   {RGB}{235, 80,  43}
\definecolor{korange2}   {RGB}{245, 100,  63}
\definecolor{kyelloworange}   {RGB}{255, 210,  110}
\definecolor{kyelloworange2}   {RGB}{240, 170,  90}
\definecolor{kred}   {RGB}{204,  102, 153}
\definecolor{kpurple}   {RGB}{153,  61, 190}
\definecolor{kpurplelight}   {RGB}{213,  161, 230}

\usepackage{cleveref}
\usepackage{subfigure} 
\usepackage{lipsum}
\definecolor{red}{rgb}{1.0, 0, 0}
 \usepackage{gensymb}
\allowdisplaybreaks

\newcommand{\bra}[1]{\ensuremath{\langle #1 |}}   
\newcommand{\ket}[1]{\ensuremath{| #1 \rangle}}   
\newcommand{\sprod}[2]{\ensuremath{\left\langle #1 |%
                     #2 \right\rangle}}  

\newcommand{\diag}{\text{diag}}
\newcommand{\BR}{\text{BR}}

\newcommand{\parenbar}[1]{\overset{
            \raisebox{-0.15em}{\scalebox{.4}{\textbf{(}}}
            \raisebox{-0.3em}{{\hspace{.03em}--\hspace{.05em}}}
            \raisebox{-0.15em}{\scalebox{.4}{\textbf{)}}}} {#1}}

\renewcommand{\vec}[1]{{\mathbf{#1}}}
\newcommand{\iso}[2]{{\ensuremath{{}^{#2}}\ensuremath{\rm #1}}}

\newcommand{\im}{\textrm{Im}}
\newcommand{\re}{\textrm{Re}}
\newcommand{\eV}{\,\mathrm{eV}}

\newcommand{\MeV}{\,\mathrm{MeV}}

\newcommand{\ie}{{i.e.}}
\newcommand{\eg}{{e.g.}}

\newcommand{\Neff}{{N_\textrm{eff}}}
\newcommand{\mnusum}{{\sum m_{\nu}}}
  
\title{Sterile Neutrinos}
\author{Basudeb Dasgupta \\
  \it\normalsize Tata Institute of Fundamental Research, \\[-0.1cm]
  \it\normalsize Homi Bhabha Road, Mumbai, 400005, India. \\[-0.1cm]
  \tt\normalsize \href{mailto:bdasgupta@theory.tifr.res.in}
                      {bdasgupta@theory.tifr.res.in} \\[0.4cm]
  Joachim Kopp \\
  \it\normalsize Theoretical Physics Department, CERN, Geneva, Switzerland and \\[-0.1cm]
  \it\normalsize Johannes Gutenberg University Mainz, 55099 Mainz, Germany \\[-0.1cm] 
  \tt\normalsize \href{mailto:jkopp@cern.ch}{jkopp@cern.ch}
}
\date{\today}

\begin{document}

%

\maketitle

\begin{abstract}
  Neutrinos, being the only fermions in the Standard Model of Particle Physics
  that do not possess electromagnetic or color charges, have the unique
  opportunity to communicate with fermions outside the Standard Model through
  mass mixing.  Such Standard Model-singlet fermions are generally referred to
  as ``sterile neutrinos''.  In this review article, we discuss the theoretical
  and experimental motivation for sterile neutrinos, as well as their
  phenomenological consequences.  With the benefit of hindsight in 2020,
  we point out potentially viable and interesting ideas.
   We focus in particular on sterile neutrinos
  that are light enough to participate in neutrino oscillations, but we also
  comment on the benefits of introducing heavier sterile states.  We discuss
  the phenomenology of eV-scale sterile neutrinos in terrestrial experiments
  and in cosmology, we survey the global data, and we highlight various
  intriguing anomalies. We also expose the severe tension that exists between
  different data sets and prevents a consistent interpretation of the global
  data in at least the simplest sterile neutrino models.  We discuss
  non-minimal scenarios that may alleviate some of this tension. We briefly review the status of keV-scale sterile neutrinos as dark matter and the possibility of explaining the matter--antimatter asymmetry of the Universe through leptogenesis driven by yet heavier sterile neutrinos.
\end{abstract}

\tableofcontents

\chapter{Introduction}

When Wolfgang Pauli proposed the existence of neutrinos in 1930, he called them
a ``desperate remedy'' and considered them undetectable \cite{Pauli:1930:LTC,
Brown:1978:IN}.  Nowadays, particle physics has come a long way: experimentalists detect
neutrinos on a daily basis, and theorists introduce new ``invisible'' particles
at the same rate, showing sometimes the same desperation as Pauli, but with much
fewer qualms. Among the most
popular invisible particles found in the literature are sterile neutrinos:
Standard Model singlet fermions that interact with ordinary matter only through
mixing with the neutrinos. (The denomination ``neutrino'' refers to their
fermionic nature and their electric charge neutrality, the adjective
``sterile'' indicates their neutrality under weak interactions.)

Sterile neutrinos are the topic of this review article, with particular
focus on relatively light (eV-scale) sterile states. Sterile neutrinos are
theoretically well-motivated, with the main arguments
in their favor being the following:
\begin{enumerate}
  \item Singlet fermions naturally appear in the \textbf{dark sector}.
    There is overwhelming evidence for the existence
    of new, extremely weakly interacting particles in the Universe that constitute
    about 80\% of its total matter content.  Given strong experimental constraints
    on this ``dark matter'', it very likely does not carry Standard Model
    gauge charges. In most models, the dark matter particle comes with
    a ``dark sector'' of additional Standard Model singlet particles. Any dark sector
    fermion will mix with neutrinos, unless the corresponding coupling --
    a Yukawa coupling of the form
    \begin{align}
      \mathcal{L} \supset y \, (i \sigma^2 H^*) L N
      \label{eq:neutrino-portal-1}
    \end{align}
    is forbidden by extra symmetries.\footnote{In our notation, $L$ is a
      Standard
    Model lepton doublet, $N$ is the sterile neutrino, $H$ is the Higgs doublet,
    $\sigma^2$ is a Pauli matrix, and $y$ is a dimensionless
    coupling constant. All fermions are interpreted as Weyl spinors.
    See \cref{sec:theory-motivation} for details.}
    In fact, the dark matter itself could be in the form of sterile neutrinos.

  \item The \textbf{neutrino portal coupling} from \cref{eq:neutrino-portal-1}
    is the \emph{only} possible renormalizable coupling between SM particles
    and new singlet fermions.  Observational constraints restrict dark matter
    from interacting too strongly with the Standard Model, but they also
    require that such interactions cannot be too weak, otherwise it would be
    difficult to understand how dark matter is produced in the early Universe.
    At the renormalizable level, there are only three possible ``portal''
    interactions: \cref{eq:neutrino-portal-1} for fermions, the ``kinetic
    mixing portal'' $\epsilon F_{\mu\nu} F'^{\mu\nu}$ for gauge bosons, and
    the ``Higgs portal'' $\lambda \, (\phi^\dag \phi) (H^\dag H)$ for scalars.
    (Here, $F^{\mu\nu}$ and $F'^{\mu\nu}$ are the field strength tensors of
    electromagnetism and of a new dark sector $U(1)'$ symmetry, respectively;
    $\phi$ is a dark sector scalar field; and $\epsilon$, $\lambda$ are
    dimensionless coupling constant.) This shows how generic the neutrino
    portal interaction is.

  \item Sterile neutrinos generically appear in models that explain the
    \textbf{smallness of neutrino masses}.  This is particularly true
    for the famous ``seesaw mechanism'' and its variants. We will dive deeper
    into this topic in \cref{sec:theory-motivation}.
\end{enumerate}

This review is organized as follows: in \cref{sec:sbl-anomalies}, we will
introduce various anomalies observed in neutrino oscillation experiments that
have sparked interest in eV-scale sterile neutrinos for more than two decades
already. We will then focus on more theoretical aspects in
\cref{sec:theory-motivation} and discuss sterile neutrinos from a
model-building point of view.  \Cref{sec:pheno} will be devoted to the rich
phenomenology of light sterile neutrinos in terrestrial experiments, in
particular oscillation experiments, precision measurements of beta decay
kinematics, and neutrinoless double beta decay searches.  We will see that the
oscillation anomalies introduced earlier cannot be consistently explained in
the simplest ``$3+1$'' sterile neutrino models, but we will also review several
scenarios involving decaying sterile neutrinos that have the potential to
alleviate this tension.  In \cref{sec:cosmology} we will review sterile
neutrino cosmology.  We will in particular show how Big Bang nucleosynthesis
(BBN), the Cosmic Microwave Background (CMB), and large scale structure (LSS)
set tight constraints on these models, ruling out the simple $3+1$ scenario,
but leaving some room for extended models. We then turn our attention to heavier sterile neutrinos. In \cref{sec:keVchapt} we review keV-scale sterile neutrino dark matter, and in \cref{sec:leptogenesis} the possibility to generate the observed matter--antimatter asymmetry due to heavy sterile neutrinos.

\chapter{Experimental Hints for Sterile Neutrinos?}
\label{sec:sbl-anomalies}

The possibility that more than three light neutrino species participate in
neutrino oscillations has been discussed for several decades already.  When a
deficit of atmospheric muon neutrinos was first observed by the Kamiokande
\cite{Hirata:1992ku, Fukuda:1994mc} and IMB \cite{Casper:1990ac,
BeckerSzendy:1992hq} experiments, the possibility that this deficit was due to
oscillations of muon neutrinos ($\nu_\mu$) into sterile neutrinos ($\nu_s$) was
widely discussed, see for instance refs.\ \cite{Akhmedov:1992mm, Liu:1997yb}.
Similarly, oscillations of active neutrinos into sterile neutrinos have also
been considered as an explanation of the solar neutrino problem \cite{Caldwell:1993,
Peltoniemi:1993ec, Fuller:1995tz, GomezCadenas:1995sj, Okada:1996kw,
Dvali:1998qy, Giunti:2000wt, GonzalezGarcia:2001uy}.

It has been firmly established that both atmospheric and solar neutrino
oscillations are due to transitions among active neutrino flavors
\cite{Fukuda:2000np, Ahmad:2001an}. However, several neutrino oscillation
anomalies remain, which cannot be explained in the standard three flavor
framework, thus keeping the experimental motivation for sterile neutrinos
alive.  In the following, we will discuss these
anomalies one by one, starting with the LSND and MiniBooNE experiments, which
employ accelerator-based neutrino beams, and continuing with experiments using
nuclear reactors and intense samples of radioactive isotopes as sources.
The discussion of these experiments will therefore also exhibit the main
experimental techniques employed to search for sterile neutrinos
in oscillations.

\section{LSND: Neutrinos from Stopped Pion Decay}
\label{sec:lsnd}

\begin{figure}
  \centering
  \includegraphics[width=\textwidth]{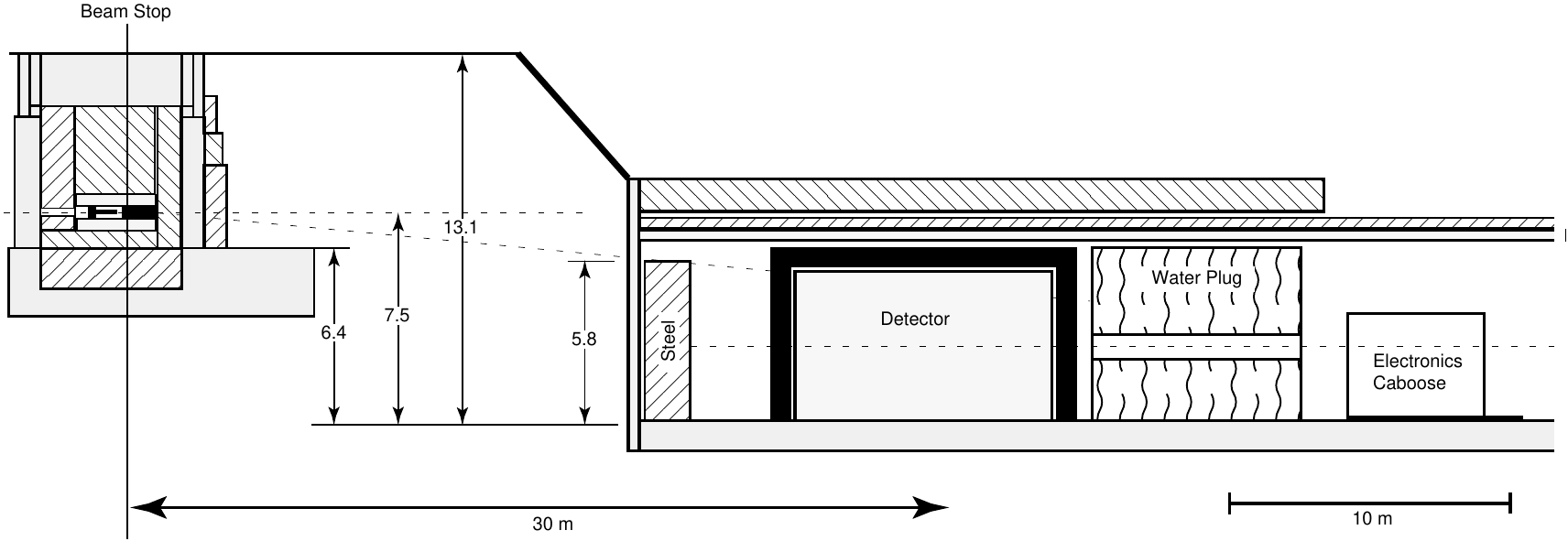}
  \caption{The layout of the LSND experiment. Figure taken from
    ref.~\cite{Athanassopoulos:1996ds}.}
  \label{fig:lsnd-layout}
\end{figure}

The Liquid Scintillator Neutrino Detector (LSND) \cite{Athanassopoulos:1996ds} had
been operating at Los Alamos National Laboratory from 1993 through 1998. As
schematically shown in \cref{fig:lsnd-layout}, the experiment consisted of
a cylindrical detector, filled with 167~metric tons of mineral oil
(chemical composition $\sim \text{CH}_2$), doped with \SI{0.031}{g/l} of scintillator.
The active target was surrounded by a veto system to suppress cosmogenic backgrounds.

The detector was exposed to the neutrinos produced at the Los Alamos Meson Physics
Facility (LAMPF).  The main neutrino production target (called the A6 target) consisted
of a water tank exposed to an $\sim \SI{800}{MeV}$ proton beam, followed by a copper beam stop.
In the beam stop, the positive pions ($\pi^+$) produced in the target were brought to a stop
before decaying mostly via $\pi^+ \to \nu_\mu + (\mu^+ \to e^+ \nu_e \bar\nu_\mu)$.
Negatively charged pions were rapidly absorbed by nuclei without producing neutrinos.
The neutrino detector was placed \SI{29.7}{m} downstream of the beam stop.  Besides the
primary A6 target, the LAMPF complex included also other targets (A1 and A2) used to
produce pion and muon beams for other experiments.  These targets, located upstream
of A6, also contribute to the neutrino flux observed by LSND, but only at the percent
level.

The LSND beam was compoased of $\nu_\mu$, $\bar\nu_\mu$, and $\nu_e$, but as
most $\pi^-$ were absorbed via the strong interaction before decaying, it
contained almost no electron anti-neutrinos ($\bar\nu_e$), making it
ideally suited for a $\bar\mu_\nu \to \bar\nu_e$ oscillation search.  $\bar\nu_e$
interactions ($p + \bar\nu_e \to n + e^+$) are tagged via the coincident observation
of the prompt positron signal and the delayed signal from neutron capture on another
proton, $n + p \to d + \gamma(\SI{2.2}{MeV})$.

\begin{figure}
  \centering
  \begin{tabular}{cc}
    \includegraphics[width=0.47\textwidth]{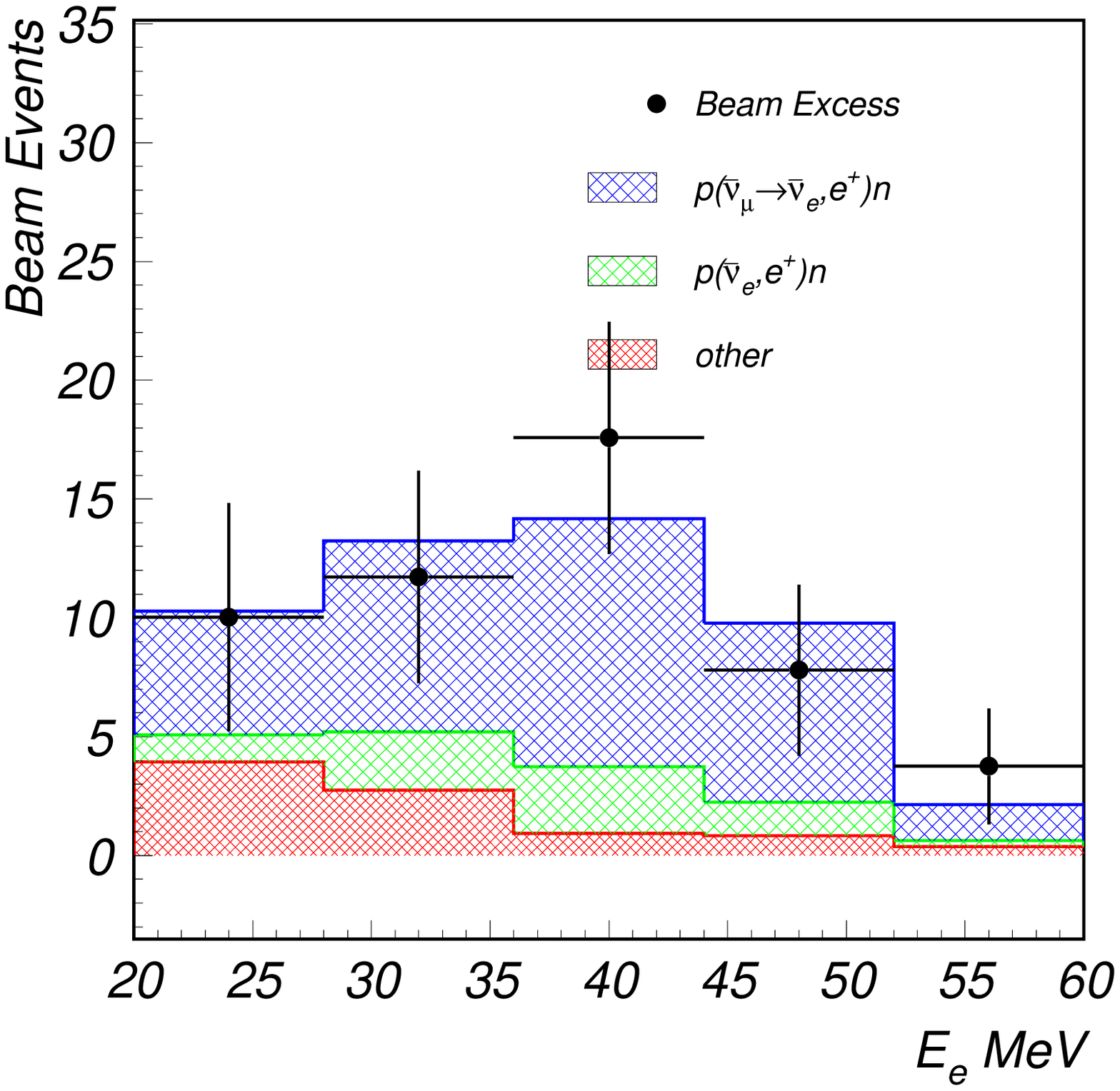} &
    \includegraphics[width=0.47\textwidth]{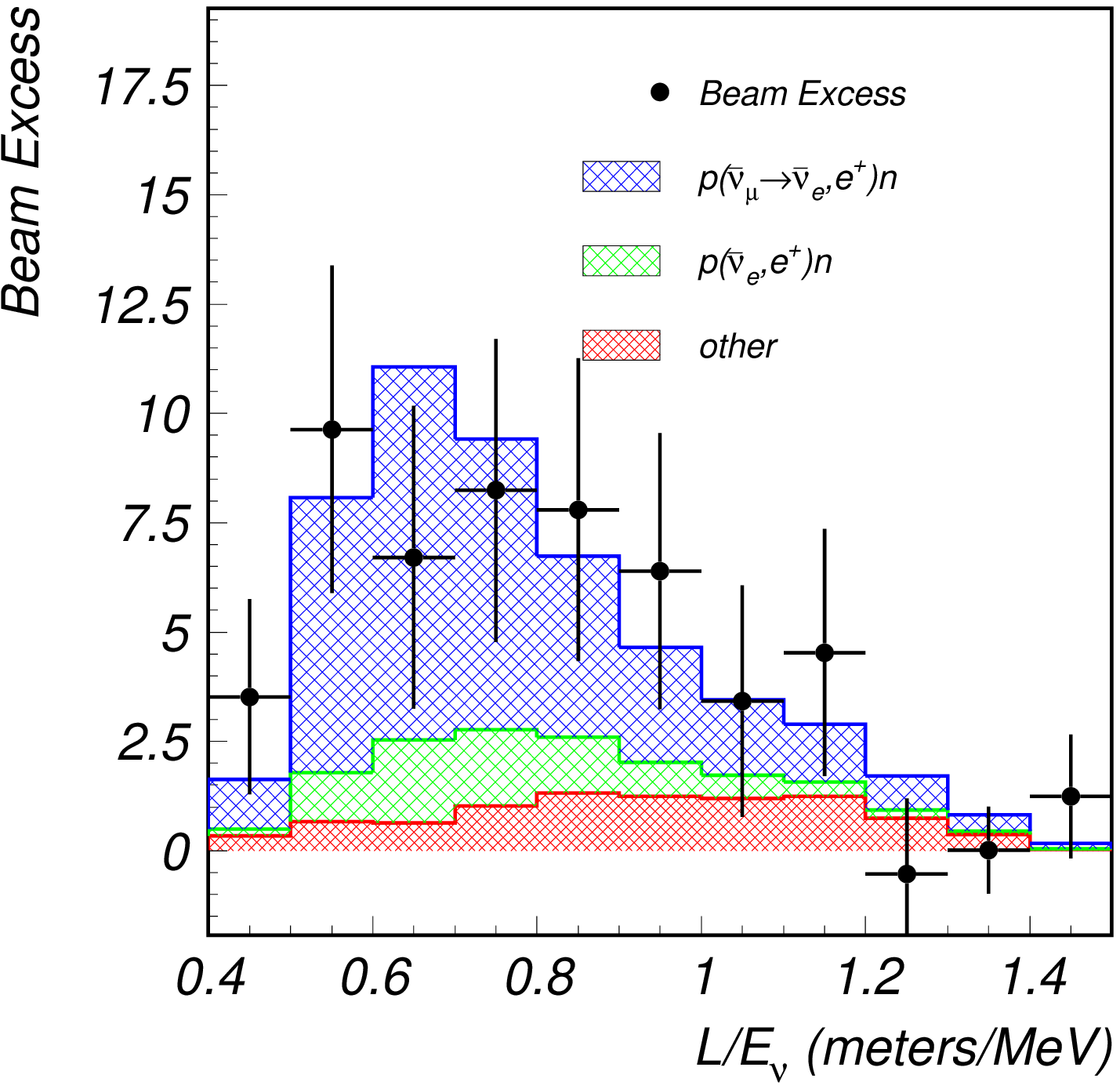}
  \end{tabular}
  \vspace{-0.2cm}
  \caption{The distribution of observed events in LSND as a function of
    positron energy $E_e$ (left) and as a function of $L/E_\nu$ (baseline over
    reconstructed neutrino energy, right). Green histograms show the
    background from the intrinsic $\bar\nu_e$ contamination in the beam (mostly
    due to the decay of $\mu^-$ from $\pi^-$ decay in flight), while red
    histograms indicate all other backgrounds, such as misidentified $\nu_\mu$
    and $\bar\nu_\mu$. Blue histograms correspond to the predicted signal
    from (2-flavor) neutrino oscillations. Figure taken
    from~\cite{Aguilar:2001ty}.}
  \label{fig:lsnd-data}
\end{figure}

The combined results from six measurement campaigns (1993--1998) are shown
in \cref{fig:lsnd-data} as a function of positron energy (left panel) and
as a function of $L/E_\nu$ (right panel). Here $L$ is the source--detector distance
(the baseline) and $E_\nu$ is the neutrino energy. The motivation for showing
the data as a function of this quantity lies in the fact that the phase of
neutrino oscillations evolves with $L/E_\nu$.  In fact, LSND observes
a significant ($3\sigma$) excess compared to the estimated background
(red and green histograms), and the most frequently entertained explanation
for this excess is in terms of oscillations between SM neutrinos and extra,
sterile neutrinos (blue histogram). However, as we will argue below, this
explanation is very difficult to reconcile with results from other experiments
so that, at the time of this writing, the LSND excess remains unexplained
more than 20~years after LSND stopped taking data.

The main backgrounds in LSND are due to the intrinsic
$\bar\nu_e$ contamination of the beam (arising mainly from the decay of
$\mu^-$ produced by $\pi^-$ that decay before being absorbed) and due to
the misidentification of $\nu_\mu$ and $\bar\nu_\mu$ as $\bar\nu_e$.

While no satisfactory explanation of LSND in terms of systematic errors
or Standard Model uncertainties has been found yet, it is nevertheless
instructive to consider a few explanation attempts, and the reasons for
their failure:
\begin{itemize}
  \item {\bf Decay in flight.} $\pi^-$ decaying in flight before they can be
    absorbed in matter are an obvious source of $\bar\nu_e$.  For this reason,
    the LSND collaboration has carefully modeled the LSND target hall
    (including also the aforementioned A1 and A2 targets). Thanks to the
    nearly hermetic shielding of this area, it is difficult to imagine
    where these estimates could have gone wrong.

  \item {\bf Accidental backgrounds.} The signature of a CC $\nu_e$ interaction,
    consisting of a positron and a neutron, can be mimicked by random
    coincidences between cosmic-ray induced neutrons and cosmic-ray induced
    electrons or positrons, or by random coincidence between
    cosmic-ray induced neutrons and electrons generated by CC $\nu_e$
    interactions. (Unlike $\bar\nu_e$, $\nu_e$ are abundant in the LSND beam.)
    However, these backgrounds can be estimated straightforwardly, and it
    turns out that they are negligible because the probability for
    both temporal and spatial coincidence between a random neutron
    and a random electron or positron is tiny.

  \item {\bf Knock-on neutrons.} If a $\nu_e$ interacts with a carbon
    nucleus inside the LSND detector via quasi-elastic scattering, it will
    convert into an electron and eject a proton.  On its way out of the
    nucleus, this proton may transfer its energy to a neutron, so that the
    final state will be $e^- + n$. This final state is indistinguishable
    from the $e^+ + n$ signature of a $\bar\nu_e$ interaction. As $\nu_e$ are
    abundant in the LSND beam, even a small probability for the production
    of such knock-on neutrons could be problematic. However, energetics
    come to the rescue: the energy required to liberate a neutron from
    a carbon nucleus is so high that events of this type would mostly fall
    below the \SI{20}{MeV} cut which LSND impose on the reconstructed
    positron energy.  The same is true for other nuclei that are present in
    the LSND scintillator (and are further suppressed due to their small
    abundance).

  \item {\bf Inhibition of $\pi^-$ capture.} As stated above, stopped
    $\pi^-$ do not contribute to LSND's neutrino flux because they
    are immediately absorbed by a nucleus. Unlike for $\pi^+$, there
    is no Coulomb barrier preventing such capture.  One may wonder
    how robust this statement is, and whether some $\mu^-$ could
    survive for long enough even after being stopped to contribute
    appreciably to the $\bar\nu_e$ flux.  This could happen for instance
    when $\mu^-$ are captured into an outer orbit of an atom,
    with a large orbital angular momentum.  The time it takes them to
    cascade down into an $s$-wave state from which they can be absorbed
    could be comparable to the muon lifetime.  However, more detailed
    estimates show that the fraction of $\mu^-$ for which  this might
    happen is too small to explain the LSND anomaly. This also because
    the number of parent $\pi^-$ that are produced in the LSND target
    is about an order of magnitude smaller than the number of $\pi^+$
    \cite{Athanassopoulos:1996ds}.
\end{itemize}

\section{MiniBooNE: A Horn-Focused Neutrino Beam}
\label{sec:mb}

The MiniBooNE experiment at Fermilab has been designed to independently test the
possibility that the LSND anomaly is explained by neutrino oscillations.\footnote{The
prefix ``Mini'' in the name refers to fact that this is a downscaled version of a
proposed experiment called ``BooNE'' (Booster Neutrino Experiment), which would
have consisted of two similar detectors placed at different baselines. In hindsight,
the second detector could have saved the community a lot of trouble. Had it been
built, the present review article might not exist.  Critics have pointed out that
the evolution of short-baseline experiments (BooNE $\to$ MiniBooNE $\to$ MicroBooNE),
when compared to the evolution of long-baseline experiments (Kamiokande $\to$
SuperKamiokande $\to$ HyperKamiokande) suggests that short-baseline oscillation
physics is moving in the wrong direction.}
Instead of a stopped pion source it employs a neutrino beam produced by pions
decaying in flight. More precisely, neutrinos are produced by directing the
\SI{8}{GeV} Booster beam at Fermilab onto a beryllium target, where
pions are copiously produced. A magnetic horn focuses pions of one polarity
in the forward direction, while simultaneously defocusing pions of the
opposite polarity. Pions are then allowed to decay in a decay tunnel, producing
a beam consisting dominantly of $\nu_\mu$ (if the horn polarity is such that
$\pi^+$ are focused) or $\bar\nu_\mu$ (if the horn polarity is such that
$\pi^-$ are focused). In the rock between the end of the decay tunnel and
the detector, muons from pion decay as well as any other beam remnants are
stopped to prevent them from contaminating the measurement.
The resulting neutrino covers the energy range
from $\sim \SI{200}{MeV}$ to $\sim \SI{1.5}{GeV}$ \cite{AguilarArevalo:2008yp}.

The detector, located \SI{541}{m} downstream of the primary pion production
target, is a spherical vessel containing \SI{818}{tons} of mineral oil
\cite{AguilarArevalo:2008qa}.  High energy charged particles lead to the
emission of both scintillation light and \v{C}erenkov radiation, which
is observed by photomultiplier tubes (PMTs) covering the surface of the detector.
The detector has been optimized to identify charged current (CC) $\nu_e$ and $\bar\nu_e$
interactions, whose signature is a single electron or positron, as well
as CC $\nu_\mu$ and $\bar\nu_\mu$ interactions, identified by the single
muon in the final state.

\begin{figure}
  \centering
  \begin{tabular}{cc}
    \includegraphics[width=0.48\textwidth]{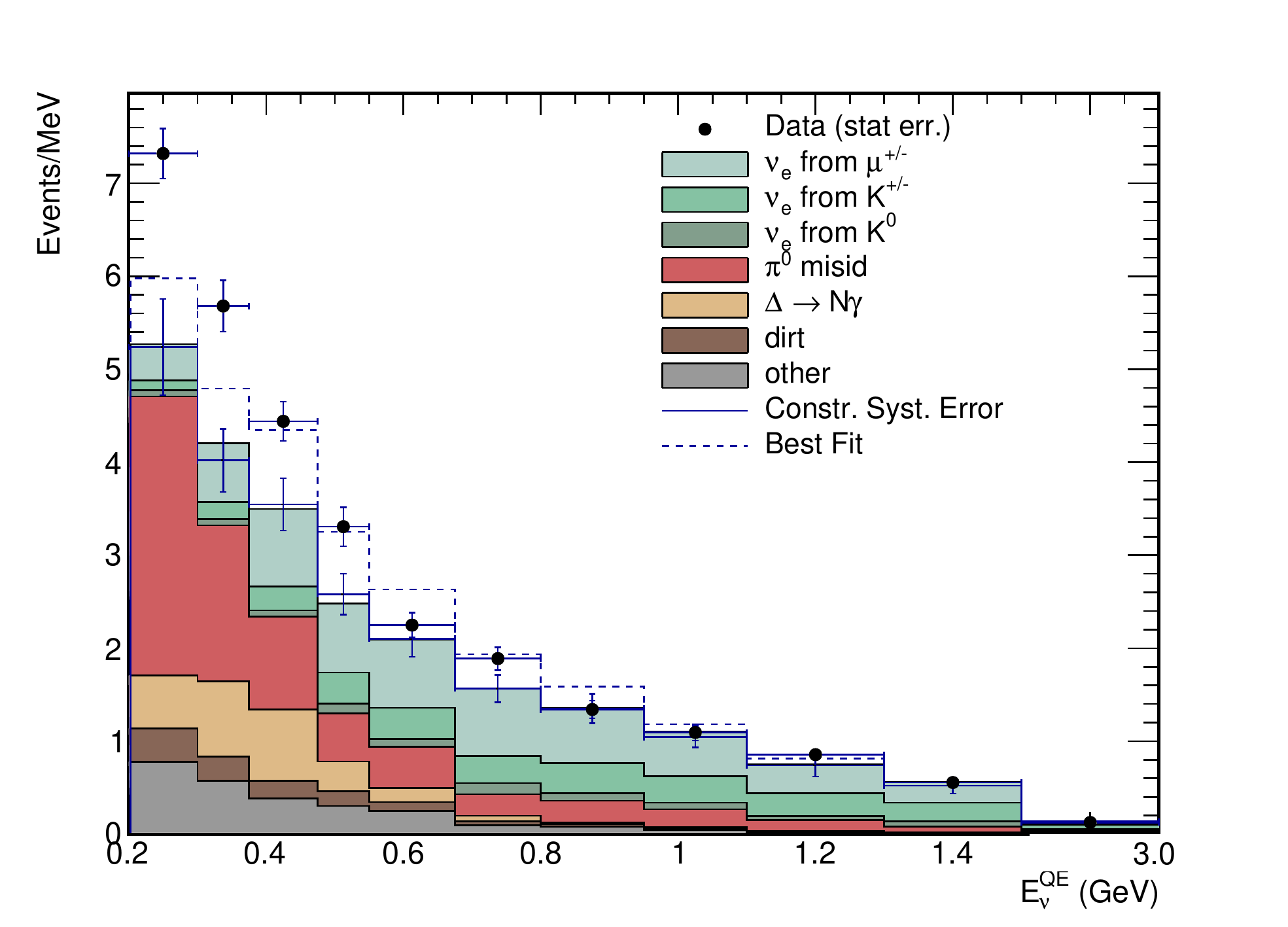} &
    \includegraphics[width=0.48\textwidth]{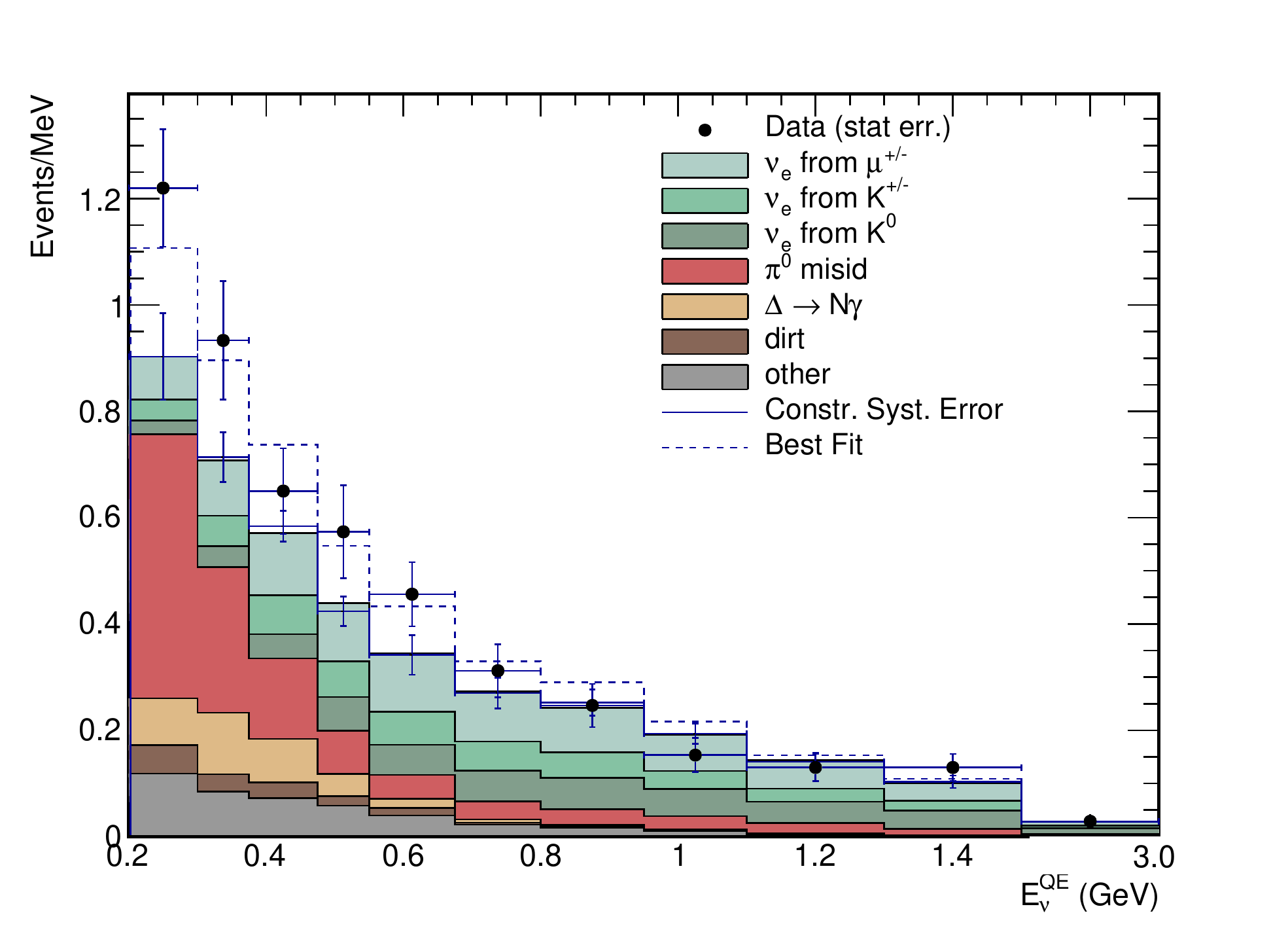}
  \end{tabular}
  \caption{MiniBooNE observed event spectra in neutrino mode (left) and in
    anti-neutrino mode (right). Black data points correspond to the observed
    event rates, shaded histograms indicate the (largely data-driven) background
    predictions, and the blue dashed histogram illustrates the best oscillation
    fit. Plots taken from refs.~\cite{Aguilar-Arevalo:2020nvw} (left) and
    \cite{Aguilar-Arevalo:2018gpe} (right).}
  \label{fig:mb-data}
\end{figure}

MiniBooNE has collected a neutrino flux corresponding to $18.75 \times 10^{20}$
protons on target in neutrino mode ($\pi^+$ focused) and $11.27 \times 10^{20}$
protons on target in anti-neutrino mode ($\pi^-$ focused).  The resulting
event spectra and predicted backgrounds are shown in \cref{fig:mb-data}.
A clear excess is observed, with a significance of $4.7\sigma$ in neutrino mode
and of $4.8\sigma$ when neutrino and anti-neutrino mode data are combined
\cite{Aguilar-Arevalo:2020nvw}. Like the LSND excess, the MiniBooNE
excess can be explained in terms of active-to-sterile neutrino
oscillations when considered in isolation (blue dashed histograms in
\cref{fig:mb-data}), but this explanation runs into severe difficulties
when fitted together with data from other experiments (see \cref{sec:4f-osc}).

While backgrounds in MiniBooNE are manifold, many background contributions
can be estimated using data-driven methods.  An irreducible source of background
arises from the intrinsic contamination of the beam by $\nu_e$ and $\bar\nu_e$
produced in kaon or muon decays (turquoise histograms in \cref{fig:mb-data}.
Kaons are produced in the primary target,
and albeit the production rate is much lower than for pions, their larger
branching ratio to $\nu_e$ / $\bar\nu_e$ makes them non-negligible.  Muons
are produced in pion decay, and while most of them are stopped in the rock
separating the decay tunnel from the detector, some of them decay already
in the decay tunnel, leading to an extra flux of high energy neutrinos.
The intrinsic $\nu_e$ / $\bar\nu_e$ contamination of the MiniBooNE beam
was determined based on external measurements and on data from the SciBooNE
detector located in the same beam, upstream of MiniBooNE~\cite{Cheng:2011wq}.

A second source of background arises from $\pi^0$ production in neutral current
(NC) neutrino interactions, followed by the decay $\pi^0 \to \gamma\gamma$ (red
histograms in \cref{fig:mb-data}).  As MiniBooNE cannot distinguish
photons from electrons or positrons, this process can mimic the signal if one
of the photons leaves the fiducial detector volume before converting to
a visible $e^+ e^-$ pair, if one photon is absorbed by a nucleus before converting, or if the
two photons are too close to each other to be separately identified.
To make the last condition more precise, note that, to reject $\pi^0$ events
MiniBooNE reconstructs candidate events assuming a two-photon topology, and
then imposes an invariant-mass cut on the two reconstructed
photons~\cite{Aguilar-Arevalo:2018gpe}.  The $\pi^0$
background is predicted based on the $\pi^0$ production cross section
measured in situ in events in which the two photons are successfully
separated.  Therefore, the $\pi^0$ background prediction is independent
of the large uncertainties in the $\pi^0$ production cross section.

Similar to the $\pi^0$ background, also the background from the NC production
of a $\Delta$ resonance, followed by the decay of that resonance via photon
emission, can be estimated from data. In particular, this is done using
the much more numerous events in which the $\Delta$ resonance instead emits
a pion upon decaying.

Finally, MiniBooNE expects a few background events from neutrino interactions
outside the active detector volume (``dirt background'').

Let us once again discuss several proposed explanations of the MiniBooNE
anomaly within the Standard Model:
\begin{itemize}
  \item {\bf Neutral pions.}  If the NC $\pi^0$ background was larger than
    predicted by roughly a factor two, it could explain
    the observed excess quite well.  Given that the prediction for the $\pi^0$
    background is data-driven, this seems unlikely. In particular, it would
    imply that MiniBooNE's detector simulation would need to be off by a factor
    two when estimating the probability that the two photons from $\pi^0$ are
    not separately reconstructed.  In this case, one would expect most excess
    events to originate close to the boundary of the fiducial volume as this is
    where the probability for missing one of the two photons from $\pi^0 \to
    \gamma\gamma$.  MiniBooNE have recently published the spatial distribution
    of excess events in ref.~\cite{Aguilar-Arevalo:2020nvw}, showing that the
    excess events are instead evenly distributed throughout the detector.

  \item {\bf Decays of heavier mesons to photons.} In principle, heavier neutral
    mesons such as the $\eta$ contribute to di-photon production in the detector
    in the same way as $\pi^0$'s do. Once again, if the decay products are reconstructed
    as a single photon -- either because one of them is lost, or because their
    invariant mass is small -- this could mimic a $\nu_e$ signal. However, note
    that for decays of heavy particles like the $\eta$, the likelihood of the
    two photons having a small invariant mass is much smaller.

  \item {\bf Single photons.} As mentioned above, the largest contribution to
    single-photon events is due to production of an $s$-channel $\Delta(1232)$
    resonance.  By far the dominant decay mode of the $\Delta(1232)$ is to a pion and
    a nucleon, so the rate of $\Delta(1232)$ production can in principle be measured in
    single pion events. There are, however, several complications: first,
    this measurement is not background-free as there are other processes that
    can yield single pions, such deep-inelastic scattering events or the
    decays of heavier resonances. These must be separated from true $\Delta \to N \pi$
    events.  Second, pions produced inside a nucleus can suffer final-state
    interactions as they leave the nucleus. These interactions can either lead
    to the re-absorption of the pion, or to the excitation of additional
    resonances.  When neglected, both effects would lead to an under-prediction
    of the single-photon background to MiniBooNE's $\nu_e$ appearance search
    \cite{Ioannisian:2019kse, Giunti:2019sag}:
    the first one because the number of observed $\Delta \to N \pi$ decays would
    be lower by an $\mathcal{O}(1)$ factor than the true rate of $\Delta$ production;
    the second one because the excitation of a secondary resonance due to final-state
    interactions increases the overall probability that one of the resonances
    decays to a photon. Therefore, MiniBooNE carefully model these effects
    \cite{Aguilar-Arevalo:2020nvw}.
    Comparisons with ab~initio calculations suggest that this modeling
    is roughly correct~\cite{Hill:2010zy, Lalakulich:2012cj, Alvarez-Ruso:2016ikb}.

    It is worth mentioning that the branching ratio of the
    $\Delta$ resonance to photons -- a crucial ingredient to MiniBooNE's
    data-driven background estimate -- has never been directly measured.
    Instead, it is calculated based on the measured photo-excitation rate of
    the resonance.  The error on this branching ratio is therefore of order
    20\% (the Particle Data Group quotes the range 0.55--0.65\% for $\BR(\Delta
    \to N\gamma)$ \cite{Tanabashi:2018oca}). The errors on the radiative decay
    modes of heavier hadronic resonances are even larger.  To the best of our
    knowledge, the impact of these uncertainties on MiniBooNE's background prediction
    has never been studied.

    Finally, radiative decays of heavy hadronic resonances are not the only
    source of single-photon events in MiniBooNE. Another important
    contribution is initial and final-state radiation in NC neutrino--nucleon
    scattering, which, however, appears to also be too small to explain the
    excess~\cite{Hill:2010zy, Wang:2013wva, Wang:2014nat, Garvey:2014exa,
    Wang:2015ivq}.  
\end{itemize}

\section{The Reactor Anomaly: Neutrinos from Nuclear Reactors}
\label{sec:reactor}

\subsection{Reactor Neutrino Fluxes}
\label{sec:reactor-fluxes}

Nuclear reactors have been used as neutrino sources ever since the
Reines \& Cowan experiment that first provided experimental evidence
for the existence of neutrinos \cite{Cowan:1956a, Cowan:1956b}.
Besides the long-baseline experiments Chooz, Double Chooz, RENO,
Daya Bay, and KamLAND that contributed to the precision
measurement of the neutrino oscillation parameters \cite{Apollonio:2002gd,
KamLAND:2008ee, Abe:2014bwa, An:2016ses, RENO:2015ksa}, a large number
of experiments was and is still being carried out at short baselines ($\lesssim
\SI{100}{m}$), where no oscillations are expected in the standard three-flavor
oscillation framework. (Of course, the values of the standard three-flavor
  oscillation parameters were not known yet at the time many of these
experiments were done.)

\begin{figure}
  \centering
  \includegraphics[width=\textwidth]{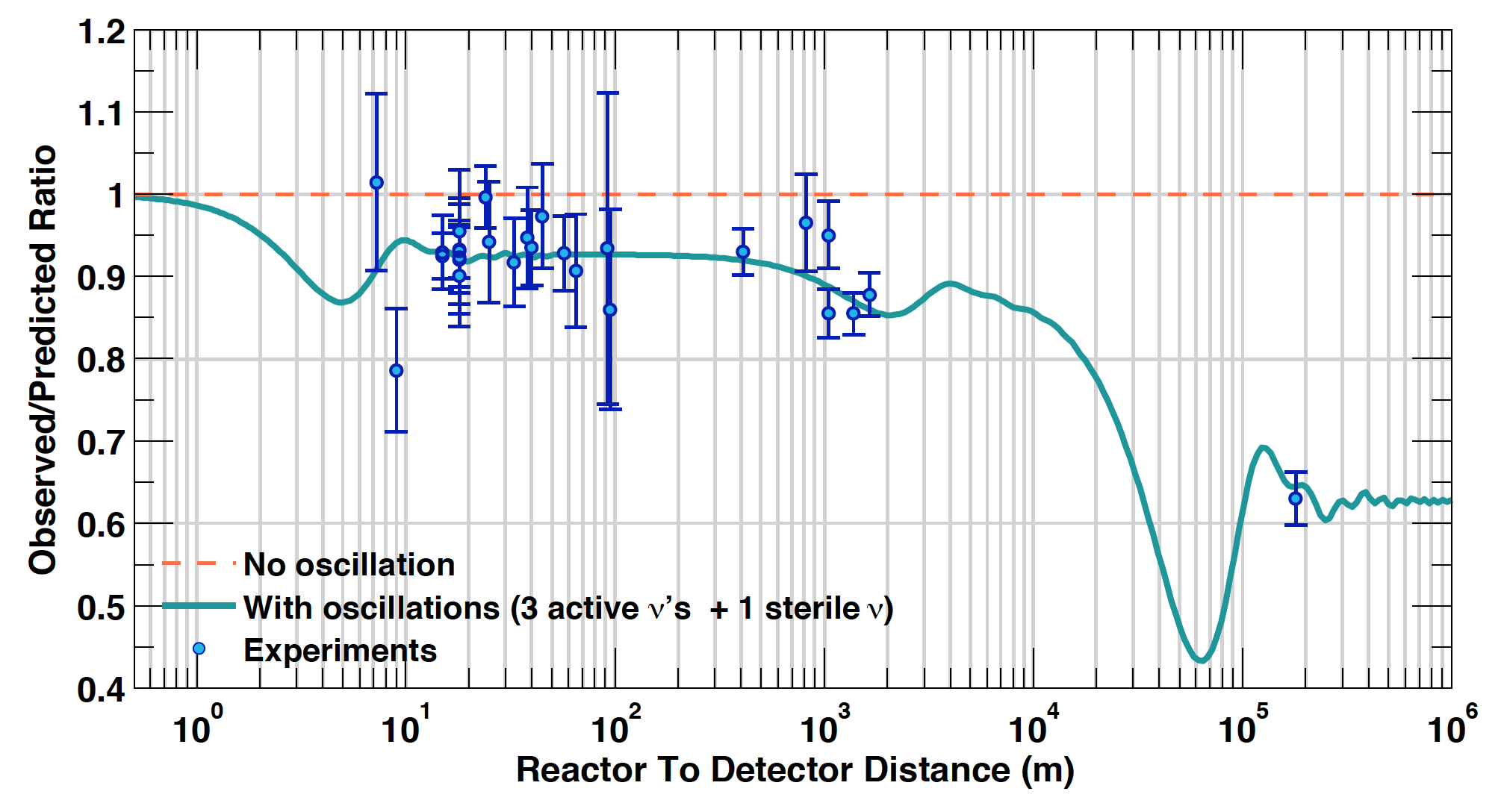}
  \caption{Comparison of measured and predicted $\bar\nu_e$ event rates from
    nuclear reactors.  Results from a long list of past and
    present experiments are shown as a function of baseline. The event deficit
    compared to the prediction is clearly visible.  Figure courtesy of
    Thierry Lasserre (updated from ref.~\cite{Mention:2011rk}).}
  \label{fig:raa}
\end{figure}

Until 2011, results from such short-baseline reactor experiments were in good
agreement with the then state-of-the-art theoretical predictions by
Schreckenbach et al.\ \cite{VonFeilitzsch:1982jw, Schreckenbach:1985ep,
Hahn:1989zr}.  However, they are in disagreement with newer, more detailed,
theoretical predictions by Mueller et al.\ \cite{Mueller:2011nm} and by Huber
\cite{Huber:2011wv}.  This few percent disagreement, which has a significance
of about $3\sigma$, has become known as the reactor anti-neutrino anomaly
\cite{Mention:2011rk}.  The effect is illustrated in \cref{fig:raa}, which
compares the measured reactor neutrino event rates to the prediction.

As the Huber and Mueller predictions lead to similar results even though they
are based on different theoretical approaches, and as they address several
well-known shortcomings of the Schreckenbach et al.\ predictions, there is
general agreement on the fact that the Huber/Mueller predictions are indeed
superior to their predecessors.  However, it is still unclear whether they are
accurate enough to make authoritative statements at the few percent level
\cite{Fallot:2012jv, Hayes:2013wra, Dwyer:2014eka, Fang:2015cma, Hayes:2015yka,
  Huber:2015ouo, Novella:2015eaw, Ankowski:2016oyj, Huber:2016fkt,
  Hayes:2016qnu, Giunti:2016elf, An:2017osx, Giunti:2017nww, Hayes:2017res,
  Giunti:2017yid, Gebre:2017vmm, Sonzogni:2017voo, Mohanty:2017iyh, Ma:2018izf,
Giunti:2019qlt, Li:2019quv, Adey:2019ywk, Estienne:2019ujo, Hayen:2019eop}.
One reason for concern is that none of the flux predictions successfully
accounts for an observed ``bump'' in the spectrum around \SI{5}{MeV}
\cite{An:2015nua,An:2016srz,Abrahao:2017zvl,Kwon:2017nzf}.  This highlights
the fact that there are features in the reactor spectrum that are still poorly
understood and raises the question to what extent they can be trusted in
other parts of the spectrum.

The main complication in predicting reactor neutrino fluxes is the vast number
of beta decay channels ($\sim 6\,000$) that contribute to the final spectrum.
Many of these decays involve very short-lived, neutron-rich isotopes
which cannot be studied in laboratory experiments.  Consequently, their
properties (decay energy, spin and parity, etc.) are poorly known.
The problem is exacerbated by the fact that the most short-lived (and therefore
least understood) nuclei tend to have the largest $Q$-values and therefore
produce neutrinos with the highest energies. As the cross section for
the detection reaction in a reactor neutrino experiment, $\bar\nu_e + p \to e^+ + n$,
increases quadratically with the neutrino energy and has a threshold at $E_\nu
\simeq \SI{1.8}{MeV}$, these neutrinos contribute
disproportionately to the observed event rate.  To deal with the fact
that not all beta decays relevant to reactor neutrino spectra are
well-understood, Mueller et al.\ begin their calculation by determining
the \emph{electron} spectrum from those beta decays on which reliable nuclear
data is available.  The resulting electron spectrum is then compared to the
measured electron spectrum from uranium and plutonium fission.  The difference --
which is presumed to be due to additional, not well understood, beta decays,
is then parameterized in terms of a few ``effective'' beta decays.
Each effective beta decay is described by an allowed beta decay spectrum
whose parameters are determined by a fit to the data \cite{Mueller:2011nm}.
Once these parameters have been determined, it is also possible to compute
the neutrino spectra from the effective decays.
The calculation by Huber relies entirely on fitted effective beta decays,
resorting to nuclear data tables only for quantities whose uncertainties
are relatively small such as effective nuclear charges~\cite{Huber:2011wv}.

One source of uncertainty in all calculations is the
contribution from non-unique first forbidden beta decays, that is from
decays whose initial and final states do not obey spin and parity
selection rules and can only occur when the final state particles
carry non-zero orbital angular momentum, when extra photons are emitted,
or when the finite nuclear size is taken into account.
The label ``non-unique'' implies that several different QFT operators
(with different Lorentz structures) contribute to the decay amplitude.
As the relative importance of these different operators is crucial 
for determining the neutrino spectrum, but is typically not
known, a large uncertainty ensues. The standard procedure is therefore
to treat non-unique forbidden decays as if they were allowed decays, but
to assign them a systematic uncertainty of 100\%.  The problem is that
there is no guarantee that the error is not much larger than this.

\subsection{Measurements of Isotope-Dependent Fluxes}
\label{sec:reactor-isotopes}

A promising method for resolving the reactor neutrino anomaly has been
pioneered by the Daya Bay Collaboration \cite{An:2017osx}, and has
been exploited also by RENO~\cite{RENO:2018pwo}. The method is based
on the fact that nuclear reactor fuel contains four main fissile isotopes:
\iso{U}{235}, \iso{U}{238}, \iso{Pu}{239}, and \iso{Pu}{241}. Each of them
has its own spectrum of primary fission products and therefore its own
neutrino spectrum.   In Daya Bay,
the dominant contribution to the neutrino flux comes from the decay chains
of the fission products of \iso{U}{235}, followed by \iso{Pu}{239};
\iso{U}{238} and \iso{Pu}{241} are subdominant. By exploiting the
fact that the isotopic composition of the fuel changes with time --
an effect known as burn-up -- Daya Bay is able to measure the neutrino
spectra from the four fissile isotopes separately. $\bar\nu_e$
oscillations into sterile states, as well as most other new-physics explanations
of the anomaly, would lead to a similar deficit of
observed events for all four fissile isotopes. If, on the other hand,
the reactor anomaly is due to misprediction of fluxes, it is
unlikely that the systematic bias is the same for all isotopes, hence
different deficits would be expected for different  isotopes.

Indeed, Daya Bay and RENO find 
that the rate of neutrino events from \iso{U}{235} is 7.8\% ($2\sigma$)
smaller than predicted, while the rate of \iso{Pu}{239}-induced events
is in excellent agreement with the prediction.  Taken at face value, this
result disfavors the sterile neutrino explanation of the reactor anomaly.
However, it should be taken with a grain of salt as the analysis presented
in \cite{An:2017osx} neglects two important effects: non-equilibrium
corrections and non-linear isotopes.

Non-equilibrium corrections arise
from beta decay chains that involve isotopes whose lifetimes are comparable
to or longer than the typical reactor fuel cycle of $\sim \SI{3}{yrs}$.
An example is the decay chain $\iso{Sr}{90} \to \iso{Y}{90} \to \iso{Zr}{90}$.
The first of these decays has a half life of \SI{29}{yrs}, which means that
production and decay of \iso{Sr}{90} will never reach equilibrium
in a nuclear reactor. Throughout the fuel cycle, the \iso{Sr}{90} abundance
will gradually increase, and so will the neutrino flux from its decay.
(Note that neutrinos from $\iso{Sr}{90} \to \iso{Y}{90}$ are unobservable in
Daya Bay because their endpoint energy of \SI{0.55}{MeV} is below the
threshold of inverse beta decay (the detection reaction in Daya Bay) at
\SI{1.8}{MeV}. The neutrinos from the decay
$\iso{Y}{90} \to \iso{Zr}{90}$ (half-life \SI{2.7}{days}) are, however,
observable.)

Non-linear isotopes are isotopes that can ``escape'' their normal beta decay
chain by capturing a neutron \cite{Huber:2015ouo}. An example is the decay chain
$\iso{Zr}{99} \to \iso{Nb}{99} \to \iso{Mo}{99} \to \iso{Tc}{99}$. Here, neutron
capture on \iso{Mo}{99} converts it to the stable isotope \iso{Mo}{100},
while neutron capture on the very long-lived
\iso{Tc}{99} ($t_{1/2} = \SI{2e5}{yrs}$) produces the short-lived \iso{Tc}{100}.
The resulting modifications to the neutrino flux depend on the neutron flux in
the nuclear reactor, and therefore on burn-up.

The analysis in ref.~\cite{An:2017osx} includes a time-averaged non-equilibrium
correction, but neglects the time-dependence.  Non-linear isotopes are not
considered at all.  This has been improved upon in ref.~\cite{Adey:2019ywk},
where again a larger deficit is found for neutrinos originating from \iso{U}{235}
fission than for those from \iso{Pu}{239} fission.  However, within the
quoted uncertainties, the results of ref.~\cite{Adey:2019ywk} are also
compatible with an equal flux deficit for all isotopes.

\subsection{Unexplained Spectral Features}
\label{sec:danss-neos}

Because of the unexplained reactor neutrino anomaly, the focus of reactor neutrino
physics in recent years has increasingly shifted towards analyses and experiments
which are independent of theoretical flux predictions. This is achieved by
comparing the fluxes measured at a given baseline to measurements at a different
baseline, either from the same experiment or from a different one.

One experiment claiming a signal using this approach is
Neutrino-4~\cite{Serebrov:2018vdw, Serebrov:2020kmd}.
Using a segmented and movable detector, this experiment is able to
measure neutrino fluxes and spectra at distances from \SI{6.4}{m} to \SI{11.9}{m}
from the compact core of a \SI{100}{MW} research reactor.  Neutrino-4 claim a
$4.6\sigma$ preference for $\bar\nu_e$ disappearance into a sterile neutrino,
based on an analysis of the $\bar\nu_e$ event rate on a $9 \times 23$ grid
in energy $E$ and baseline $L$.  In each energy bin, event rates are normalized
to the distance-averaged rate in that bin, making the analysis largely
independent of the theoretical flux prediction.  Neutrino-4 results
have been criticized
in refs.~\cite{Almazan:2020drb, Coloma:2020ajw}, highlighting in particular the fact
that the log-likelihood
distribution in an experiment like Neutrino-4 does not obey Wilk's
theorem, as assumed by the Neutrino-4 collaboration. It is therefore
likely that the statistical significance of the Neutrino-4 result has
been overestimated.  The authors of ref.~\cite{Almazan:2020drb}
also list a number of possible systematic effects that could mimic
an oscillation signal. The Neutrino-4 collaboration has
replied to this criticism in ref.~\cite{Serebrov:2020yvp}.

STEREO and PROSPECT have also released results from their own
searches for sterile neutrinos at short baseline~\cite{AlmazanMolina:2019qul,
Andriamirado:2020erz}.  Both experiments are installed at research reactors,
and both of them used fixed (non-movable) detectors which, however, offer
spatial resolution in order to track possible oscillations as a function
of both baseline and energy. Neither experiment reports a signal,
and both of them rule out the Neutrino-4 best-fit point as well as
a large portion of Neutrino-4's $2\sigma$ preference region, though
not all of it.

Limits have also been set by the NEOS experiment~\cite{Ko:2016owz}, who
normalize their data to the neutrino spectrum measured at Daya Bay~\cite{An:2016srz}
to be largely independent of theoretical predictions. and by the DANSS
experiment~\cite{Danilov:2019aef} using a movable detector at a power reactor.
The sensitivity of NEOS and DNASS is worse than that of STEREO, PROSPECT, and
Neutrino-4 at the Neutrino-4 best fit $\Delta m_{41}^2 \sim \SI{7}{eV^2}$, but
better at lower $\Delta m_{41}^2$.  Note that a hint for an oscillation pattern
visible in earlier DANSS data \cite{danss-moriond17, danss-solvay17} has now
disappeared.

\section{The Gallium Anomaly: Neutrinos from Intense Radioactive Sources}
\label{sec:gallium}

The reactor anomaly becomes even more intriguing when
viewed in the context of yet another anomaly, coming from experiments with
neutrinos from intense radioactive source.  Measurements using such
sources were carried out in the 1990s to demonstrate the performance of
the radiochemical detection method for solar neutrinos. The experiments
in question, GALLEX~\cite{Hampel:1997fc, Kaether:2010ag} and
SAGE~\cite{Abdurashitov:1998ne, Abdurashitov:2005tb}, used gallium
as the active component in the target material are therefore collectively
referred to as ``gallium experiments''. A total of four measurement
campaigns have been carried out, and in all of them the observed number
of events falls below the expectation. When combined, this deficit
has a significance of slightly below $3\sigma$~\cite{Acero:2007su,
Giunti:2010zu, Kostensalo:2019vmv, Kostensalo:2020hbc}.

\chapter{Theoretical Models with Sterile Neutrinos}
\label{sec:theory-motivation}

The simplest theoretical framework for interpreting the various short-baseline
oscillation anomalies discussed in \cref{sec:sbl-anomalies} is the so-called
``$3+1$'' scenario, which augments the Standard Model by one additional
neutrino flavor.  The corresponding fourth mass eigenstate is assumed to be
of $\mathcal{O}(\text{eV})$ to allow the fourth neutrino to participate in
oscillations.  As the LEP measurement of the invisible decay width of the $Z$
boson constrains the number of light, weakly interacting neutrino species
to be three \cite{ALEPH:2005ab,Tanabashi:2018oca}, the fourth neutrino flavor
cannot couple to SM weak interactions, so it must be a sterile neutrino.
Of course, in the same spirit, the model can also be extended by several new
neutrino flavor eigenstates. Scenarios of this type are referred to as
$3+n$ models. In this section, we will discuss how sterile neutrinos
naturally appear in many extensions of the Standard Model, and how they
can be embedded in others.  In doing so, we address critics who claim
that neutrino physics is merely about solving three-dimensional eigenvalue
problems. In fact, we are able to handle matrices much
larger than $3 \times 3$.

\section{The Neutrino Portal}
\label{sec:neutrino-portal}

It is often argued that the observation of neutrino oscillations -- and thus
of non-zero neutrino mass -- is the first evidence for physics beyond the
SM of particle physics. Indeed, in the Lagrangian of the original
SM, only left-handed neutrino fields appear.  A single Weyl
spinor was thus sufficient to describe each neutrino flavor, whereas for all other fermions,
two Weyl spinors (corresponding to left-handed and right-handed polarizations)
were required.  Introducing neutrino masses suggests including new
Weyl fermions $N^\beta$, to allow for a Yukawa coupling of the form%
\footnote{We use here two-component notation for the fermion fields, i.e.\ each of
the two $SU(2)$ components of $L^\alpha$, as well as $N^\beta$, should be
interpreted as a two-component spinor, which transforms as $(\tfrac{1}{2},0)$
under the Lorentz group. The contraction of two such two-component
spinors $\psi$ and $\chi$ is defined as $\psi\chi \equiv \psi^a \chi_a \equiv
\psi_a \varepsilon^{ab} \chi_b$, where $a$, $b$ are spinor indices
and $\varepsilon^{ab}$ is the totally antisymmetric tensor in two
dimensions.}
\begin{align}
  \mathcal{L}_\text{Yukawa} \supset
    -y^{\alpha\beta} (i\sigma^2 H^*) L^\alpha N^\beta + h.c.\,,
  \label{eq:L-Yukawa}
\end{align}
where $H$ is the SM Higgs doublet, $L^\alpha = (\nu^\alpha, e^\alpha)^T$
are the SM lepton doublets, $\sigma^2$ is the second Pauli matrix,
and $\alpha$, $\beta$ are flavor
indices. Summation is implied over $SU(2)$ indices of $H$ and $L$, as well as
over $\alpha = e, \mu, \tau$ and $\beta = 1\dots
n$, with $n$ being the number of $N^\beta$ fields. Note that $n$
can be different from three -- the only requirement is $n \geq 2$,
as otherwise there would be two or more exactly massless neutrino states left,
in conflict with the observed oscillation patterns. After the Higgs
field acquires its vacuum expectation value $v$, the Yukawa couplings from
\cref{eq:L-Yukawa} yield mass terms of the form
\begin{align}
  \mathcal{L}_\text{mass} \supset
    -M_D^{\alpha\beta} \, \nu^\alpha N^\beta + h.c. \,,
  \label{eq:nu-s-mass-mixing}
\end{align}
with $M_D^{\alpha\beta} \equiv y^{\alpha\beta} v / \sqrt{2}$. (The subscript
$D$ stands for Dirac mass term.)
Looking at the transformation properties of the fields appearing in
\cref{eq:L-Yukawa} under the SM gauge symmetries, it straightforward to verify
that $(i\sigma^2 H^*)$ and $L^\alpha$ both transform as doublets under $SU(2)_L$
and carry opposite hypercharges of $+\tfrac{1}{2}$ and $-\tfrac{1}{2}$,
respectively.  Thus, the combination $(i\sigma^2 H^*) L^\alpha$ is a total
singlet under the SM symmetries, and so the $N^\beta$ must be total singlets
as well.  The $N^\beta$ are therefore called ``sterile neutrinos'' --
``neutrinos'' because they do not couple to the strong and electromagnetic
interactions, and ``sterile'' because they also do not couple to the weak
interaction. In this context, the SM neutrino fields $\nu^\alpha$ are
referred to as ``active neutrinos''.

We see that a sterile neutrino is a fairly generic addition to the SM
or to most of its extensions: it is just a fermion that is uncharged any
any of the SM gauge groups.  Sterile neutrinos are therefore commonplace
in models with ``dark'' or ``hidden'' sectors. This includes many models
of dark matter, even though in these models, the coupling constants
$y_{\alpha\beta}$ are often taken to be negligibly small.
In fact, a coupling to the operator $(i\sigma^2 H^*) L^\alpha$ is the only
direct renormalizable coupling that a SM singlet fermion can have with the SM.
This type of operator is therefore also called the ``neutrino portal''.

\section{Sterile Neutrinos and the Seesaw Mechanism}
\label{sec:seesaw}

In the following, we introduce a number of neutrino mass models that contain
sterile neutrinos. We begin with the generic type-I seesaw
(\cref{sec:typeI-seesaw}), which features three very heavy sterile neutrinos
with very small mixing angles. We then discuss the Neutrino Minimal Standard
Model ($\nu$MSM, \cref{sec:nuMSM}), in which the mass of one of the sterile
neutrinos is lowered, albeit mixing angles are still small.  Phenomenologically
more interesting for oscillation experiments is the Inverse Seesaw model
(\cref{sec:inverse-seesaw}) in which both sterile neutrino masses as well as
mixing angles are in the experimentally accessible range.  We finally introduce
an extended seesaw model (\cref{sec:extended-seesaw}) which is not only
experimentally testable, but also generates the light mass scale for the
sterile neutrino in a natural way.

\subsection{The Generic Type-I Seesaw Scenario}
\label{sec:typeI-seesaw}

The fact that the $N$ fields in \cref{eq:L-Yukawa,eq:nu-s-mass-mixing} are
total SM singlets implies that, besides the standard Dirac mass term from
\cref{eq:nu-s-mass-mixing}, they also admit a Majorana mass term of the form
\begin{align}
  \mathcal{L}_\text{$N$-mass} = -\frac{1}{2} M_M^{\alpha\beta} N^\alpha N^\beta + h.c. \,,
  \label{eq:L-Majorana}
\end{align}
To understand the phenomenological consequences of the mass terms in
\cref{eq:nu-s-mass-mixing,eq:L-Majorana}, it is convenient to arrange the
$\nu$ and $N$ fields into a vector $n \equiv (\nu^e \dots \nu^\tau, N^1 \dots N^n)^T$
and write the mass term in block matrix notation:
\begin{align}
  \mathcal{L}_\text{mass} \supset -\frac{1}{2} n^T M n
                          \equiv  -\frac{1}{2} n^T \begin{pmatrix}
                                                     0     &  M_D \\
                                                     M_D^T &  M_R
                                                   \end{pmatrix} n  +  h.c.\,.
  \label{eq:L-mass-matrix}
\end{align}
Here, $M_D = (M_D^{\alpha\beta})$ is the general complex Dirac mass matrix from
\cref{eq:nu-s-mass-mixing}, while $M_M = (M_M^{\alpha\beta})$ is the complex
symmetric Majorana mass matrix from \cref{eq:L-Majorana}.

In the limit where the eigenvalues of $M_M$ in \cref{eq:L-mass-matrix} are
significantly larger than those of $M_D$, we recover the seesaw mechanism
\cite{Minkowski:1977sc, Mohapatra:1979ia, Yanagida:1979as, GellMann:1980vs},
with three neutrino mass eigenvalues of order $\|M_D\|^2/\|M_M\|$,\footnote{We
use the notation $\| \cdot \|$ to denote the Euclidean matrix norm. For a matrix
$A = (a_{ij})$ it is given by $\|A\| \equiv \sqrt{\sum_{ij} |a_{ij}|^2}$.
However, for the order-of-magnitude estimates that we are interested in in this
section, other matrix norms are equally suitable. For instance, for our purposes,
we could also define $\| A \|$ to correspond to the largest eigenvalue, or
to the largest entry of $A$.} and $n$ mass
eigenvalues of order $\|M_M\|$.  Choosing $\|M_M\| \gg \|M_D\|$ is motivated by
the fact that $M_M$ is not protected by any symmetry, whereas the entries of
$M_D$ cannot be significantly larger than the electroweak scale as this would
require Yukawa coupling $\gg 1$.  For $\|M_D\| \sim \SI{100}{GeV}$ and $\|M_M\|
\sim \SI{e14}{GeV}$, the light neutrino mass eigenvalues are of order
\SI{0.1}{eV}, so the seesaw mechanism offers an explanation for the observed
smallness of neutrino masses. Adopting instead $\|M_D\| \sim \SI{100}{keV}$
(comparable to the electron mass), the required scale for $\|M_M\|$ is lowered
to \SI{100}{GeV}.  While the seesaw mechanism is perhaps the most widely
discussed application of sterile neutrinos, the singlet states appearing in it
are currently unobservable in practice. Even when their masses are within reach
of current collider experiments, their extremely small mixing $\sim \|M_D\| /
\|M_M\|$ with active neutrinos prevents their efficient production.  It is of course
possible to invoke fine-tuning to make at least one of the sterile states
observable.  Another possibility is to lower also $\|M_M\|$ to the eV-scale.
For instance, for $\|M_D\| \simeq \SI{0.3}{eV}$ and $\|M_M\| \simeq
\SI{1.5}{eV}$, we find light neutrino masses around \SI{0.06}{eV}, sterile
neutrinos at the eV-scale, and active--sterile mixing of order
20\%~\cite{Barry:2011wb}.  Both fine-tuning and lowering $\|M_D\|$ would, however,
run counter to the main motivation for the seesaw mechanism, namely explaining
the smallness of neutrino masses. We will therefore not discuss this
possibility further here.

\subsection{The Neutrino Minimal Standard Model ($\nu$MSM)}
\label{sec:nuMSM}

As the short-baseline anomalies require only one sterile neutrino
at the eV-scale, one may ask the question whether it is possible to
lower only one of the eigenvalues of the Majorana mass matrix
$M_M$ in \cref{eq:L-mass-matrix} to the eV-scale, while
keeping the others super-heavy.  This could be motivated by a
symmetry~\cite{Shaposhnikov:2006nn, Lindner:2010wr}.
A similar idea has been widely explored
in the context of the ``Neutrino Minimal Standard Model'' ($\nu$MSM)
\cite{Asaka:2005an}, even though there the motivation for
lowering one of the eigenvalues of $M_M$ was to obtain a dark matter
candidate at the keV-scale. Following ref.~\cite{Barry:2011wb},
we collectively denote the three Standard Model (active) neutrinos $\nu_a$, the
light sterile neutrino $\nu_s$, and the two heavy sterile states $N$.
In the basis $(\nu_a, \nu_s, N)$, the mass matrix then reads
\begin{align}
  M = \begin{pmatrix}
           0     &    0     &    0     & M_s^{1} & M_D^{11} & M_D^{12} \\
           0     &    0     &    0     & M_s^{2} & M_D^{21} & M_D^{22} \\
           0     &    0     &    0     & M_s^{3} & M_D^{31} & M_D^{32} \\
        M_s^{1}  & M_s^{2}  & M_s^{3}  & \mu_S   &     0    &    0     \\
        M_D^{11} & M_D^{21} & M_D^{31} &    0    & M_M^{1}  &    0     \\
        M_D^{12} & M_D^{22} & M_D^{32} &    0    &     0    & M_M^{2}
      \end{pmatrix} \,.
  \label{eq:mass-matrix-nuMSM}
\end{align}
Here, we have already used the freedom to re-define the sterile states
such that the lower right-hand $3 \times 3$ block of $M$ is diagonal.
We assume the hierarchy $\|M_s\| \ll \mu_S \ll M_M^1, M_M^2$ and, as a first
step, integrate out the two heavy states $N$:
\begin{align}
  M_{4\times4} = \begin{pmatrix}
                   -M_D M_M^{-1} M_D^T  &  M_s \\
                        M_s^T           &  \mu_S
                 \end{pmatrix} \,.
\end{align}
Applying the seesaw formula a second time, we find for the light neutrino
masses
\begin{align}
  M_\nu \simeq -M_D M_M^{-1} M_D^T - M_S \mu_S^{-1} M_S^T \,.
\end{align}
and for the light sterile state
\begin{align}
  m_s \simeq \mu_s \,. 
\end{align}
Though here we have downplayed the role of the two heavy states $N$, we will see in \cref{sec:leptogenesis} that they can have a phenomenologically important role in certain leptogenesis models.

\subsection{The Inverse Seesaw}
\label{sec:inverse-seesaw}

Sterile neutrinos with masses below the electroweak scale and with
not too small ($\gtrsim 10^{-2}$) mixing angles could be related to
the generation of neutrino masses in the context of the
inverse seesaw mechanism \cite{Mohapatra:1986aw, Mohapatra:1986bd,
GonzalezGarcia:1988rw}.  The main idea behind the inverse seesaw is
to invoke lepton number to protect the active neutrino masses. This is achieved
by making the singlet fermions $N$ Dirac spinors. In other words,
one introduces two sets of sterile neutrinos, $N^\beta$ and $N'^\beta$,
all of which carry lepton number.
Arranging the $\nu^\alpha$, $N^\beta$, and $N'^\beta$ fields into
a vector $n \equiv (\nu^e \dots \nu^\tau, N^1 \dots N^n, N'^1 \dots N'^n)^T$,
the neutrino mass term in an inverse seesaw scenario is
\begin{align}
  \mathcal{L}_\text{mass} \supset -\frac{1}{2} n^T M n
                          \equiv  -\frac{1}{2} n^T \begin{pmatrix}
                                                     0     & M_D    & 0 \\
                                                     M_D^T & 0      & M_D' \\
                                                     0     & M_D'^T & \mu
                                                   \end{pmatrix} n  +  h.c.\,.  
  \label{eq:inverse-seesaw}
\end{align}
The terms coupling $\nu$ to $N$ (with the general complex $3 \times n$
mass matrix $M_D$), and the terms coupling $N$ to $N'$
(with the general complex $n \times n$ mass matrix $M_D'$) respect lepton
number.  The Majorana mass term $-\frac{1}{2} \mu^{\beta\gamma} N'^\beta
N'^\gamma$, on the other hand, violates lepton number. In the spirit
of 't~Hooft naturalness, $\|\mu\|$ is assumed to be much smaller
than $\|M_D\|$ and $\|M_D'\|$ because in the limit $\|\mu\| \to 0$,
lepton number becomes a conserved quantity.  In this limit, it is
easy to see that the mass matrix in \cref{eq:inverse-seesaw} has at least
$n$ zero eigenvalues.
For non-zero $\|\mu\|$, these eigenvalues are lifted
to $\sim \|\mu\| \cdot \|M_D'\|^2 / (\|M_D\|^2 + \|M_D'\|^2)$, and the
active--sterile mixing angles are of order $\|M_D\| / \|M_D'\|$.
Given that neutrino masses are mainly suppressed by the natural
smallness of $\|\mu\|$, these mixing angles can be sizable.

\subsection{An Extended Seesaw Model}
\label{sec:extended-seesaw}

The variants of the seesaw mechanism introduced so far share the feature
that, in order to obtain eV-scale sterile states, this scale must
already be present in the parameters of the Lagrangian.  These models
do not explain why such a low scale appears in the Lagrangian and therefore
run counter to the original motivation of the seesaw mechanism. We now discuss
a seesaw model in which the eV-scale arises naturally. In the following
discussion, we once again draw heavily from ref.~\cite{Barry:2011wb}.
The Minimal Extended Type-I Seesaw Model discussed there is based on
augmenting the SM with three heavy right-handed neutrinos $N$ and one
singlet fermion $S$. The neutrino mass terms in the Lagrangian are
\begin{align}
  \mathcal{L} \supset -M_D^{\alpha\beta} \nu^\alpha N^\beta
                      -M_S^\alpha S N^\alpha
                      -\frac{1}{2} M_M^{\alpha\beta} N^\alpha N^\beta \,.
  \label{eq:L-extended-seesaw}
\end{align}
This structure can be achieved for instance if the $N^\alpha$ are true
singlet fields under all symmetries, while $S$ carries quantum numbers
under a new sterile sector symmetry that is broken by a Higgs mechanism.
In the basis $n = (\nu^e, \nu^\mu, \nu^\tau, N^1, N^2, N^3, S)$, the mass
term is thus
\begin{align}
  \mathcal{L} \supset -\frac{1}{2} n^T
                      \begin{pmatrix}
                        0     & M_D & 0      \\
                        M_D^T & M_R & M_S  \\
                        0     & M_S^T & 0
                      \end{pmatrix} n + h.c. \,,
  \label{eq:M-extended-seesaw}
\end{align}
with the general $3 \times 3$ complex matrix $M_D$, the complex symmetric
$3\times 3$ matrix $M_R$, and the complex 3-vector $M_S$. Under the assumption
that $\|M_R\| \gg \|M_S\|, \|M_D\|$, we first integrate out the heavy fields
$N^\beta$. This leads to an effective mass term in the basis $n' = (\nu^e,
\nu^\mu, \nu^\tau, S)$:
\begin{align}
  \mathcal{L} \supset -\frac{1}{2} n'^T
                      \begin{pmatrix}
                        M_D M_R^{-1} M_D^T        &  M_D M_R^{-1} M_S \\
                        M_S^T (M_R^{-1})^T M_D^T  &  M_S^T M_R^{-1} M_S
                      \end{pmatrix} n' + h.c. \,.
  \label{eq:M-extended-seesaw-4x4}
\end{align}
If, moreover, $\|M_S\| \gg \|M_D\|$, the seesaw formula can be applied once
again to obtain the $3 \times 3$ mass matrix of the light (SM) neutrinos,
\begin{align}
  m_\nu \simeq - M_D M_R^{-1} M_D^T
               + M_D M_R^{-1} M_S \big( M_S^T M_R^{-1} M_S \big)^{-1}
                 M_S^T (M_R^{-1})^T M_D^T \,,
  \label{eq:m-nu-extended-seesaw}
\end{align}
and the mass of the light sterile neutrino $S$:
\begin{align}
  m_s \simeq M_S^T M_R^{-1} M_S \,.
  \label{eq:m-s-extended-seesaw}
\end{align}
The mixing matrix between the active neutrinos and the light sterile
neutrino is
\begin{align}
  V \simeq \begin{pmatrix}
             (1 - \tfrac{1}{2} R R^\dag) U  &  R \\
             -R^\dag U                      & 1 - \tfrac{1}{2} R^\dag R 
           \end{pmatrix} \,,
  \label{eq:U-PMNS-extended-seesaw}
\end{align}
where $U$ is the matrix diagonalizing $m_\nu$ from \cref{eq:m-nu-extended-seesaw}
and $R = M_D M_R^{-1} M_S \big( M_S^T M_R^{-1} M_S \big)$.
With $\|M_D\|$ at the electroweak scale $\sim \SI{100}{GeV}$,
$\|M_S\| \simeq \si{TeV}$, and $\|M_R\| \simeq \SI{e14}{GeV}$,
we find active neutrino masses around \SI{0.1}{eV} and a light
sterile neutrino at \SI{10}{eV}. The mixing between the active
and sterile states is of order $\|M_D\| / \|M_S\| \sim 0.1$.

\chapter{Sterile Neutrino Phenomenology}
\label{sec:pheno}

Having shown how sterile neutrinos can arise in neutrino mass models, we now
turn our attention to their phenomenology in terrestrial experiments, including
oscillation searches, kinematic measurements of neutrino mass, searches for
sterile neutrino decay in fixed target experiments and at colliders, and neutrinoless
double beta decay.

\section{Neutrino Oscillations with More than Three Neutrino Flavors}
\label{sec:4f-osc}

When the Standard Model is extended by $n$ sterile neutrino flavors,
the formalism of neutrino oscillation remains largely the same as in
the three-flavor case (see for
instance \cite{Akhmedov:1999uz} for an introduction and
\cite{Beuthe:2001rc} for an in-depth discussion of the quantum mechanical
subtleties involved).  The weak interaction Lagrangian in the flavor basis is the same as in the Standard Model:
\begin{align}
  \mathcal{L} &= \sum_{\alpha=e,\mu,\tau} \bigg[
      \frac{g}{\sqrt{2}} \Big( \overline{\nu_{\alpha,L}} \gamma^\mu e_{\alpha,L}
                                                      W_\mu^+ + h.c. \Big)
    + \frac{g}{2\cos\theta_w} \overline{\nu_{\alpha,L}} \gamma^\mu \nu_{\alpha,L} Z_\mu
  \bigg] \,.
  \label{eq:L-weak}
\end{align}
What changes is the transformation to the mass basis,
\begin{align}
  \nu_\alpha = \sum_{j=1}^{3+n} U_{\alpha j} \nu_j \,,
\end{align}
where the sum now runs over all $3+n$ mass eigenstates, and the mixing matrix
$U$ is a $(3+n) \times (3+n)$ matrix.
An initial neutrino flavor state $\ket{\nu_\alpha}$ is created by acting
on the vacuum $\ket{0}$ with the operator $\nu_\alpha^\dag$. Therefore,
its decomposition into mass eigenstates $\ket{\nu_j}$ is given by
\begin{align}
  \ket{\nu_\alpha} = \sum_j U_{\alpha j}^* \ket{\nu_j} \,,
  \label{eq:nu-alpha}
\end{align}
Note that, as usual, the fields transform with $U$, while the states transform
with $U^*$.
After a time $T$, the neutrino state at a distance $L$ can be written as
\begin{align}
  \ket{\nu_\alpha(T,L)} = \sum_j e^{-i E_j T + i p_j L} U_{\alpha j}^* \ket{\nu_j} \,,
  \label{eq:nu-alpha-TL}
\end{align}
where $E_j$ and $p_j$ are the energy and momentum associated with the $j$-th
mass eigenstate.  If the neutrino is detected in a flavor state $\bra{\nu_\beta}$,
the corresponding oscillation probability is
\begin{align}
  P_{\alpha\beta} \equiv \big| \sprod{\nu_\beta}{\nu_\alpha(T,L)} \big|^2
    = \sum_{j,k} U_{\alpha j}^* U_{\beta j} U_{\alpha k} U_{\beta k}^*
      e^{-i (E_j - E_k) T + i (p_j - p_k) L)} \,.
  \label{eq:Pab-1}
\end{align}
In a more careful treatment, we would need to account for
the fact that neutrinos emitted from a localized source do not have
a definite momentum because of the Heisenberg uncertainty principle. Instead,
they need to be treated as wave packets with a non-zero width in momentum space;
see for instance ref.~\cite{Beuthe:2001rc}. Here, we use the Heisenberg
principle only inasmuch as we allow neutrino mass eigenstates with formally
different energy and momentum to interfere. (This would not be possible if their
energies and momenta were indeed known with infinite precision; in this
case, the precise knowledge of the event kinematics would allow us to infer
the mass of the propagating neutrino state, which would suppress oscillations.)
To proceed further from \cref{eq:Pab-1}, we use the relativistic energy momentum
relation $p_j = \sqrt{E_j^2 - m_j^2}$ to rewrite the oscillation phase approximately as
$-i (E_j - E_k) (T - L) - i \Delta m_{jk}^2 L / (2 E)$, where $E$ is a
suitably chosen average energy, which should differ from $E_j$, $E_k$, by
no more than the neutrino wave packet width in momentum space.
The first term here is of order $\Delta m_{jk}^2 / E^2 \cdot \sigma_x$
(with $\sigma_x$ the wave packet width in coordinate space) and is
therefore negligible. The second term is the well-known standard oscillation
phase, so that we recover the standard expression for the oscillation
probability,
\begin{align}
  P_{\alpha\beta} = \sum_{j,k} U_{\alpha j}^* U_{\beta j} U_{\alpha k} U_{\beta k}^*
  e^{-i \Delta m_{jk}^2 L / (2 E)} \,,
  \label{eq:Pab-2}
\end{align}
where it is now understood that $j$ and $k$ run from $1$ to $3+n$.

\subsection{$\nu_e$ Appearance Searches}

Several groups are carrying out combined analysis of all relevant experimental
data to assess the global viability of sterile neutrino models. The US-based
group \cite{Diaz:2019fwt,Moulai:2019gpi} involves members of the MiniBooNE and IceCube
collaborations and is therefore able to offer particularly accurate fits to these
data sets.  The Italian-led group \cite{Gariazzo:2017fdh} were
the first to realize the significance of the gallium anomaly, and they are the
only group so far to have carried out joint fits between short-baseline
oscillation experiments and cosmological observations~\cite{Archidiacono:2013xxa,
Archidiacono:2020yey}. The third group
\cite{Dentler:2018sju}, which emerged from the NuFit~\cite{Esteban:2018azc} and
GLoBES~\cite{Huber:2007ji} collaborations, is the one from whose results
we will draw most heavily in this review as one the present authors (JK)
is a member of this group. While there
are differences between the three groups in the data sets analyzed and the
level of detail in which each of them is treated, their results
are in broad agreement.\footnote{It is worth pointing out that some luminaries
in the field hold the view that theorists -- such as the authors of this article
-- should not be allowed to analyze data.  Thankfully, among the rank and file
of neutrino physics, global fits are generally viewed as a useful resource
for guiding experimental and theoretical work.  In the past, they were often able
to anticipate new discoveries such as the non-zero value
of $\theta_{13}$ \cite{GonzalezGarcia:2010er, Fogli:2011qn}.  We will therefore not
heed the luminaries' advice here.}

We begin here with a discussion of $\nu_\mu \to \nu_e$ appearance searches,
which are currently dominated by the signals from LSND~\cite{Aguilar:2001ty} and
MiniBooNE \cite{Aguilar-Arevalo:2013pmq, MBdataReleaseAPP} discussed in
\cref{sec:lsnd} and \cref{sec:mb}, respectively.  They are complemented by data
from the short-baseline experiments KARMEN \cite{Armbruster:2002mp}, with a
setup very similar to LSND, as well as NOMAD
\cite{Astier:2003gs} and E776 \cite{Borodovsky:1992pn} who have used high-energy
$\nu_\mu$ beams.  None of these experiments has observed an anomaly, however
their sensitivity was below that of LSND and MiniBooNE.  Strong exclusion
bounds come also from the CNGS (CERN Neutrinos to Gran Sasso) project
comprising the long-baseline detectors ICARUS \cite{Antonello:2012pq,
Farnese:2015kfa} and OPERA \cite{Agafonova:2013xsk}.

While the results we are going to show below are based on oscillation
probabilities calculated in the full four-flavor framework
(\cref{eq:Pab-2}), it is instructive to consider approximations to
this framework. In particular, in the short-baseline limit
($\Delta m_{21}^2 L / E \ll 1$, $\Delta m_{31}^2 L / E \ll 1$, where
  standard model oscillations have not developed yet, the
  probability of $\nu_\mu \to \nu_e$ conversion is
\begin{align}
  P_{\mu e}^\text{sbl} \simeq 4 |U_{e4}|^2 |U_{\mu 4}|^2
                              \sin^2 \frac{\Delta m_{41}^2 L}{4 E} \,.
  \label{eq:Pmue-sbl}
\end{align}
This is just the familiar two-flavor oscillation formula, with an
effective mixing angle $\theta_{\mu e}$ defined as
\begin{align}
  \sin^2 2\theta_{\mu e} \equiv 4 |U_{e4}|^2 |U_{\mu 4}|^2 \,.
  \label{eq:s22thmue}
\end{align}
We see that the effective mixing angle depends both on
the mixing of sterile neutrinos with electron neutrinos, $U_{e4}$, and
on their mixing with muon neutrinos, $U_{\mu 4}$.

\begin{figure}[t]
  \centering
  \begin{tabular}{cc}
    \includegraphics[width=0.46\textwidth,trim=0 0.5cm 7cm 0,clip]{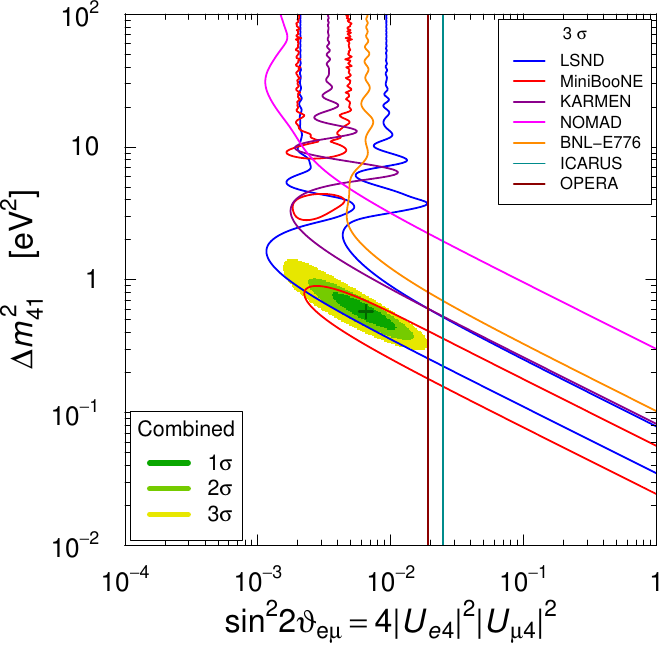} &
    \raisebox{0.12cm}{\includegraphics[width=0.49\textwidth]{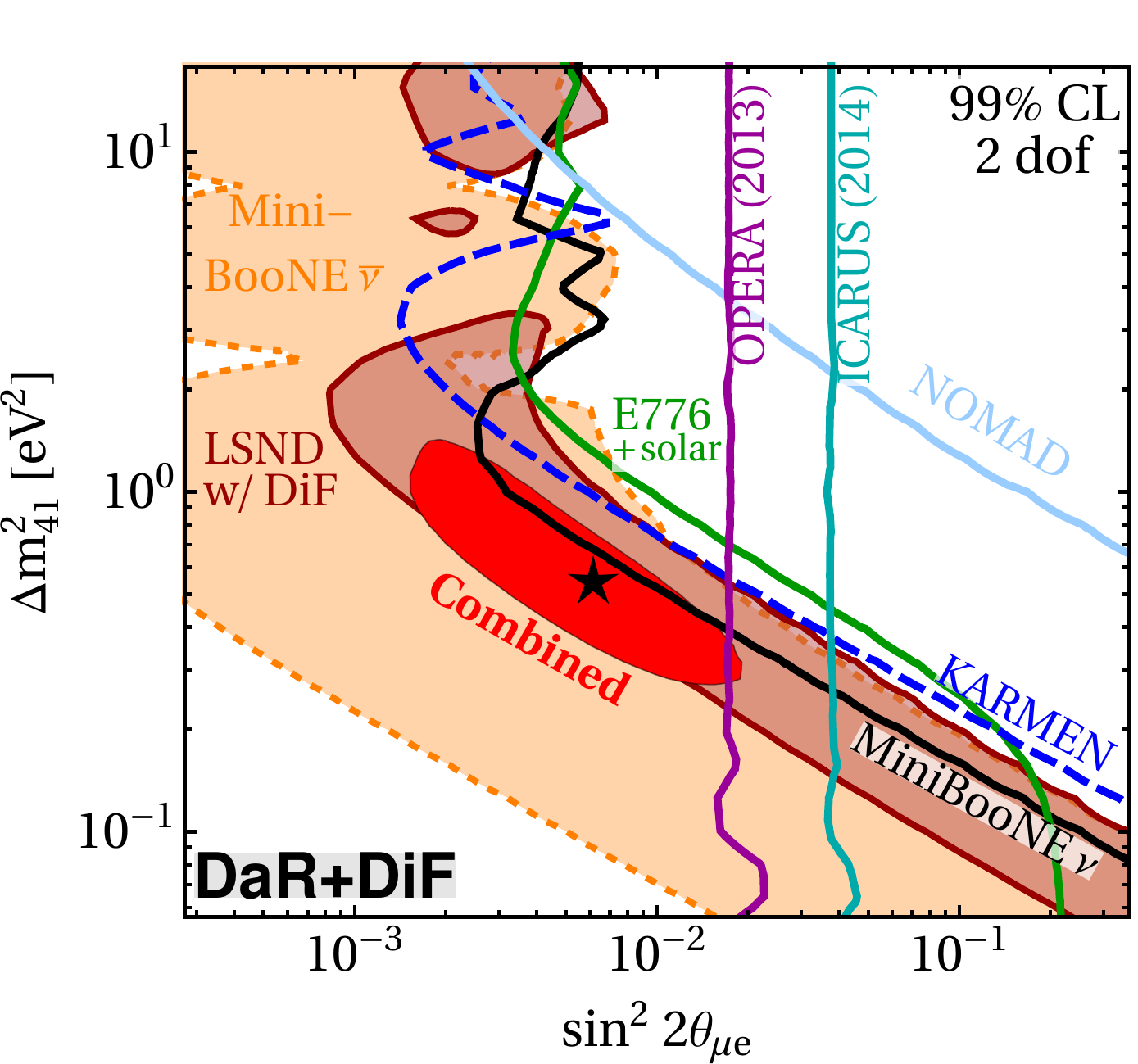}} \\
    (a) & (b)
  \end{tabular}
  \caption{Global constraints on $\nu_\mu \to \nu_e$ (and $\bar\nu_\mu \to \bar\nu_e$)
    oscillations at short baseline in the $3+1$ framework. The relevant
    parameters are $\Delta m_{41}^2$, measuring the mass of the 4th neutrino, and
    $\sin^2 2\theta_{\mu e} = 4 |U_{e4}|^2 |U_{\mu 4}|^2$ measuring the effective
    mixing, which depends on the mixing between sterile neutrinos and electron
    neutrinos ($U_{e4}$) and on the mixing between sterile neutrinos and muon
    neutrinos ($U_{\mu 4}$).  The left panel shows results from the Italian fit
    \cite{Giunti:2019aiy}, while the right panel shows the same from the
    NuFit/GLoBES fit~\cite{Dentler:2018sju}.}
  \label{fig:nu-e-app}
\end{figure}

In \cref{fig:nu-e-app}, we show the global constraints (as of spring 2018) on
$\nu_\mu \to \nu_e$ oscillations at short baseline in the $3+1$ framework,
comparing the results from refs.~\cite{Dentler:2018sju} (NuFit/GLoBES) and
\cite{Gariazzo:2017fdh} (Italy).  Both groups agree on the best fit region
around effective mixing angle $\sin^2 2\theta_{\mu e} \sim 0.01$ and mass squared
difference $\Delta m_{41}^2 \lesssim \SI{1}{eV^2}$.  The parameter region
favored by MiniBooNE and LSND is also in agreement with the null results from
all other experiments, leading us to the conclusion that the global data on
$\nu_\mu \to \nu_e$ oscillations, when viewed in isolation, is consistent and
could point towards the existence of an eV-scale sterile neutrino.

Before moving on, let us comment on the fact that, in some cases,
single-experiment exclusion contours obtained in global fits differ from those
appearing in the official experimental publications. This does \emph{not}
indicate a problem with the global fit -- in fact, all analyses entering a global
are vetted first to make sure they reproduce the official results, and many
of them are directly based on recommendations from the experimental collaborations.
However, the assumptions made in experimental analyses are not always suitable
for a global fit. For instance, in a 2-flavor oscillation framework as used in
refs.~\cite{Aguilar:2001ty,Aguilar-Arevalo:2018gpe}, MiniBooNE's background
prediction is essentially unaffected by the presence of the sterile neutrino.
In a $3+1$ model, however, non-negligible $\nu_e$ and $\nu_\mu$ disappearance
is predicted, altering the background prediction.  A detailed discussion of
these corrections, and how they enter the results shown in \cref{fig:nu-e-app}~(b)
is given in the appendix of ref.~\cite{Dentler:2019dhz}.

We comment briefly on updates to the results of refs.~\cite{Giunti:2019aiy,
Dentler:2018sju} shown in \cref{fig:nu-e-app}.  Notably, MiniBooNE have since
updated their results \cite{Aguilar-Arevalo:2018gpe, Aguilar-Arevalo:2020nvw},
increasing the statistics in neutrino-mode ($\nu_\mu$-dominated beam) from
\SI{6.46}{pot} (protons on target) \cite{Aguilar-Arevalo:2012fmn}
to \SI{18.75e20}{pot}~\cite{Aguilar-Arevalo:2020nvw}.  This has led to
an increase in the statistical significance of the anomaly from $3.8\sigma$
to $4.8\sigma$, while leaving its qualitative features unchanged.
The conclusions drawn from \cref{fig:nu-e-app} therefore remain largely
unchanged with the new data: global $\nu_e$ appearance data remain
consistent; only the exclusion of the null hypothesis
$\sin^2 2\theta_{\mu e} = 0$ becomes statistically more significant.

\subsection{$\nu_e$ Disappearance Searches}

Constraints on $\nu_e$ and $\bar\nu_e$ disappearance are driven by reactor
experiments. In fact, a large number of such experiments have been carried out
already in the 1980s and 1990s \cite{Kwon:1981ua, Zacek:1986cu, Vidyakin:1987ue, Vidyakin:1994ut,
  Kozlov:1999cs, Afonin:1988gx, Kuvshinnikov:1990ry, Declais:1994su,
Declais:1994ma, Greenwood:1996pb}, typically presenting their results as a
comparison between the total number of observed neutrino events and the (then
state-of-the-art) theoretical prediction by Schreckenbach et al.\
\cite{Schreckenbach:1985ep}. In recent years, these experiments have been
superseded by multi-baseline experiments, in particular Double Chooz
\cite{Giunti:2016elf,Abrahao:2017zvl}, RENO \cite{reno-EPS17,reno-Neutrino14},
Daya Bay \cite{An:2016ses, An:2017osx} (optimized for neutrino oscillations
with $\Delta m^2 \sim \SI{1e-3}{eV^2}$), DANSS \cite{danss-solvay17, Danilov:2019aef},
Neutrino-4 \cite{Serebrov:2018vdw, Serebrov:2020kmd}, STEREO~\cite{AlmazanMolina:2019qul},
and PROSPECT \cite{Andriamirado:2020erz} (optimized for
  oscillations with $\Delta m^2 \sim \SI{1}{eV^2}$).
By comparing fluxes and spectra at different baselines, these experiments
can search for $\bar\nu_e$ disappearance due to oscillations into sterile neutrinos
without having to rely on theoretical flux predictions. The same is true for
modern single-baseline experiments like NEOS \cite{Ko:2016owz,An:2016srz}
when analyzed by comparing their data to that of
other experiments.  To some extent, also the long-baseline experiment
KamLAND ($L \sim \SI{60}{km}$) \cite{Gando:2010aa} weighs in.

The search for $\bar\nu_e$ disappearance in Reactor experiments is complemented
by searches using intense radioactive sources \cite{Hampel:1997fc,
Kaether:2010ag, Abdurashitov:1998ne, Abdurashitov:2005tb}, solar neutrinos
\cite{Cleveland:1998nv, Abdurashitov:2009tn, Kaether:2010ag, Hosaka:2005um,
Cravens:2008aa, Abe:2010hy, sksol:nakano2016, Aharmim:2007nv, Aharmim:2005gt,
Aharmim:2008kc, Bellini:2011rx, Bellini:2008mr, Bellini:2014uqa}, and $\nu_e$
scattering on carbon ($\nu_e + \iso{C}{12} \to \text{e}^- + \iso{N}{12}$)
\cite{Reichenbacher:2005nc, Armbruster:1998uk, Conrad:2011ce, Auerbach:2001hz,
Conrad:2011ce}.

We consider again the short-baseline approximation
$\Delta m_{21}^2 L / E \ll 1$, $\Delta m_{31}^2 L / E \ll 1$, in which
the $\nu_e$ disappearance probability becomes
\begin{align}
  P_{ee}^\text{sbl} \simeq 1 - 4 |U_{e4}|^2 (1 - |U_{e4}|^2)
                           \sin^2 \frac{\Delta m_{41}^2 L}{4 E} \,.
  \label{eq:Pee-sbl}
\end{align}
Once again, we find an expression that has the form of a two-flavor
survival probability, with the effective mixing angle defined via
\begin{align}
  \sin^2 2\theta_{ee} \equiv 4 |U_{e4}|^2 (1 - |U_{e4}|^2) \,.
  \label{eq:s22thee}
\end{align}

\begin{figure}[t]
  \centering
  \begin{tabular}{cc}
   \includegraphics[width=0.45\textwidth]{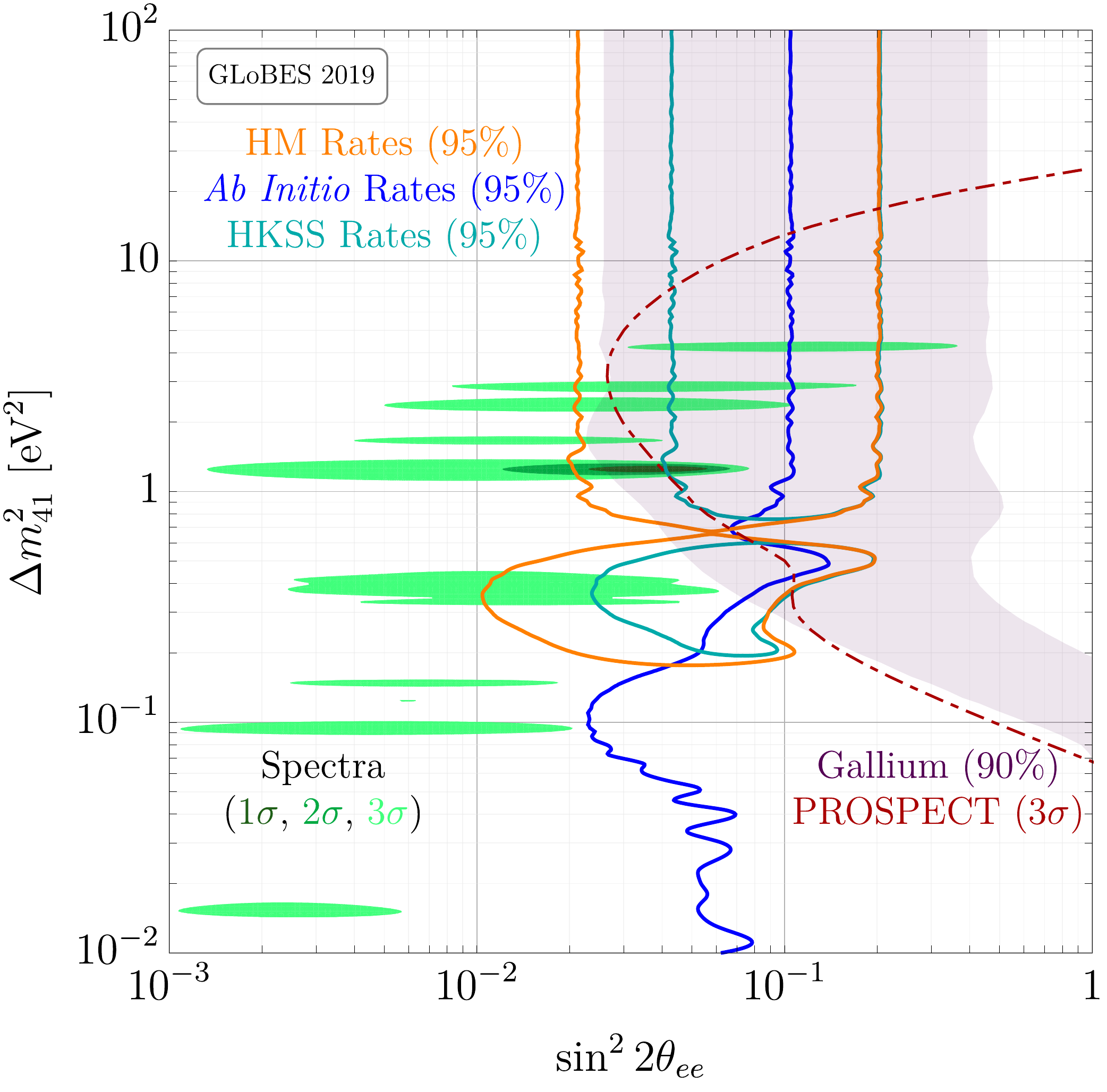} &
   \raisebox{-0.1cm}{\includegraphics[width=0.5\textwidth]{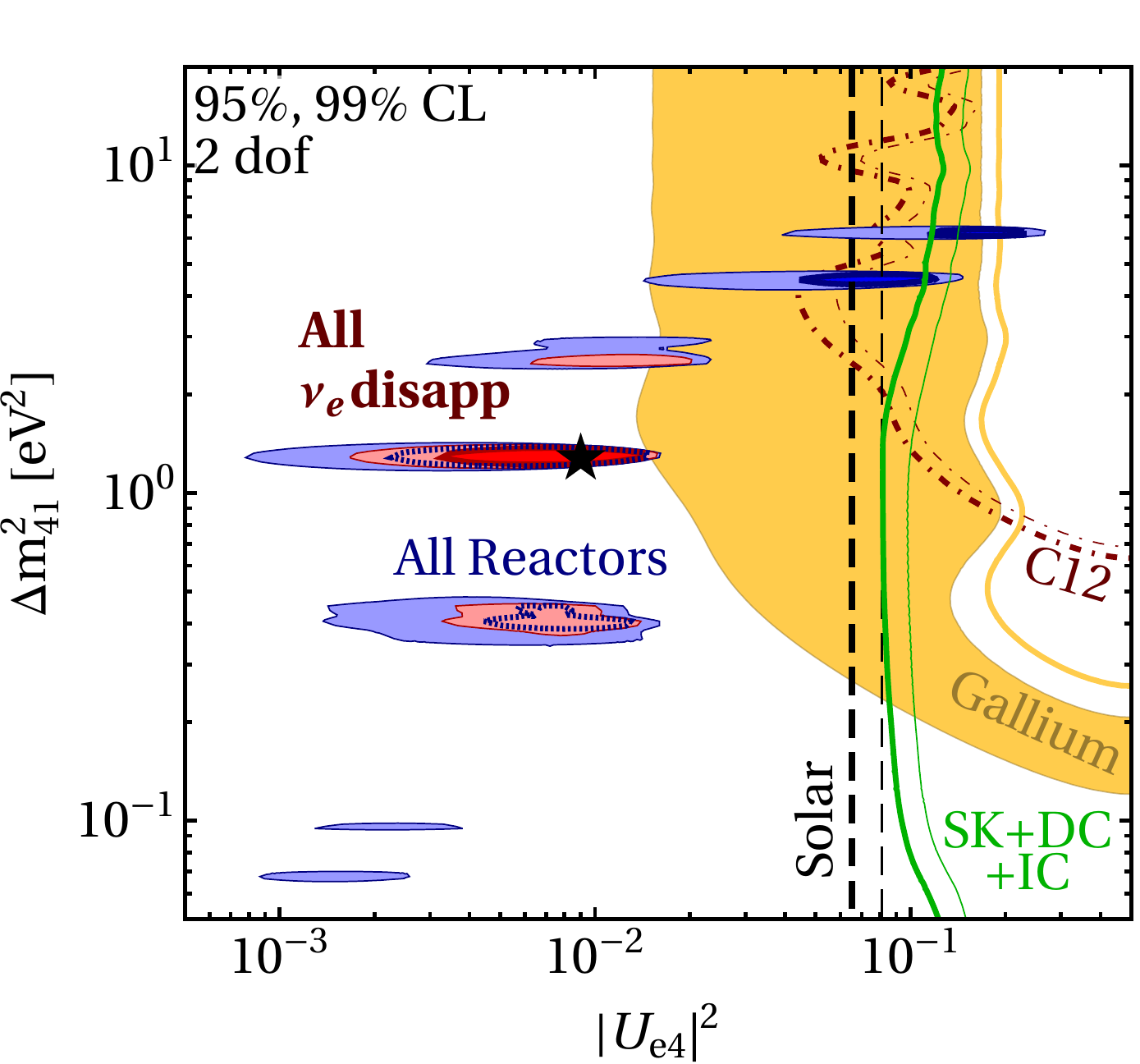}} \\
    (a) & (b)
  \end{tabular}
  \caption{Global constraints on short-baseline $\nu_e$ and $\bar\nu_e$
    disappearance, shown as a function of the mass squared difference $\Delta
    m_{41}^2$ and of the $\nu_e$--$\nu_s$ mixing.  Panel (a), taken from
    ref.~\cite{Berryman:2019hme}, displays the preferred parameter region based
    on comparing total measured reactor neutrino event rates to three different
    theoretical calculations (orange, blue and turquoise lines); it also
    displays as green filled contours the preferred regions based on reactor
    spectra alone (without considering total rate information), and in gray the
    preferred region corresponding to the ``gallium anomaly''.  The maroon dot-dashed
    curve shows the sensitivity of the PROSPECT experiment \cite{Ashenfelter:2015uxt}.
    In this panel, the horizontal axis shows $\sin^2\theta_{ee} = 4 |U_{e4}| (1 -
    |U_{e4}|^2)$. In panel (b), taken from ref.~\cite{Dentler:2018sju}, we
    compare the fit to reactor data (blue ellipses) also to constraints from
    solar neutrinos (black dashed),
    atmospheric neutrinos (green), $\nu_e$--\iso{C}{12} scattering (maroon
    dot-dashed), and to the best fit region based on a combination of all
    $\nu_e$ and $\bar\nu_e$ datasets (red ellipses, with the star showing the
    best fit point).}
  \label{fig:nue-disapp}
\end{figure}

We see in panel (a) (from ref.~\cite{Berryman:2019hme}) that different
theoretical flux predictions lead to significantly different results
concerning the significance of the reactor anomaly: the Huber--Mueller
calculation from refs.~\cite{Mueller:2011nm,Huber:2011wv} and
the newer calculation from ref.~\cite{Hayen:2019eop} both lead to a
significant apparent deficit of events, while the calculation
from ref.~\cite{Estienne:2019ujo} does not lead to a significant
anomaly.  This discrepancy between different theoretical calculations
highlights the large and difficult-to-estimate systematic uncertainties in
these calculation and prevents us from drawing firm conclusions
from the rate anomaly.  On the other hand, \cref{fig:nue-disapp}~(a)
also shows that the spectral anomalies (driven mostly by DANSS)
are robust with regard to the flux normalization, and lead to a
preference for oscillations. With the data sets used in ref.~\cite{Berryman:2019hme},
the statistical significance of this anomaly is $3\sigma$. With more
recent data from DANSS~\cite{Danilov:2019aef}, the significance decreases
to $2\sigma$~\cite{Giunti:2020uhv,Berryman:2020agd}.
\Cref{fig:nue-disapp}~(b) finally shows that constraints from
null results on solar, atmospheric, and $\nu_e$--\iso{C}{12} scattering
data do not impose any significant constraints.

\subsection{$\nu_\mu$ Disappearance Searches}

Similar to $\nu_e$ disappearance, also $\nu_\mu$ disappearance can be
described by a simple two-flavor expression in the short-baseline
limit. In complete analogy to \cref{eq:Pee-sbl,eq:s22thee}, the
$\nu_\mu$ survival probability in this approximation reads
\begin{align}
  P_{\mu\mu}^\text{sbl} \simeq 1 - 4 |U_{\mu 4}|^2 (1 - |U_{\mu 4}|^2)
                           \sin^2 \frac{\Delta m_{41}^2 L}{4 E} \,,
  \label{eq:Pmm-sbl}
\end{align}
and an effective two-flavor mixing angle can thus be defined via
\begin{align}
  \sin^2 2\theta_{\mu\mu} \equiv 4 |U_{\mu4}|^2 (1 - |U_{\mu4}|^2) \,.
  \label{eq:s22thmm}
\end{align}
Let us consider now in particular the case that $\Delta m_{21}^2 L / E \ll 1$,
$\Delta m_{31}^2 L / E \ll 1$, so that the short baseline approximations
from \cref{eq:Pmue-sbl,eq:Pee-sbl,eq:Pmm-sbl} are valid, but simultaneously
$\Delta m_{41}^2 L / E \gg 1$, so that the oscillating terms
$\sin^2 \Delta m_{41}^2 L / (4 E)$ in these equations average to $\tfrac{1}{2}$
due to the limited experimental energy resolution.  In this limit,
\cref{eq:Pmue-sbl,eq:Pee-sbl,eq:Pmm-sbl} depend
on only two parameters: $|U_{e4}|^2$ and $|U_{\mu4}|^2$.  By observing
all three oscillation channels ($\nu_e$ disappearance, $\nu_\mu$ disappearance,
and $\nu_\mu \to \nu_e$ appearance), one can thus over-constrain the system, allowing
for consistency tests.  In practice, the above limits on $\Delta m_{21}^2$,
$\Delta m_{31}^2$, and $\Delta m_{41}^2$ are simultaneously realized only
in a small region of parameter space. However, as long as measurements
at different energies are available (which is almost always the case),
the conclusion that the system can be over-constrained remains valid.

\begin{figure}
  \centering
  \begin{tabular}{c}
    \includegraphics[width=0.45\textwidth]{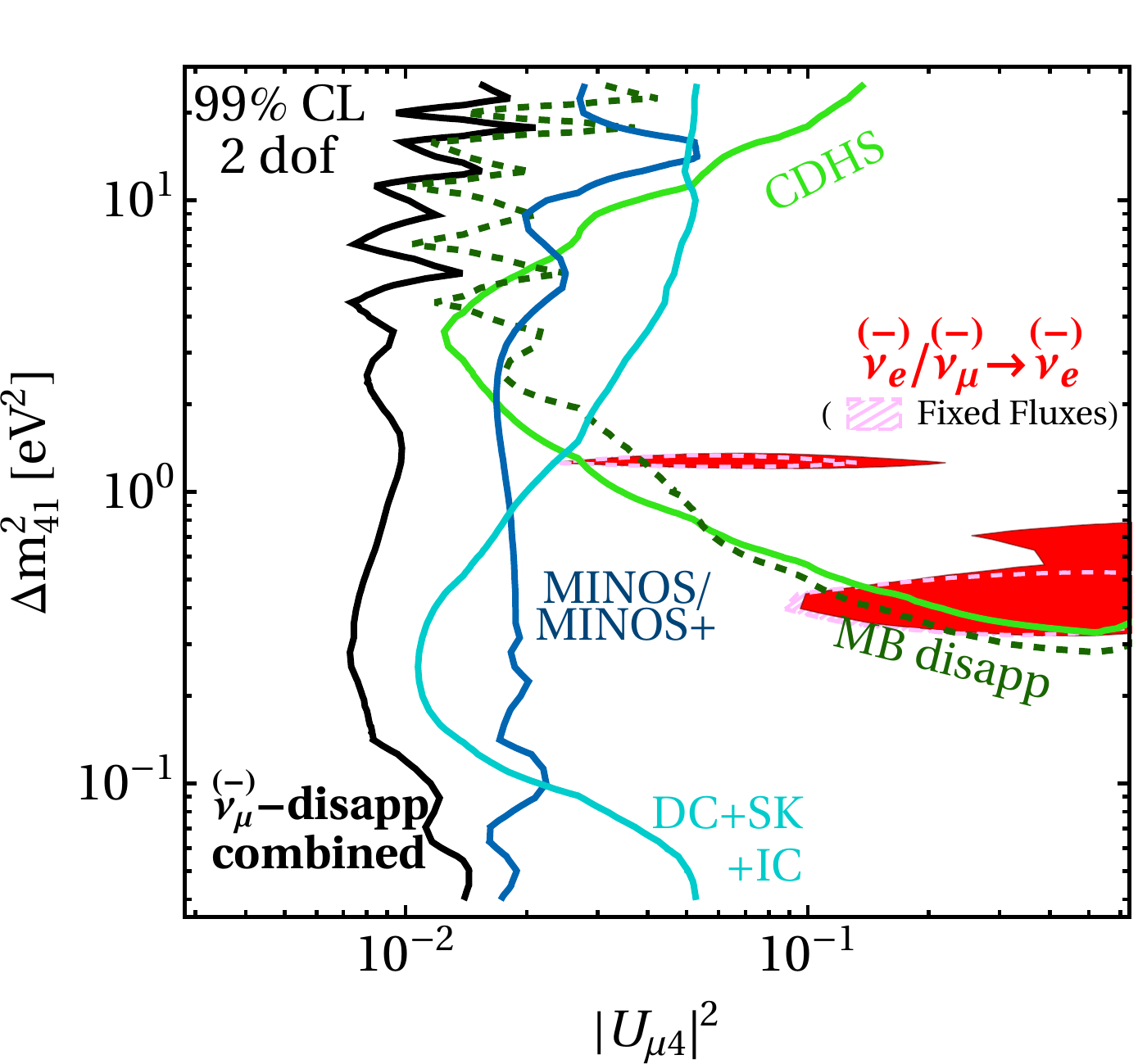}
  \end{tabular}
  \caption{Global constraints on short-baseline $\nu_\mu$ and $\bar\nu_\mu$ disappearance
  in the $3+1$ scenario, shown as a function of the $\nu_s$--$\nu_\mu$ mixing
  $|U_{\mu 4}|^2$ and the mass squared difference $\Delta m_{41}^2$.  Thick colored
  lines show exclusion limits from various searches, with the region to the left
  of the curves excluded. The red shaded region is preferred by the combination of
  $\nu_e/\bar\nu_e$ appearance and disappearance searches, where the anomalies
  are observed. There is clear tension between this data and the exclusion limits.
  Figure taken from ref.~\cite{Dentler:2018sju}.}
  \label{fig:numu-disapp}
\end{figure}

We illustrate this in \cref{fig:numu-disapp}, where we compare the
parameter region preferred by $\nu_e$ appearance experiments and $\nu_e$
disappearance experiments to the exclusion limits from $\nu_\mu$
disappearance searches in the $|U_{\mu 4}|^2$--$\Delta m_{41}^2$
plane.  Evidently, there is stark tension in the global data set:
the parameter region preferred by the short baseline anomalies is ruled out
at high significance by the null results from $\nu_\mu$ disappearance.

This tension has been quantified in ref.~\cite{Dentler:2018sju} using a
parameter-goodness-of-fit (PG) test~\cite{Maltoni:2003cu}. This test measures
the statistical ``penalty'' one has to pay for combining two data sets. It does
so by comparing the likelihood of the individual data sets at their respective
best fit points to the likelihood of the combined data set at the global best
fit point.  If the global short-baseline neutrino oscillation data were fully
consistent, the PG test should yield a large $p$-value for \emph{any} portioning
of the data into two independent subsets.  The authors of ref.~\cite{Maltoni:2003cu}
have in particular divided the data into appearance and disappearance data,
and have found a tiny $p$-value of \num{3.71e-7}.  The $p$-value remains
very small when any individual data set is removed from the fit.  Only removing
LSND has a significant impact, increasing the $p$-value to \num{1.6e-3}, which
is still small.

The consistency of the fit also does not improve when more than one
sterile neutrino is considered~\cite{Kopp:2011qd, Conrad:2012qt,
Kopp:2013vaa, Moulai:2019gpi}.  Even though the mixing matrix has
many more parameters in this case, the only qualitatively new feature
that appears in $3+2$ and $3+3$ models compared to the simple $3+1$
scenario is the possibility of CP violation at short baseline.  In the short-baseline
limit, the mixing matrix in a $3+1$ model reduces to a $2 \times 2$ matrix,
which does not admit CP violation; in a $3+2$ model, in contrast, CP violation
is possible even at short baseline. However, since
there is no tension between neutrino and anti-neutrino data, adding
CP violation does not improve the global fit.  The main source of tension
-- namely the fact that explaining $\nu_e$ appearance data requires
the sterile state(s) to have sizable mixing with muon neutrino, in
tension with $\nu_\mu$ disappearance data -- is not resolved by
including extra sterile states.

It is therefore clear that vanilla $3+n$ scenarios are not sufficient to
explain the entirety of the short-baseline anomalies.  They remain
a viable option to explain some of them, though.  In the next section,
we will comment on a number of extended sterile neutrino
models that are potentially able to relax at least some of the tension
in the global fit.

\section{Attempts to Resolve the Short-Baseline Anomalies}
\label{sec:nu-s-decay}

\begin{figure}
  \centering
  \includegraphics[width=\textwidth]{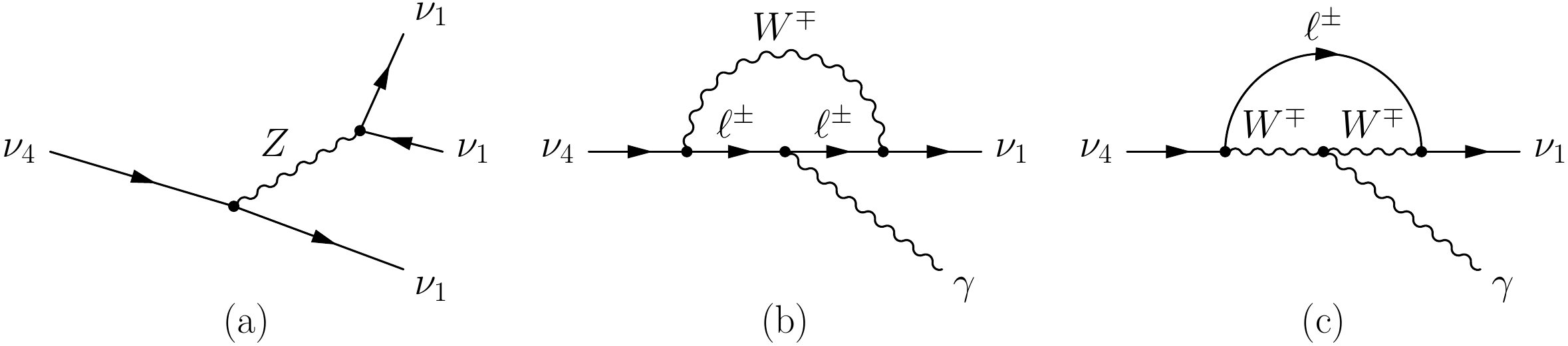}
  \caption{The leading Feynman diagrams contributing to $\nu_4$ decay in sterile
    neutrino models. The process $\nu_4 \to 3 \nu_a$ in panel (a) has the larger rate,
    while $\nu_4 \to \nu_a + \gamma$ (depicted in panels (b) and (c)) is easier to
    observe due to the emission of a monoenergetic photon.}
  \label{fig:nu-decay}
\end{figure}

The phenomenology of sterile neutrinos in oscillation experiments and elsewhere
is significantly enriched when sterile neutrinos can decay.  Without
further model ingredients, a heavy ($> \si{eV}$) neutrino state $\nu_s$ mixing with
the active neutrinos $\nu_a$ has two important decay modes: $\nu_4 \to 3 \nu_a$ and $\nu_4 \to \nu_a + \gamma$.
The corresponding leading-order Feynman diagrams are depicted in \cref{fig:nu-decay}.
The rate for $\nu_4 \to 3 \nu_a$ is~\cite{Lee:1977tib, Pal:1981rm, Barger:1995ty,
Adhikari:2016bei}
\begin{align}
  \Gamma(\nu_4 \to 3 \nu_a)
    &= \frac{G_F^2 m_4^5}{96 \pi^3} \sin^2\theta \notag\\
    &= \SI{7.0e-35}{sec^{-1}} \times
       \sin^2 \theta \times
       \bigg( \frac{m_4}{\si{eV}} \bigg)^5 \,,
  \label{eq:nu-3nu}
\intertext{while the decay rate for $\nu_4 \to \nu_a + \gamma$ is \cite{Shrock:1974nd,
Lee:1977tib, Marciano:1977wx, Pal:1981rm, Shrock:1982sc}}
  \Gamma(\nu_4 \to \nu_a \gamma)^\text{D}
  &= \frac{9 \alpha_\text{em} G_F^2 m_4^5}{512 \pi^4}
       \sum_{j=1,2,3} \bigg( 1 - \frac{m_j^2}{m_4^2} \bigg)^3
       \bigg( 1 + \frac{m_j^2}{m_4^2} \bigg)
       \bigg| \sum_{\alpha=e,\mu,\tau}
         \bigg( 1 - \frac{m_\alpha^2}{2 M_W^2} \bigg) U_{\alpha 4} U_{\alpha j}^* \bigg|^2
                                   \nonumber\\[0.2cm]
  &\simeq
     \SI{2.73e-37}{sec^{-1}} \times
       \sin^2 \theta \times
       \bigg( \frac{m_4}{\si{eV}} \bigg)^5
  \label{eq:nu-nu-gamma-Dirac}
\intertext{in the case of Dirac neutrinos, and \cite{Pal:1981rm, Shrock:1982sc, Xing:2011}}
  \Gamma(\nu_4 \to \nu_a \gamma)^\text{M}
  &= \frac{9 \alpha_\text{em} G_F^2 m_4^5}{256 \pi^4}
       \sum_{j=1,2,3} \bigg( 1 - \frac{m_j^2}{m_4^2} \bigg)^3
       \bigg\{
         \bigg( 1 + \frac{m_j^2}{m_4^2} \bigg)^2
         \bigg[ \sum_{\alpha=e,\mu,\tau}
           \bigg( 1 - \frac{m_\alpha^2}{2 M_W^2} \bigg) \im(U_{\alpha 4} U_{\alpha j}^*) \bigg]^2
                                   \nonumber\\[0.2cm]
  &\qquad
       + \bigg( 1 - \frac{m_j^2}{m_4^2} \bigg)^2
         \bigg[ \sum_{\alpha=e,\mu,\tau}
         \bigg( 1 - \frac{m_\alpha^2}{2 M_W^2} \bigg) \re(U_{\alpha 4} U_{\alpha j}^*) \bigg]^2
       \bigg\}
                                   \nonumber\\[0.2cm]
  &\simeq
     \SI{5.46e-37}{sec^{-1}} \times
       \sin^2 \theta \times
       \bigg( \frac{m_4}{\si{eV}} \bigg)^5
  \label{eq:nu-nu-gamma-Majorana}
\end{align}
for Majorana neutrinos.  In these expressions, $m_j$ ($j=1..4$) are the neutrino
mass eigenvalues,  $m_e$, $m_\mu$, and $m_\tau$
denote the charged lepton masses, $M_W$ is the $W$ boson mass, $G_F$ is the
Fermi constant, and $\alpha_\text{em}$ is the electromagnetic fine structure
constant.  The numerical approximations in the second lines of
\cref{eq:nu-nu-gamma-Dirac,eq:nu-nu-gamma-Majorana} were obtained in the limit
$m_j \ll m_4$ and $m_\alpha \ll M_W$. Moreover, in this limit, the dependence on
the mixing matrix elements can be expressed in terms of the effective mixing angle
$\sin^2\theta \equiv \sum_j |U_{s4} U_{sj}^*|^2$. Assuming that $\nu_4$ mixes
predominantly with only one of the light mass eigenstates, and that the corresponding
mixing angle is $\ll 1$, $\theta$ can be identified with that mixing angle.

From the above numerical estimates, it is clear that in simple $3+1$ or $3+n$
scenarios, neutrino decay is completely irrelevant in terrestrial experiments
unless sterile neutrinos with masses $\gtrsim \SI{100}{MeV}$ exist.  However, the
situation changes dramatically in models with extended sterile sectors.

We will now review the most important classes of such models, in particular
those which have been proposed as possible explanations for some of the
short baseline anomalies. It is worth emphasizing already here that most models
with decaying are geared towards the MiniBooNE anomaly and cannot explain all
anomalies simultaneously.

\subsection{Sterile Neutrino Decay to Photons}

\subsubsection{Sterile Neutrino Production in the Target, followed by
  $\nu_4 \to \nu\gamma$ in the detector}

The decay of sterile neutrinos to photons can be boosted compared to
\cref{eq:nu-nu-gamma-Dirac,eq:nu-nu-gamma-Majorana} if neutrinos possess
transition magnetic moments, that is couplings of the form
\begin{align}
  \mathcal{L}_\text{magn.\ moment} \supset
  \frac{1}{\Lambda} \bar\nu_s \sigma^{\alpha\beta} \nu_a F_{\alpha\beta} \,.
  \label{eq:mm}
\end{align}
Here, $\nu_4$ is the mostly sterile mass eigenstate, $\nu_a$ denotes
one of the light neutrino mass eigenstates, and $F_{\alpha\beta}$ is the
electromagnetic field strength tensor. In fact, the loop diagrams shown
in \cref{fig:nu-decay} (b) and (c) generate precisely such an operator,
with a suppression scale of order $\Lambda \sim M_W^2 / (m_4 \sin\theta)$. In extensions
of the $3+1$ model, however, $\Lambda$ can be smaller, leading to an enhanced decay
rate \cite{Fischer:2019fbw}
\begin{align}
  \Gamma(\nu_4 \to \nu_a\gamma) \simeq \frac{m_4^3}{4 \pi \Lambda^2}
                          \simeq \SI{1.2e4}{sec^{-1}} \times
                                 \bigg( \frac{m_4}{\SI{100}{MeV}} \bigg)^3
                                 \bigg( \frac{\SI{1e5}{TeV}}{\Lambda} \bigg)^2 \,.
  \label{eq:nu-decay-rate-mm}
\end{align}
It is thus possible for neutrinos with $\mathcal{O}(\SI{100}{MeV})$ masses to
decay on their way from the neutrino source to the detector.

The authors of ref.~\cite{Fischer:2019fbw} consider this scenario
particularly in the context of the MiniBooNE experiment. In MiniBooNE, the heavy mass
eigenstate $\nu_4$ could be copiously produced in the target via kaon decay,
$K \to \ell \nu_4$
(where $\ell$ is a charged lepton) if their mixing with active neutrinos
is of order $10^{-11} \lesssim |U_{\ell 4}|^2 \lesssim 10^{-7}$.  Some of them
would decay inside the detector, where the final state photon could be easily
misreconstructed as an electron, thus faking a charged current $\nu_e$ interaction.
Constraints on this scenario arise from the angular distribution of events
in MiniBooNE and from experiments studying the kinematics of kaon decay.

\subsubsection{Sterile Neutrino Production in the Detector, Followed by the
  Decay $\nu_4 \to \nu_a\gamma$}

As an alternative to the production of $\nu_4$ at the target station of an
accelerator-based neutrino beam, heavy sterile neutrinos can also be produced
in neutral current $\nu_\mu$--nucleus interactions in the detector
\cite{Gninenko:2009ks, Gninenko:2010pr}. Strong constraints on this
scenario were obtained by the ISTRA+ experiment in Protvino, Russia,
searching for the anomalous decay $K^- \to \mu^- + (\nu_4 \to \nu_a \gamma)$.
In particular, the squared mixing matrix element $|U_{\mu 4}|^2$ describing
the mixing between $\nu_4$ and $\nu_\mu$ is constrained to be $\lesssim
10^{-4}$--$10^{-5}$ for $m_4$ between 30 and \SI{80}{MeV}, and
for $\nu_4$ lifetimes between \SI{1e-11}{sec} and \SI{1e-9}{sec}.
Note that such short lifetimes, while needed to explain the MiniBooNE and LSND
anomalies with $\nu_4 \to \nu_a \gamma$ decays inside the detector, cannot be
realized without further ingredients such as large transition magnetic moments,
see \cref{eq:nu-nu-gamma-Dirac,eq:nu-nu-gamma-Majorana}.

\subsection{Sterile Neutrino Decay to Dark Photons}

Exploring alternative decay modes of heavy sterile neutrinos,
the authors of refs.~\cite{Bertuzzo:2018itn, Ballett:2018ynz} extend the SM gauge
group by an extra, ``dark'' $U(1)'$ factor. The corresponding gauge
boson $A'$ (the ``dark photon'') couples directly to the sterile flavor 
eigenstate $\nu_s$ via a minimal
gauge interaction of the form
\begin{align}
  \mathcal{L} \supset g' A'^\mu \bar\nu_s \nu_s \,,
\end{align}
where $g'$ is the $U(1)'$ gauge coupling constant.  Couplings to the
standard model can be induced by a gauge kinetic mixing term of the form
\begin{align}
  \mathcal{L} \supset -\frac{\varepsilon}{4} F^{\mu\nu} F'_{\mu\nu} \,,
\end{align}
where $F^{\mu\nu}$ is the electromagnetic field strength tensor,
$F'^{\mu\nu}$ is the $U(1)'$ field strength tensor, and $\varepsilon$ is
a small dimensionless parameter. We refer the reader to
refs.~\cite{Jaeckel:2010ni, Essig:2013lka, Fabbrichesi:2020wbt}
for detailed reviews on dark photon physics.

If the mostly sterile neutrino mass eigenstate is heavier than the dark photon,
$m_4 > m_{A'}$, decays of the form $\nu_4 \to \nu_a A'$ are possible, where $\nu_a$
is again one of the light neutrino mass eigenstates. The dark photon will subsequently
decay via its $\varepsilon$-suppressed coupling to the electromagnetic
current. For $2 m_e < m_{A'} < 2 m_\mu$, the only allowed decay channel is
$A' \to e^+ e^-$.  As long as $m_{A'} \gg m_e$, the electron--positron pair
will be strongly boosted in the forward direction.  In detectors with
limited track resolution -- such as MiniBooNE -- it is therefore likely to
be reconstructed as a single electron, mimicking the signature of a charged
current $\nu_e$ interaction.

In view of this possibility, the authors of
ref.~\cite{Bertuzzo:2018itn} propose to explain the MiniBooNE anomaly by
assuming heavy sterile neutrinos are produced in the detector via
neutral current neutrino--nucleus scattering of the form $\nu_\mu + A \to \nu_4 + X$,
with the target nucleus $A$ and the hadronic interaction product(s) $X$.
The subsequent decay $\nu_4 \to \nu_a + (A' \to e^+ e^-)$ can explain the
observed event excess for sterile neutrino masses around $m_4 \sim \SI{100}{MeV}$,
dark photon masses $m_{A'} \sim \SI{30}{MeV}$,
active-to-sterile neutrino mixings of order $|U_{\mu 4}|^2 \sim \num{1e-8}$,
and kinetic mixing of order $\varepsilon \sim 10^{-4}$.

If, on the other hand, $m_A' > m_4$, the sterile neutrinos will decay via an
off-shell $A'$~\cite{Ballett:2018ynz}. For $m_4 < 2 m_\mu$, the only decay mode
will be $\nu_4 \to \nu_a e^+ e^-$.  Once again, this is an interesting process
that can be searched for in neutrino experiments.  In the specific case of
MiniBooNE, the $e^+ e^-$ pair can mimic a charged current $\nu_e$ interaction,
thus possibly explaining the observed low-energy excess.

An important consideration for explanations of the MiniBooNE anomaly invoking
heavy sterile neutrino decays to photons or or $e^+ e^-$ pairs is the
angular distribution of detected events.  In particular, the decay products of
boosted $\nu_4$ are predominantly emitted in the forward direction.  A true
$\nu_e$ appearance signal, on the other hand, leads to a more isotropic
distribution of the $e^+$ or $e^-$ produced in CC interactions.
Current MiniBooNE data appears more consistent with the latter hypothesis
than with a strongly forward-peaked distribution~\cite{Aguilar-Arevalo:2020nvw}.

\subsection{Active Neutrinos from Sterile Neutrino Decay}

In the previous sections, we have discussed scenarios in which sterile neutrino
decay yields electromagnetically interacting, and thus easy to observe, products.
However, even active neutrinos produced in sterile neutrino decay can have
important phenomenological consequences.  In particular, their energy spectrum
will be shifted to lower energies compared to the bulk of the beam.  Moreover, their
flavor composition may be significantly altered.

The authors of refs.~\cite{PalomaresRuiz:2005vf, Moss:2017pur,
Dentler:2019dhz, Hostert:2020oui}
consider the decay $\nu_i \to \nu_j + \phi$, where $i,j = 1\dots4$, and $\phi$
is a new scalar or pseudoscalar boson.  In \cite{Dentler:2019dhz, Hostert:2020oui},
also the subsequent decay $\phi \to \nu_j \nu_k$ is considered.
In the presence of neutrino decay, the
evolution of a neutrino ensemble is most easily described in the density matrix
formalism.  Let us therefore introduce the energy ($E$) and time ($t$)
dependent neutrino density matrix $\hat\rho_\nu(E, t)$, the corresponding
anti-neutrino density matrix $\hat\rho_{\bar\nu}(E, t)$, and the scalar density
functional $\hat\rho_\phi(E, t)$.  It is understood that $\hat\rho_\nu(E, t)$
and $\hat\rho_{\bar\nu}(E, t)$ are $n \times n$ matrices in flavor space, where
$n$ is the number of neutrino flavors. The evolution equations read
\cite{GonzalezGarcia:2005xw, Moss:2017pur, Dentler:2019dhz}
\begin{align}
  \frac{d\hat\rho_\nu(E, t)}{dt}
    &= - i [\hat{H}, \hat\rho_\nu]
       - \frac{1}{2} \big\{ \tfrac{m_4}{E} \hat\Gamma, \rho \big\}
       + \mathcal{R}_\nu[\hat\rho_\nu, \hat\rho_\phi, E, t] \,,
                     \label{eq:nu-decay-eom-1} \\
  \frac{d\hat\rho_\phi(E, t)}{dt}
    &= - \tfrac{m_\phi}{E} \Gamma_\phi \rho_\phi
       + \mathcal{R}_\phi[\hat\rho_\nu, E, t] \,,
                     \label{eq:nu-decay-eom-2}
\end{align}
where $\hat{H} \equiv \tfrac{1}{2 E} \diag(0, \Delta m_{21}^2, \Delta m_{31}^2, \dots)$
is the standard neutrino oscillation Hamiltonian in the mass basis.
The second term
in \cref{eq:nu-decay-eom-1} describes neutrino decay. It depends on the decay
operator $\hat\Gamma \equiv \sum_j \tfrac{m_j}{E} \Gamma_j \hat\Pi_j$, which contains the
total rest frame decay width $\Gamma_j$ of the $j$-th neutrino mass eigenstate,
as well as the projection operator $\hat\Pi_j \equiv \ket{\nu_j}\bra{\nu_j}$
onto that mass eigenstate.  The third term in \cref{eq:nu-decay-eom-1}
accounts for the flux of daughter neutrinos from the decay processes
$\nu_i \to \nu_j + \phi$ and $\phi \to \nu_j \nu_k$.  Neglecting the masses of
the daughter neutrinos compared to the parent neutrinos, It is given by
\begin{align}
\begin{split}
  \mathcal{R}_\nu[\hat\rho_\nu, \hat\rho_\phi, E, t]
    &= \sum_i \hat{\Pi}_{Fi} \int_{\frac{E}{1-m_\phi^2/m_i^2}}^{\infty} \! dE_i \sum_k
       \bigg[
         \hat\rho_{\nu,ii}(E_i, t) \, \frac{d\Gamma^\text{lab}(\nu_i \to \nu_k \phi)}{dE_k}
       + \hat{\rho}_{\bar\nu,ii}(E_i, t) \,
              \frac{d\Gamma^\text{lab}(\bar{\nu}_i \to \nu_k \phi)}{dE_k}
       \bigg] \\
    &+ \hat{\Pi}_{F\phi} \sum_{k, j} \int_E^{\infty} \! dE_\phi \,
       \hat\rho_\phi(E_\phi, t) \frac{d\Gamma^\text{lab}(\phi \to \nu_k \bar{\nu}_j)}{dE_k} \,,
\end{split}
\label{eq:R-nu}
\end{align}
where $d\Gamma^\text{lab}(X \to Y) / dE_k$ is the differential decay width for the various
decays $X \to Y$ \emph{in the laboratory frame}, and $\hat{\Pi}_{Fj}$,
$\hat{\Pi}_{F\phi}$ are projection operators onto the final states of the
respective decays.  The integral runs over all parent energies $E_j$ that lead
to daughter neutrinos with energy $E$.  The sum over $i$ runs over all parent
neutrino mass eigenstates, while the sums over $k$ and $j$ run over daughter
mass eigenstates.  The second term in square brackets in \cref{eq:R-nu}
is only present if neutrinos are Majorana particles. For Dirac particles,
lepton number violating decays of the form $\bar{\nu}_j \to \nu_k \phi$
are obviously forbidden. In analogy to \cref{eq:R-nu}, also the evolution
equation for $\phi$, \cref{eq:nu-decay-eom-2} contains a decay term that depends
on the total $\phi$ decay width $\Gamma_\phi$, and a regeneration term
\begin{align}
\begin{split}
  \mathcal{R}_\phi[\hat\rho_\nu, E, t]
    &= \sum_i \int_{E}^{E \, m_i^2/m_\phi^2} \! dE_i \sum_k \bigg[
         \hat\rho_{\nu,ii}(E_i, t) \, \frac{d\Gamma(\nu_i \to \parenbar{\nu}_k \phi)}{dE_\phi}
       + \hat{\rho}_{\bar\nu,ii}(E_4, t)
         \frac{d\Gamma(\bar{\nu}_i \to \parenbar{\nu}_k \phi)}{dE_\phi}
       \bigg] \,.
\end{split}
\end{align}

Applying this formalism in the context of specific experiments, it has been
noted already in ref.~\cite{PalomaresRuiz:2005vf} that the active neutrinos
from the decay of heavier, sterile, neutrinos can explain the LSND anomaly
because their flavor composition can be chosen such that it is dominated
by electron neutrinos, in agreement with the LSND data. In ref.~\cite{Dentler:2019dhz},
this study has been extended to the full set of short baseline anomalies,
noting in particular the excellent agreement with MiniBooNE.  There are several
reasons for this good agreement: the first one is again the flavor composition
of the daughter neutrinos from $\nu_4$ and $\phi$ decays, which can be chosen
to be dominated by $\nu_e$. Second, the daughter neutrinos  tend to accumulate
at the lower end of the spectrum, exactly where MiniBooNE sees its excess.
Third, the required mixing $|U_{\mu 4}|^2$ between sterile neutrinos and
$\nu_\mu$ is much smaller in the neutrino decay scenario than in $3+1$
scenarios with oscillations only, where the signal is suppressed not only by
$|U_{\mu 4}|^2$, but by the product $|U_{\mu 4}|^2 |U_{e4}^2$.  Fourth, the
sterile neutrino can be heavier (up to $\sim \SI{1}{keV}$, making it easier to
avoid cosmological constraints as well as $\nu_\mu$ disappearance constraints
from oscillation experiments. For instance, as noted in \cite{Moss:2017pur},
the IceCube constraint on $|U_{\mu 4}|^2$ is significantly weakened in presence
of substantial neutrino decay. The reason is that IceCube's sensitivity comes
from Earth matter effects on the neutrino ensemble (in particular from a
Mikheyev--Smirnov--Wolfenstein (MSW) resonance between active and sterile
neutrinos), and these matter
effects cannot develop if the sterile neutrino decays before traveling through
a significant amount of matter.   Finally, the model has the ``secret
interactions'' mechanism (see \cref{sec:non-standard-cosmo}) built-in, guaranteeing
compatibility with the cosmological constraint on the effective number of
neutrino species, $N_\text{eff}$.   A combined explanation of LSND and
MiniBooNE, as well as the reactor and gallium anomalies may also be possible
\cite{Dentler:2019dhz, Hostert:2020oui}, but requires some extensions on the
particle physics side as well as somewhat non-standard cosmological evolution to
avoid constraints that arise because the model predicts active neutrinos to
free-stream less than in the standard $\Lambda$CDM scenario, see
\cref{sec:cosmo-observables,sec:non-standard-cosmo}

\subsection{Invisibly Decaying Sterile Neutrinos}

Several authors have considered the possibility of completely invisible sterile
neutrino decay modes, in particular the decay into a second, lighter, sterile
neutrino and a new scalar or pseudoscalar
\cite{
  Gariazzo:2014pja,  
  Denton:2018aml}.   
In oscillation experiments, such a scenario would lead to disappearance
signals identical to those in a $3+1$ model, up to corrections of order
$|U_{\alpha 4}|^4$, where $\alpha$ is the disappearing active neutrino flavor.
These corrections arise if the $\nu_4$ decays before reaching the detector
because in this case the small admixture of active neutrinos to the mostly
sterile mass eigenstate $\nu_4$ does no longer contribute to the detected flux
as it has decayed. For the same reason, also appearance signals will be
modified (though in general not completely repressed) if $\nu_4$ decays before
reaching the detector.  Of course, if the daughter fermion produced in the
decay has a non-negligible mixing with active neutrinos, the appearance of
these daughter particles will lead to new signals -- though the decay then
shouldn't be called invisible any more.

\subsection{Active Neutrino Decays into Sterile Neutrinos}

In the same way as sterile neutrinos can decay into active ones, also
active neutrinos can decay into sterile neutrinos, provided the sterile
neutrino and the other decay products are sufficiently light.
Due to the small masses of active neutrinos, the decay lengths for such
decays are typically too long to be relevant to terrestrial experiments.

They may, however,  be relevant to astrophysical neutrinos. In fact, active
neutrino decay has been originally proposed in ref.~\cite{Bahcall:1972my} as a
possible (but ultimately unsuccessful \cite{Acker:1993sz}) solution of the
solar neutrino problem.  Nevertheless, active neutrino decay is still
interesting as a possible subdominant process for solar neutrinos
\cite{Berezhiani:1991vk, Berezhiani:1992xg, Bandyopadhyay:2001ct,
Bandyopadhyay:2002qg, Berryman:2014qha}.
Moreover, it could be relevant to high-energy astrophysical neutrinos
\cite{Denton:2018aml, deSalas:2018kri} and to cosmology \cite{Escudero:2019gfk,
Escudero:2020ped}.

\section{Constraints from Beta Decay Kinematics}
\label{sec:kinematics}

\begin{figure}
  \centering
  \includegraphics[width=0.6\textwidth]{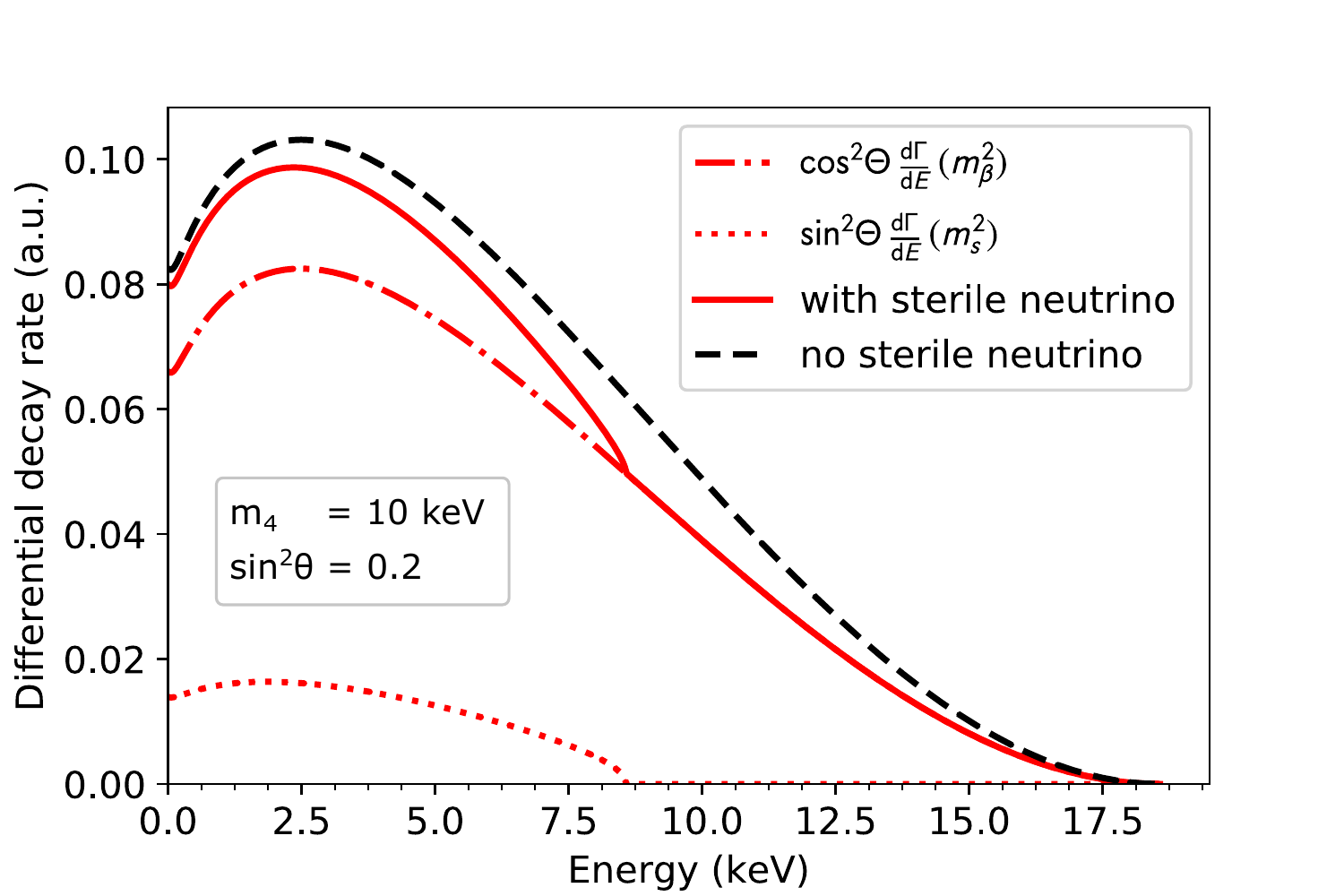}
  \caption{Electron spectrum from tritium beta decay in the SM (dashed black) and
  in presence of a \SI{10}{keV} sterile neutrino (solid red).  The $\nu_s$--$\nu_e$ mixing
  angle $\Theta$ has been chosen unphysically large, $\sin^2 \Theta = 0.2$ to better
  illustrate the effect.  The dash-dotted and dotted red curves show the
  contributions from the light mass eigenstates and from the heavy mass eigenstate,
  respectively.  Figure adapted from ref.~\cite{Mertens:2018vuu}.}
  \label{fig:katrin}
\end{figure}

If additional neutrino mass eigenstates exist with masses $\lesssim \si{MeV}$,
these states can be emitted in nuclear beta decays through their mixing with
the $\nu_e$ flavor eigenstate.  As the shape of the $e^\pm$ spectrum from
beta decay depends on the mass of the emitted neutrino state, models with
sterile neutrinos can be tested by precisely measuring this spectrum.  In particular,
in a $3+n$ model, the experimentally observed spectrum will be a superposition of $3+n$
spectra (one for each neutrino mass eigenstate), weighted with the mixing
matrix elements $|U_{ej}|^2$.   This is illustrated in \cref{fig:katrin}
(adapted from ref.~\cite{Mertens:2018vuu}) for the case of tritium beta decay
(SM endpoint energy $Q = \SI{18.59}{keV}$) and for sterile neutrinos with mass
$m_4 = \SI{10}{keV}$ and with unphysically large but illustrative mixing
$\sin^2 \Theta = 0.2$. A pronounced kink is visible in the spectrum at energy
$Q - m_4$, corresponding to the endpoint energy for decays into an $e^- +
\nu_4$ final state.

Such kinks have been searched for in the beta decay spectra of numerous isotopes,
and stringent constraints have been imposed on the $\nu_s$--$\nu_e$ mixing
$|U_{e4}|^2$~\cite{Atre:2009rg,
Dragoun:2015oja, 
deGouvea:2015euy,
Bryman:2019ssi, 
Bolton:2019pcu,
Giunti:2019fcj, 
Aker:2020vrf}. 
The KATRIN collaboration is planning to install a dedicated detector called
TRISTAN to enhance their sensitivity especially in the keV mass range which is
interesting for sterile neutrino dark matter~\cite{Mertens:2018vuu, 
Barry:2014ika, 
Benso:2019jog}. 

\begin{figure}
  \centering
  \includegraphics[width=\textwidth]{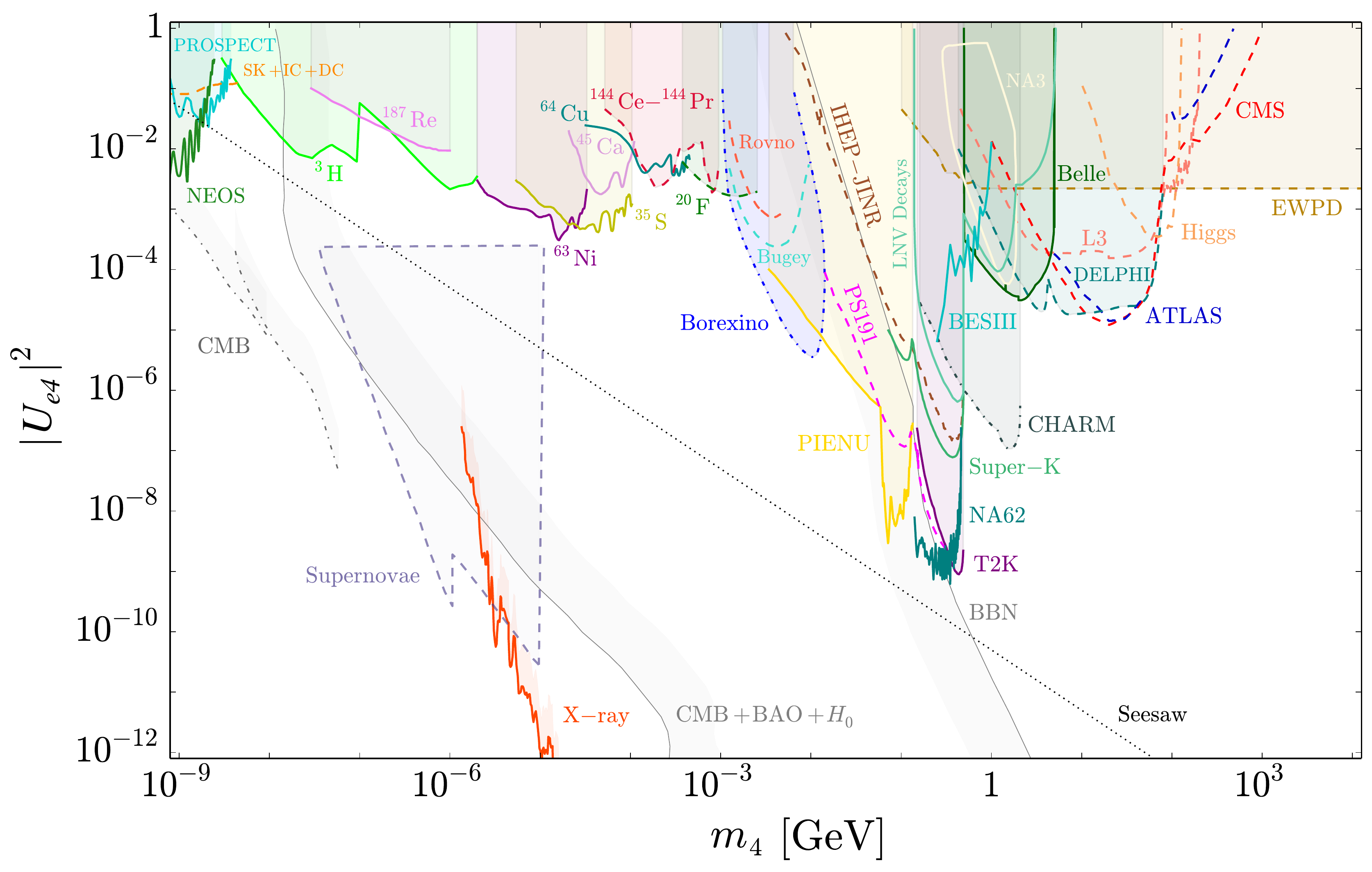}
  \caption{Constraints on the mixing of electron neutrinos with a sterile neutrino
  as a function of the sterile neutrino mass. See text for details. Figure adapted
  from ref.~\cite{Bolton:2019pcu}.}
  \label{fig:beta-decay-limits-nue}
\end{figure}

A summary of current constraints is shown in \cref{fig:beta-decay-limits-nue},
adapted from ref.~\cite{Bolton:2019pcu}.  Exclusion regions from beta decay
kinematics, shown as colored regions in the top part of the plot and labeled
with the respective isotopes, dominate at masses between $m_4 \sim \SI{10}{eV}$
and $m_4 \sim \SI{1}{MeV}$~\cite{Hiddemann:1995ce, Kraus:2012he,
  Belesev:2013cba, Abdurashitov:2017kka, Calaprice:1983qn, Holzschuh:2000nj,
  Derbin:1997ut, Holzschuh:1999vy, Schreckenbach:1983cg, Derbin:2018dbu,
Galeazzi:2001py}.  In this mass range, the mixing matrix element $|U_{e4}|^2$
is constrained to be below well below $10^{-2}$, with constraints reaching well
below $10^{-3}$ at keV-scale masses. This implies in particular that the
oscillation anomalies in the $\nu_e$ sector -- namely the reactor and gallium
anomalies -- cannot be explained by sterile neutrinos with masses above $\sim
\SI{10}{eV}$.

Nuclear beta decay constraints are less important at very low $m_4 \lesssim
\SI{10}{eV}$, where the kink in the beta decay spectrum moves too close to the
endpoint to be discernible. In this mass range, the limit is indeed dominated
by oscillation searches, with \cref{fig:beta-decay-limits-nue} showing in
particular the limits from the reactor neutrino experiments PROSPECT
\cite{Ashenfelter:2018iov} and NEOS \cite{Ko:2016owz}, and by studies of
atmospheric neutrinos in SuperKamiokande (SK), IceCube (IC) and DeepCore (DC)
\cite{Dentler:2018sju}.

\Cref{fig:beta-decay-limits-nue} also shows the generic, theoretically
expected value for $|U_{e4}|^2 \sim m_\nu / m_4$ (with $m_\nu = \SI{0.05}{eV}$)
in type-I seesaw models.  We see that none of the laboratory constraints
reach this parameter region, with the exception of oscillation searches at very
low $m_4$, where the main motivation for the type-I seesaw -- explaining the
smallness of neutrino masses -- is lost.

For masses above $\sim \si{MeV}$, where sterile neutrino production in nuclear
decays becomes kinematically forbidden, one can instead use the lepton spectra
from meson decays to set constraints. As can be seen from
\cref{fig:beta-decay-limits-nue}, the strongest limits on $\nu_e$--$\nu_s$
mixing are obtained from studies of the decay $\pi \to e \nu$ in the
appropriately named PIENU experiment \cite{Aguilar-Arevalo:2017vlf,
Bryman:2019bjg}, and of $K^+ \to e \nu$ in NA62~\cite{CortinaGil:2017mqf,
Abada:2016plb}.

In a similar way, also mixing of sterile neutrinos with $\nu_\mu$ is
constrained.  The relevant constraints, shown in the top panel of
\cref{fig:beta-decay-limits-numu-nutau} are from
PIENU~\cite{Aguilar-Arevalo:2019owf} and PSI~\cite{Daum:1987bg} looking for
modifications of $\pi \to \mu \nu$ decay, and from
KEK~\cite{Hayano:1982wu,Yamazaki:1984sj}, E949~\cite{Artamonov:2014urb}, and
NA62~\cite{CortinaGil:2017mqf, Lazzeroni:2019} using kaon decays.  In addition,
precision studies of the muon decay spectrum~\cite{Shrock:1981wq} are relevant
for sterile neutrino masses just below \SI{100}{MeV}.  Once again, searches
using modified  decay spectra lose their sensitivity at low $m_4$. Hence, a
vast range of sterile neutrino masses is left unconstrained by laboratory
searches in \cref{fig:beta-decay-limits-numu-nutau}. At very small masses, $m_4
\lesssim \SI{100}{eV}$, it is once again oscillation experiments that set limits,
in particular IceCube~\cite{TheIceCube:2016oqi}, DeepCore~\cite{Aartsen:2017bap}, and
Super-Kamiokande~\cite{Abe:2014gda} using disappearance of atmospheric
$\nu_\mu$, CDHS~\cite{Dydak:1983zq},
CCFR~\cite{Stockdale:1984cg}, and MiniBooNE~\cite{AguilarArevalo:2009yj,
Cheng:2012yy}, using disappearance of beam $\nu_\mu$, NO$\nu$A
\cite{Adamson:2017zcg} looking for a deficit in neutral current events, and
MINOS/MINOS+~\cite{Adamson:2017uda} using both $\nu_\mu$ disappearance and
neutral current events.

Constraints at masses $m_4 \gtrsim \SI{100}{MeV}$ will be discussed in
\cref{sec:decay-constraints}, and cosmological constraints (labeled CMB,
CMB+BAO+$H_0$ and BBN in \cref{fig:beta-decay-limits-nue} will be the topic of
\cref{sec:cosmology}.

Note that, at masses of order keV, sterile neutrinos are an interesting dark
matter candidate \cite{Dodelson:1993je,Shi:1998km,Boyarsky:2009ix,Adhikari:2016bei,
Abazajian:2017tcc}, see \cref{sec:keVchapt}. If they are indeed abundant
in the Universe, very stringent constraints can be set by looking for the
radiative decay $\nu_4 \to \nu_{1,2,3} \gamma$, which leads to a monochromatic
peak in the astrophysical X-ray flux.  As such a peak is easy to distinguish
from astrophysical backgrounds and foregrounds, the sensitivity is superb, as
shown by the jagged red curve labeled ``X-ray'' in
\cref{fig:beta-decay-limits-nue} \cite{Ng:2019gch, Roach:2019ctw}.
We emphasize that this limit only
applies if the heavy mass eigenstate $\nu_4$ is abundant in the Universe, with
its abundance matching the observed dark matter abundance.  Models that avoid
cosmological limits by strongly suppressing the $\nu_4$ abundance therefore
avoid this limit as well.

\section{Constraints from Sterile Neutrino Decay}
\label{sec:decay-constraints}

While the main focus of this review is on very light ($\lesssim \text{few eV}$)
sterile neutrinos, we also comment briefly on searches for heavier sterile
neutrinos.  In this section, we closely follow ref.~\cite{Bolton:2019pcu},
which offers a much more extensive review on constraints on heavy sterile
neutrinos.  We have already seen in \cref{sec:kinematics}
and \cref{fig:beta-decay-limits-nue} that nuclear beta decay spectra
place strong limits at masses up to about \SI{1}{MeV}.

\begin{figure}
  \centering
  \includegraphics[width=0.95\textwidth]{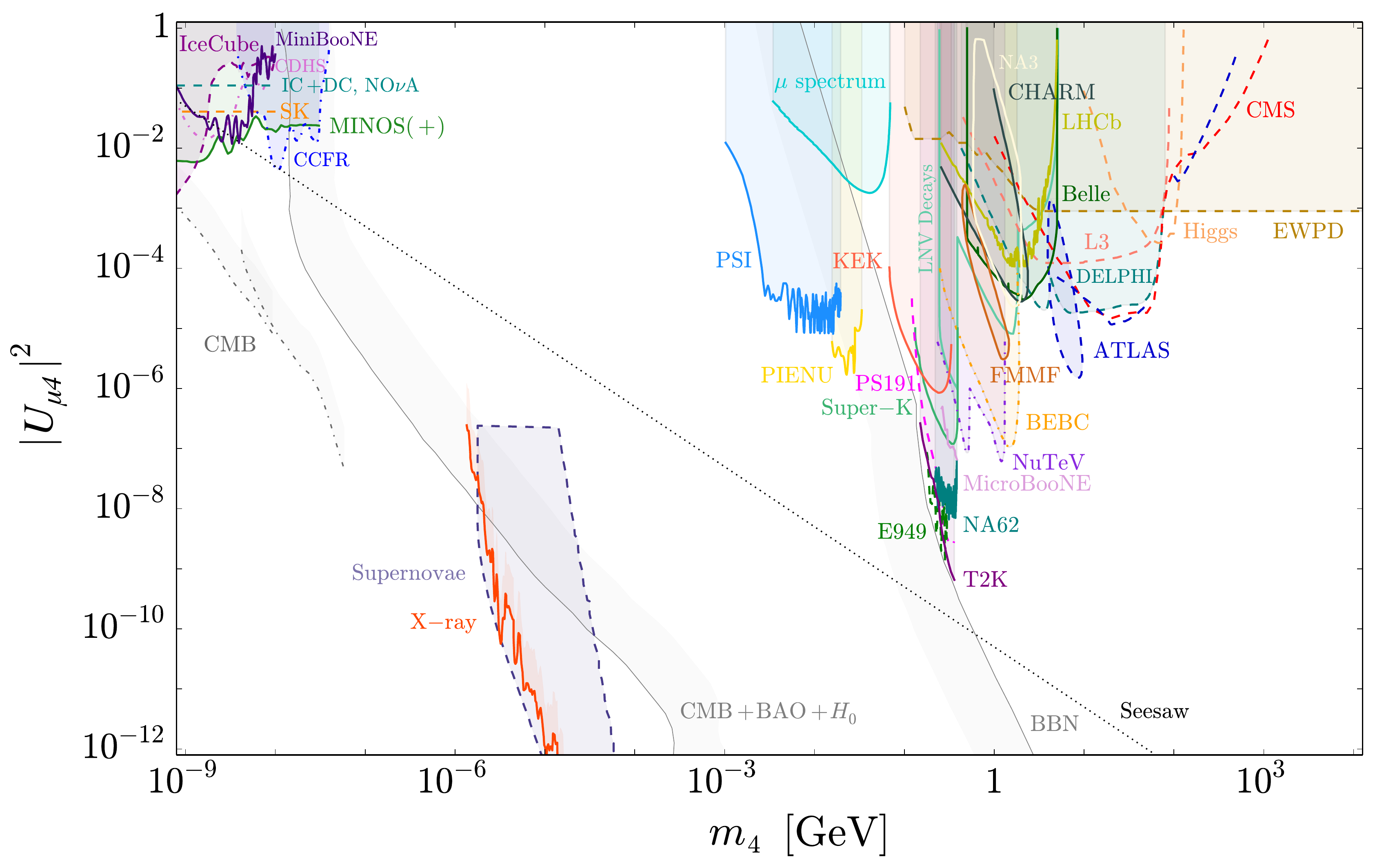} \\[0.3cm]
  \includegraphics[width=0.95\textwidth]{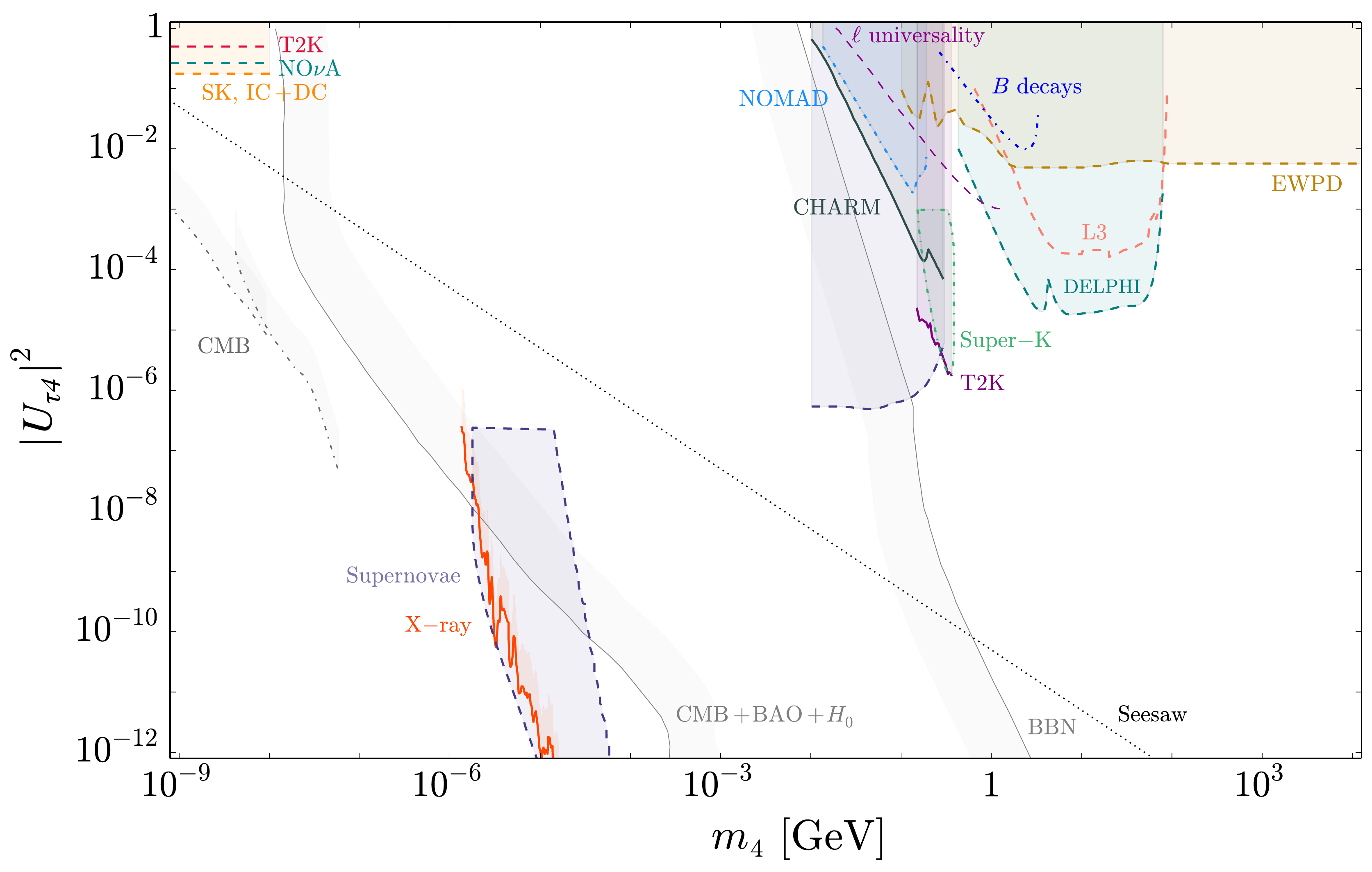}
  \caption{Constraints on the mixing of muon neutrinos (top) and tau neutrinos
  (bottom) with a sterile neutrino
  as a function of the sterile neutrino mass. See text for details. Figure adapted
  from ref.~\cite{Bolton:2019pcu}.}
  \label{fig:beta-decay-limits-numu-nutau}
\end{figure}

An even more important role is played over a wide mass range by decays of
the sterile neutrinos themselves.  At $\SI{1}{MeV} \lesssim m_4 \lesssim
\SI{10}{MeV}$, strong constraints on $\nu_e$--$\nu_s$ mixing
are obtained from searches for the decay $\nu_4 \to e^+ e^- \nu$ in the
reactor neutrino experiments Rovno \cite{Derbin:1993wy} and Bugey
\cite{Hagner:1995bn}. In the same mass range, Borexino
sets highly competitive limits by looking for modifications of the
solar \iso{B}{8} neutrino spectrum \cite{Bellini:2013uui}.
Indeed, if a fraction of the $\iso{B}{8}$
neutrino flux is carried by the heavy mass eigenstate $m_4$, the
decay $\nu_4 \to e^+ e^- \nu$ occurring along the way from the Sun to the
detector makes the spectrum softer than in the SM. 

Sterile neutrino decays have also been searched for in
Belle~\cite{Liventsev:2013zz}, Super-Kamiokande~\cite{Coloma:2019htx},
NA3~\cite{Badier:1986xz}, CHARM~\cite{Bergsma:1985is, Vilain:1994vg,
Orloff:2002de}, NOMAD~\cite{Astier:2001ck}, PS191~\cite{Bernardi:1987ek},
IHEP-JINR~\cite{Baranov:1992vq,Aguilar-Arevalo:2017vlf},
T2K~\cite{Abe:2019kgx}, E949~\cite{Artamonov:2014urb},
MicroBooNE~\cite{Abratenko:2019kez}, NuTeV~\cite{Vaitaitis:1999wq},
BEBC~\cite{CooperSarkar:1985nh}, and FMMF~\cite{Gallas:1994xp}. In these
experiments, $\nu_4$ may be produced in meson decays, and the most relevant
decay modes are $\nu_4 \to \ell \pi$ and $\nu_4 \to \ell^+ \ell^- \nu$.  (Here,
$\ell = e, \mu$ denotes a charged lepton.) The primary mesons originate from
$e^+e^-$ collisions in Belle, from the collisions of cosmic rays with the
Earth's atmosphere in Super-Kamiokande, from pions hitting a target in NA3, and
from protons hitting a target in CHARM, NOMAD, PS191, IHEP-JINR, T2K, E949,
MicroBooNE, NuTeV, BEBC, and FMMF.
An interesting possibility is that heavy sterile neutrino decays may actually
help explain some anomalies observed in semi-leptonic $B$ meson
decays~\cite{Cvetic:2017gkt}. (The bottom panel of \cref{fig:beta-decay-limits-numu-nutau}
shows only the upper limit on $|U_{\tau 4}|^2$ from this study.)
Note that not all experiments listed here
are sensitive to mixing with all active neutrino flavors, hence only subsets of
them appear in each of the three plots of
\cref{fig:beta-decay-limits-nue,fig:beta-decay-limits-numu-nutau}.

Of course, collider searches at LEP and LHC also weigh in on sterile neutrino
constraints, especially at masses $\gtrsim \si{GeV}$: the LEP experiments
L3~\cite{Adriani:1992pq, Achard:2001qv} and DELPHI~\cite{Abreu:1996pa} have
searched for $\nu_4$ production in $Z$ decays, followed by $\nu_4$ decay into a
SM lepton and an electroweak gauge or Higgs boson.  They also contribute
significantly to electroweak precision data (EWPD), which would also be affected by
the active neutrinos mixing with new sterile states
\cite{Abada:2007ux, Fernandez-Martinez:2016lgt, Blennow:2016jkn,
delAguila:2008pw, Akhmedov:2013hec, deBlas:2013gla, Antusch:2014woa,
Blennow:2016jkn, Flieger:2019eor}. In ATLAS~\cite{Aad:2019kiz, Aad:2015xaa} and
CMS~\cite{Sirunyan:2018mtv, Sirunyan:2018xiv}, the most important production
mode is $W \to \ell \nu_4$, and the most important decay channels are
$\nu_4 \to \ell \ell \nu$ and $\nu_4 \to \ell j j$.  An interesting alternative
way of constraining sterile neutrinos at the LHC is looking for Higgs boson
decays to sterile neutrinos \cite{Das:2017zjc}.  Note that ATLAS and CMS
sensitivity is best for lepton number violating decays, which exist only if
neutrinos are Majorana particles. Therefore, the ATLAS and CMS constraints
shown in \cref{fig:beta-decay-limits-nue} only apply to this case.

If neutrinos are Majorana particles, searches for lepton number violating
decays mediated by sterile neutrinos, such as $K^+ \to \ell^+ \ell'^+ \pi^-$ or
$B^+ \to \ell^+ \ell'^+ K^-$ lead to additional constraints.  In
\cref{fig:beta-decay-limits-nue,fig:beta-decay-limits-numu-nutau}, these
constraints are combined into the exclusion curves labeled ``LNV
Decays''~\cite{Atre:2009rg, Helo:2010cw, Kovalenko:2009td, Abada:2017jjx},
except for the BESIII limit on $D^+ \to \ell^+ \ell'^+ \pi^-$ and $D^+ \to
\ell^+ \ell'^+ K^-$~\cite{Ablikim:2019gvd} and the LHCb
limit~\cite{Aaij:2014aba} on $B^- \to \mu^- \mu^- \pi^+$, which are shown
separately.

\section{Sterile Neutrinos and Neutrinoless Double Beta Decay}
\label{sec:0nu2beta}

Searches for neutrinoless double beta decay (``$0\nu2\beta$ decay'')
are among the most sensitive
probes of neutrino masses.  As such, they also offer exquisite sensitivity
to extra neutrino states more massive than the Standard Model neutrinos.

\begin{figure}
  \centering
  \includegraphics[width=11cm]{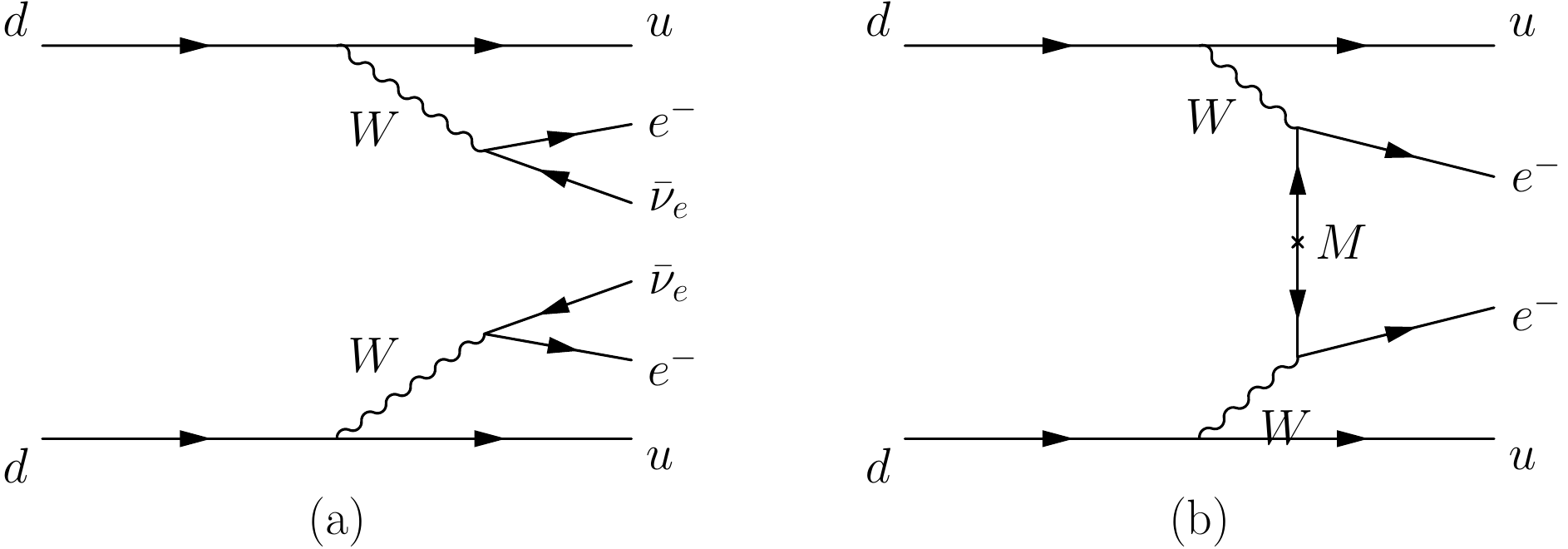}
  \caption{The Feynman diagrams for (a) two-neutrino double beta decay
    and (b) neutrinoless double beta decay.}
  \label{fig:0nu2beta-diagrams}
\end{figure}

Double beta decay occurs for neutron-rich or
proton-rich isotopes whose single beta decay
is forbidden, while two simultaneously occurring beta transitions are allowed.
This typically happens when the would-be daughter nucleus of single beta
decay has odd numbers of protons and neutrons (``odd--odd nucleus''),
while the daughter state of double beta decay has even numbers of
both protons and neutrons (``even--even nucleus''). As nucleons are
wont to combine into tightly bound pairs, odd--odd configurations are
energetically unfavorable.  The Feynman diagram for conventional
double decay accompanied by the emission of two neutrinos is shown
in \cref{fig:0nu2beta-diagrams}~(a), while the diagram for
neutrinoless double beta decay is shown in
\cref{fig:0nu2beta-diagrams}~(b).

The crucial difference is that,
in the second diagram, the neutrino lines from the two decays have
been connected, which is only possible if neutrinos are Majorana particles.

We can read off from \cref{fig:0nu2beta-diagrams}~(b) that the rate
for neutrinoless double beta decay will be of order
\begin{align}
  \Gamma_{0\nu2\beta} \propto G_F^4 |\tilde{M}_{0\nu2\beta}|^2 \,
    \big|\sum_j U_{ej}^2 m_j\big|^2  p_e^2 \,,
  \label{eq:0nu2beta-rate}
\end{align}
where we have omitted $\mathcal{O}(1)$ factors.
The factor $G_F^4$ corresponds to the four weak interaction vertices,
$\tilde{M}_{0\nu2\beta}$ is a nuclear matrix element describing
the overlap of the initial and final state nuclear wave functions,
and the final-state electron momentum $p_e$ is used here as a proxy
for the typical energy scales in the problem. It is introduced to
give $\Gamma_{0\nu2\beta}$ the correct dimensions. For neutrino
physics, the most important factor in \cref{eq:0nu2beta-rate} is
the \emph{effective mass}
\begin{align}
  m^\text{eff}_{0\nu2\beta} \equiv \sum_j U_{ej}^2 m_j \,.
  \label{eq:meff-0nu2beta}
\end{align}
Here, the leptonic mixing matrix elements $U_{ej}$ stem from the
$W e \nu$ vertices, and the neutrino mass eigenvalues $m_j$
appear due to the mass insertion on the neutrino line. The necessity
of a mass insertion can be understood because connecting the outgoing
neutrino line of one beta decay to an incoming anti-neutrino line for
the second beta decay is only possible through a Majorana mass
term.

\begin{figure}
  \centering
  \includegraphics[width=\textwidth]{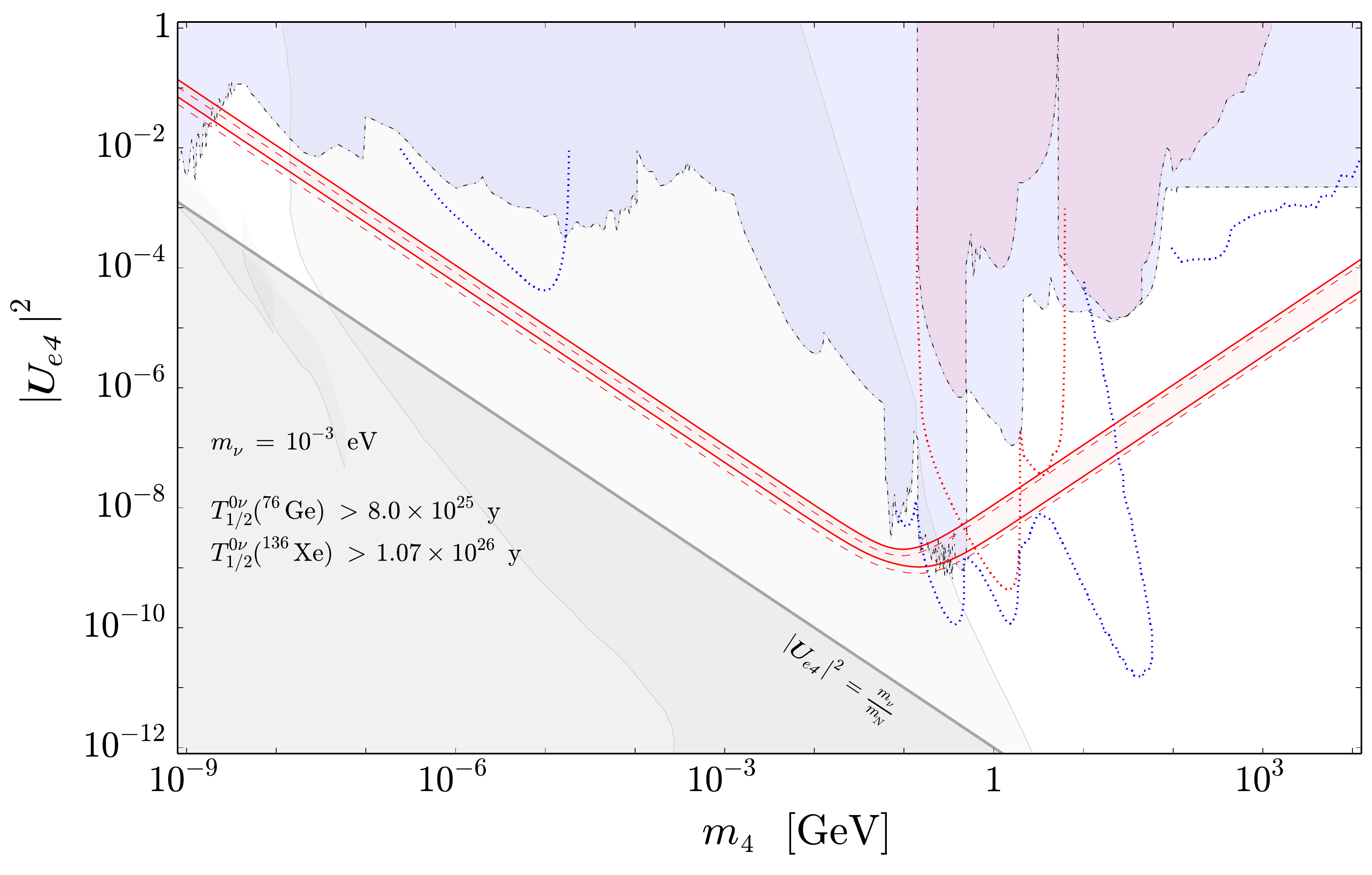}
  \caption{Neutrinoless double beta decay constraints on the mixing
    between sterile neutrinos and electron neutrinos, $|U_{e4}|^2$
    in a $3+1$ model.  We show 90\% CL constraints from
    \iso{Ge}{76} (red dashed) \cite{Agostini:2018tnm}
    and \iso{Xe}{136} (red solid) \cite{KamLAND-Zen:2016pfg}. The
    width of the bands indicates the theoretical uncertainty, which
    arises from uncertainties in the nuclear matrix elements
    $\tilde{M}_{0\nu2\beta}$ in \cref{eq:0nu2beta-rate}.
    We also show the parameter region generically favored by
    type-I seesaw scenarios (gray shaded area at the bottom),
    and we compare to the limits from beta decay kinematics
    discussed in \cref{sec:kinematics} (shaded areas at the top).
    This plot has been adapted from ref.~\cite{Bolton:2019pcu}.}
  \label{fig:0nu2beta}
\end{figure}

If there are more than three mass eigenstates, the additional
mostly sterile ones need to be included in the sum in
\cref{eq:meff-0nu2beta}.  Hence, any sterile Majorana neutrino that
mixes into the $\nu_e$ flavor eigenstate alters the rate of
neutrinoless double beta decay.  Given the current 90\% confidence level
constraints on the $0\nu2\beta$ half-lives of \iso{Ge}{76} ($T_{1/2}^{0\nu} >
\SI{8.0e25}{yrs}$)~\cite{Agostini:2018tnm} and \iso{Xe}{136} ($T_{1/2}^{0\nu} >
\SI{1.07e26}{yrs}$)~\cite{KamLAND-Zen:2016pfg}, the authors of
ref.~\cite{Bolton:2019pcu} have derived limits on the mixing between
sterile neutrinos and electron neutrinos, $|U_{e4}|^2$.  These limits
are shown in \cref{fig:0nu2beta} and compared to the constraints
from beta decay kinematics discussed in \cref{sec:kinematics}.
We see that $0\nu2\beta$ decay dominates the limit on $|U_{e4}|^2$
over a vast mass range. Only at very low (eV-scale) masses,
oscillation experiments are more sensitive. At masses
larger than the typical momentum transfers in nuclear processes
(few hundred MeV), the sensitivity of $0\nu2\beta$ decay searches
deteriorates because the contributions of the heavy mass eigenstate
to the diagrams of \cref{fig:0nu2beta-diagrams}~(b) becomes suppressed.

In spite of its excellent sensitivity, however, $0\nu2\beta$ decay
searches do not quite reach the parameter region favored by the generic
type-I seesaw, shown in gray at the bottom of \cref{fig:0nu2beta}.
It is also important to keep in mind that $0\nu2\beta$ decay is only
sensitive to sterile neutrinos mixing with electron neutrinos,
and even that only in the case of Majorana neutrinos.
The mixings with muon and tau neutrinos, as well as scenarios
with Dirac neutrinos, remain unconstrained.

\chapter{eV Sterile Neutrinos in Cosmology}
\label{sec:cosmology}

Neutrino cosmology represents one of the most successful meeting points between the fields of particle physics and cosmology. Naturally, this subject has been covered in great detail in the literature. Previous reviews by Dolgov~\cite{Dolgov:2002wy} and by Pastor and Lesgourgues~\cite{Lesgourgues:2006nd}, as well as the excellent monograph by Lesgourgues, Mangano, Miele, and Pastor~\cite{Lesgourgues:2018ncw}, cover much of the classic material on the topic, and part of our discussion will follow their treatment quite closely. However, we will cover sterile neutrino interactions in more detail than previous reviews.

Neutrino oscillation data require that at least two of the mostly active neutrino mass eigenstates must be massive. Depending on the unknown ordering of the masses, the active masses are denoted as
\begin{align}
&m_1=m_{0}\hspace{2.7cm}({\rm NO})\hspace{1.25cm} m_3=m_{0}\hspace{4.1cm}({\rm IO})\,,\\
&m_2=\sqrt{m_{0}^{2}+\Delta m^2_\textrm{sol}}\hspace{0.78cm}({\rm NO})\hspace{1.25cm} m_1= \sqrt{m_{0}^{2}+\Delta m^2_\textrm{atm}}\hspace{2.1cm}({\rm IO})\,,\\
&m_3= \sqrt{m_{0}^{2}+\Delta m^2_\textrm{atm}}\hspace{0.6cm}({\rm NO})\hspace{1.25cm}m_{2}= \sqrt{m_0^{2}+\Delta m^2_\textrm{atm}+\Delta m^2_\textrm{sol}}\hspace{0.6cm}({\rm IO})\,.
\end{align}
Here, we have expressed the three neutrino masses in terms of the yet undetermined lightest neutrino mass, $m_{0}$, and  the two known mass-square differences $\Delta\,m^{2}_\text{atm} = |m_{3}^{2}-m_{1}^{2}|\approx 2.5\times10^{-3}$\,eV$^{2}$ and $\Delta\,m^{2}_\text{sol} = m_{2}^{2}-m_{1}^{2}\approx 7.4\times10^{5}$\,eV$^{2}$~\cite{Esteban:2020cvm,deSalas:2020pgw,Capozzi:2017ipn}. While solar neutrino data dictate $m_{2}>m_{1}$, the sign of $\Delta\,m^{2}_\text{atm}$ is not yet known. Normal ordering (NO) refers to the case $m_{3}>m_{1,2}$, whereas in inverted ordering $m_{3}<m_{1,2}$. The sum of neutrino masses is $\sum m_{\nu}\geq0.058$\,eV, and the minimum possible masses are approximately $0$\,eV, $8.4$\,meV, and $0.05$\,eV. Laboratory limits on the maximum value of $m_{0}$ are $\sim1$\,eV~\cite{Aker:2019uuj}, and thus the maximum allowed values of $m_1$, $m_2$, $m_3$ are also $\sim1$\,eV.

Cosmology is sensitive to neutrinos in a way that is complementary to laboratory searches. It is less sensitive to individual masses and mixings, but is more directly affected by the absolute mass scale, possible new self-interactions, etc. We present a picture of how neutrino properties affect cosmology and explain some of the recent developments. Readers familiar with basic cosmology can skip \cref{sec:cosmorev}. We do not review cosmological perturbation theory, for which the reader may refer to~\cite{Lesgourgues:2006nd, Lesgourgues:2018ncw}.

\section{Standard Cosmology}
\label{sec:cosmorev}

Nearby galaxies, on average, appear to be receding away from us at a velocity $\vec{v}$ proportional to their distance $\vec{r}$ from us. This is expressed as Hubble's law
\begin{align}
&\vec{v} = H_0 \vec{r}\,, 
\end{align}
where $H_0 = 100\,h\,{\rm km\,s^{-1}Mpc^{-1}}$ is the local Hubble expansion parameter. The precision in the knowledge of $H_{0}$ has improved from tens-of-percent to the percent-level over the past few decades. The latest determinations of $H_0$ are up to 1\% precise, but span a very wide range $\approx \mbox{(68-74)}\,\textrm{km}\,\textrm{s}^{-1}\,\textrm{Mpc}^{-1}$~\cite{Aghanim:2018eyx, Riess:2019cxk, Abbott:2017xzu, Wong:2019kwg, Freedman:2020dne}. At the time of writing this review, in the midst of the COVID-19 pandemic, there is a significant disagreement between the various precise measurements of this parameter. This so-called Hubble tension is undoubtedly the most discussed anomaly in cosmology at the moment~\cite{Riess:2020sih}, as recently analyzed in this lockdown perspective~\cite{Efstathiou:2020wxn}. Remarkably, light relics such as sterile neutrinos may reduce this tension, providing a rather topical motivation for this review of sterile neutrinos in cosmology.

A nonzero $H_{0}$ leads to the natural conclusion that our Universe is expanding. This is expressed in terms of the background metric of the Universe,
\begin{align}
&ds^2=dt^2-a(t)^2\left[\frac{dr^2}{1-Kr^2}+r^2d\Omega^2\right]\,,
\end{align}
where the time-dependent scale factor $a(t)$ encodes the expansion of the Universe. It converts the coordinate distance $dr$ between what would be two relatively stationary points in a static Universe to a physical distance $r^\text{phys}=a(t) \, r$. The discrete curvature parameter $K$ can be $+1$, 0, or $-1$, corresponding to a closed, flat, or open Universe. Our discussion will be confined to $K=0$, corresponding to a flat Universe, as predicted by inflation and consistent with observations. The time-dependent expansion rate is given by $H={\dot{a}}/{a}$, whose value at the present time $t=t_0$ is the local Hubble parameter $H_0$. We use the convention that $a_0= a(t_0)=1$ and note that the redshift corresponding to a time $t$ in the past is then simply $z=a(t)-1$. One can use the redshift, scale factor, or time interchangeably to refer to an epoch. To paraphrase the luminaries, cosmology is a search for the few parameters that dictate the time evolution of the scale factor $a(t)$.

An expanding Universe must have been smaller and hotter in the past. Physical cosmology is built upon extending this argument and applying the known laws of physics to such a Universe.  It turns out that a very economical model, called the $\Lambda{\rm CDM}$ model -- including two novel components, dark matter and dark energy in addition to all known matter, and simple initial conditions inspired by inflation -- describes all available cosmological data adequately. These data come mainly from observations of the abundance of light nuclei, the properties of the cosmic microwave background (CMB), and the abundance and distribution of galaxies and clusters. 

Whether and when something happened in the Universe, depends on two key quantities: the Hubble expansion rate and the rate for the concerned physical process $\Gamma$. As a thumb rule, all physical processes that occur slower than the Hubble rate are inefficient, whereas those that are faster can be important. Usually both of these rates fall with time at late times, and their relative strength and scaling decides when the process stops. This ``freeze-out'' epoch is estimated by the condition
\begin{align}
  H\simeq\Gamma\,.
  \label{eq:H-eq-Gamma}
\end{align}
For some processes, $\Gamma/H$ increases with time initially, then reaches at maximum, and ultimately decreases again.  The ``freeze-in'' epoch, i.e., the time $t$ at which such a process first becomes efficient, can be estimated by \cref{eq:H-eq-Gamma} as well.

The Hubble expansion rate $H$ depends on the energy density $\rho$ of the Universe through the Friedmann equation
\begin{align}
  H^{2} = \frac{8\pi}{3} G_{N} \rho \,.
\end{align}
The energy density $\rho$ is the sum of several contributions, each of which evolves with time $t$ and temperature $T$ in its own characteristic fashion. For instance, the energy density of relativistic ($T \gg m$) particle species (``radiation'') in equilibrium scales as $\rho_\text{rad} \propto T^4$ in the absence of a chemical potential, while that of a non-relativistic species out of chemical equilibrium (``matter'') scales as $\rho_\text{mat} \propto T^3$.

To see this, consider the phase space occupation of each degree of freedom of a particle species in approximate equilibrium, i.e., a species whose interactions with the thermal plasma satisfy $\Gamma/H\gg1$.  It is simply given by the Fermi--Dirac or Bose--Einstein distribution,
\begin{align}
  f_\textrm{FD,BE}=\frac{1}{\exp[(E-\mu)/T]\pm1}\,,
  \label{eq:f-FD-BE}
\end{align}
with a multiplicative correction $1+\delta$, where $|\delta|\ll1$ encodes perturbations. Integrating the phase space density $f(p) = (1 + \delta) f_\textrm{FD,BE}$, one can compute the number density $n=\int d^{3}{\bf p}/(2\pi)^{3}\,f(p)$, the energy density $\rho=\int d^{3}{\bf p}/(2\pi)^{3}\,E\,f(p)$, etc. For example, every degree of freedom of a relativistic boson at zero chemical potential, \eg, the photon, in equilibrium contributes an energy density $\rho_\text{rad} = \pi^{2}T^{4}/30$. Similarly, the number density of such a radiation-like bosonic degree of freedom scales as $n_\text{rad} = \zeta(3) \, T^{3}/\pi^{2}$, where $\zeta$ denotes the Riemann Zeta-Function. For relativistic fermions, \eg, neutrinos, the energy density is smaller by a factor of $7/8$, and the number density by a factor $3/4$ compared to relativistic bosons. The number of degrees of freedom, $g$, corresponding to each particle species is the number of distinct quantum states that are thermally accessible at that temperature. For instance, photons have $g_{\gamma}=2$ for the two transverse polarizations, electrons have $g_{e}=4$ (two helicities, particles and anti-particles), and neutrinos have $g_{\nu}=2$ (as the right-handed state for the particle and the left-handed state for the anti-particle are never produced).  For a relativistic species with non-vanishing chemical potential $\mu$, a useful result is that $n_\textrm{rad}-\bar{n}_\textrm{rad} \approx T^{3}(\xi+\xi^{3}/\pi^{2})/6$, where $\xi=\mu/T$ is the familiar fugacity parameter from statistical mechanics which controls the particle number asymmetry.

Analogously, the number density of a matter-like species per degree of freedom in equilibrium is $n_\text{mat} \approx (mT/2\pi)^{3/2}\,\exp(-m/T)$, whereas $\rho_\text{mat} \approx m\,n + (3/2)\,T\,n$, etc. The temperature $T$ obeys $a\,T\approx\,{\rm constant}$ due to conservation of the entropy density ($s \propto g\,T^{3} \propto 1/a^{3}$), assuming that the number of degrees of freedom remains unchanged as the Universe cools. Thus, after a non-relativistic species stops being in chemical equilibrium, that is, after number-changing processes have frozen out, its number density and energy density scale as $a^{-3}$.  Whereas, for a relativistic species the latter continues to decrease as $a^{-4}$. These results can be obtained simply from thermodynamics in an expanding Universe, giving $H$ in terms of temperature. While the energy density of the Universe is dominated by radiation, one finds
\begin{align}
  H \simeq 1.66 \sqrt{g_{\star}} \, \frac{T^{2}}{M_\text{Pl}} \,,
\end{align}
where $g_{\star}(T)$ is the effective number of relativistic degrees of freedom at a temperature $T$ that make up the total radiation density.  Technically, one should distinguish two kinds of $g_{\star}$, one affecting the total energy density as above and its cousin ($h_{\star}$) encoding the entropy density instead. When there are no decoupled particles, such as neutrinos below temperatures of $\sim \SI{1.2}{MeV}$, these are the same.

As an example, consider a light particle species (for instance a sterile neutrino) which interacts with the thermal plasma via a mediator particle of mass $M$, with an effective fine-structure constant $\alpha$. While the new species is relativistic, its cross section is $\sigma \simeq \alpha^{2} T^{2} / M^{4}$ for a massive mediator ($M \gg T$) and $\alpha^{2}/T^{2}$ for a  light mediator ($M \ll T$). The particle's interaction rate is $\Gamma \sim  n \sigma v$, where $n \sim T^3$ is the number density of the species that it interacts with, and $v$ is the relative velocity of the two interacting particles.  Thus, at sufficiently large $T$ one expects $\Gamma \simeq \alpha^{2} T$ to be smaller than $H \simeq T^{2} / M_\text{Pl}$, as also at low $T$ where  $\Gamma \simeq \alpha^{2} T^{5} / M^{4}$.  This implies that the particle species under consideration is out of equilibrium at very early times, then becomes strongly coupled to the thermal plasma at $T \sim \alpha^2 M_\text{Pl}$, before freezing out again at $T \sim [M^4 / (\alpha^2 M_\text{Pl})]^{1/3}$.

\vspace{1em}
Let us finally discuss inhomogeneities, parameterized by the correction factor $1+\delta$ which we mentioned already below \cref{eq:f-FD-BE}.  Consider the process of tiny initial inhomogeneities accreting more and more matter due to gravitational collapse, thereby ultimately forming galaxies and clusters. The fractional over-density, $\delta\rho/\bar{\rho}$, encoded in the amplitude $\delta_k$ of its $k^\textrm{th}$ comoving sub-horizon Fourier mode, evolves as
\begin{align}
  \ddot\delta_k + 2H\dot\delta_k = (4\pi G_N\bar{\rho} - c_s^2 k^2)\delta_k \,.
\end{align}
For sufficiently small lengths scales (large $k$), the term in parentheses on the right-hand side is negative, and therefore $\delta_k$ can have an exponentially growing solution when $H \simeq 0$. If, on the other hand, $H \gg c_{s} k$ one would expect such growth to be significantly hampered by stretching of the scale factor.

Similar comparisons, involving the relevant competing rates, can be invoked to give a fair idea of the effectiveness of various physical processes at various epochs. This kind of intuition is captured quantitatively within the Boltzmann equations in an expanding Universe. They describe essentially everything in cosmology -- from the formation of nuclei and atoms to the seeding of galaxies and clusters of galaxies.

\subsection{Minimal $\Lambda$CDM Model}

In the minimal cosmological model, the Universe has the following components:
\begin{itemize}
  \item Radiation, with an energy density $\rho_{r}$ that scales as $1/a^{4}$. This includes all particles in the Standard Model of Particle Physics that are relativistic, i.e., photons at all epochs and other particles at temperatures higher than their respective masses.

  \item Baryons, with an energy density $\rho_{b}$ that scales as $1/a^{3}$. This includes all Standard Model particles that are non-relativistic and are coupled to photons. (Note that, for simplicity and because their mass density is low, electrons and positrons are included here even though they are leptons.)

  \item Cold Dark Matter (CDM), a novel matter component that is not coupled to photons and is non-relativistic through most of cosmological history, with an energy density $\rho_{c}$ scaling as $1/a^{3}$. Any hypothesized dark matter particles, such as axions or WIMPs, must exhibit this behavior on the relevant scales and epochs.

  \item Cosmological Constant, with an energy density $\rho_{\Lambda}$ that does not change with $a$. This hypothesized component is often generalized to Dark Energy.
\end{itemize}
This parameterization is most useful around the epoch of recombination at around $T \sim \SI{1}{eV}$, where most of these components play a non-negligible role: neutrinos and photons count as radiation, whereas electrons, ions, and atoms count as baryons, and there is empirical evidence for both dark matter and dark energy. Using the same parameterization at other epochs requires careful bookkeeping that maps the actual physical constituents of the Universe to this parameterization. As an important example, neutrinos act as radiation when the temperature of the Universe is $\gtrsim\eV$, but eventually at least two of the active neutrino mass eigenstates become non-relativistic and count towards dark matter instead.  

The energy densities in each cosmological component $i$ has a well-defined dependence on time, so for phenomenological purposes it is more convenient to work with their fractional contributions to the energy density today, defined by
\begin{align}
  \Omega_{i} = \frac{\rho_{i}(t_{0})}{\rho _\text{crit}}
  \qquad\text{with}\quad
  \rho_\text{crit} = \frac{3 H_{0}^{2}}{8\pi G_{N}} \,.
\end{align}
Here $i = r, b, c, \Lambda$ for radiation, baryons, cold dark matter, and dark energy,
respectively.  In a flat Universe like ours, one has $\Omega_r + \Omega_b + \Omega_c + \Omega_{\Lambda} = 1$, and only three of these four parameters are independent. 

All the constituents are assumed to be statistically isotropic and homogeneous at the earliest times, with small stochastic perturbations. The perturbations are adiabatic, \ie,  fractional perturbations in each component are related to those in others, and are fully encoded in the two-point correlation function, $P(k)$, of the local fluctuation of the spatial curvature in comoving units, ${\cal R}({\bf k})$, defined by $\langle {\cal R}({\bf k}') {\cal R}^{\star}({\bf k}) \rangle=P(k)\delta^{3}({\bf k}-{\bf k}')$. One usually works with the rescaled power spectrum ${\cal P}(k) = k^{3}/(2\pi^{2}) \, P(k)$, which is parameterized as
\begin{align}
  {\cal P}(k) = A_{s}\left(\frac{k}{k_{\star}}\right)^{n_{s}-1}\,.
\end{align}
Here $A_{s}$ is the amplitude of perturbations at some arbitrary pivot scale $k_{\star}$ and $n_{s}$ is the spectral index that is expected to be close to 1 when ${\cal  P}(k)$ is nearly scale-invariant.

The $\Lambda$CDM model has one more parameter, $\tau$, which is the optical depth to reionization. While this is not really a fundamental parameter, it is practically impossible to calculate it as this would require fully tracking the reionization history through simulations. It is much easier to determine $\tau$ from the data.

In all, the model thus has six free parameters, $\{\Omega_{b}h^{2},\Omega_{c}h^{2},H_{0},A_{s},n_{s},\tau\}$. The physical energy densities $\rho_{i}$ are proportional to the combinations $\omega_{i} = \Omega_{i}\,h^{2}$, which are directly constrained by the CMB data. Note that $\Omega_r$ is conventionally not counted as one of the free parameters of the minimal model because the two largest contributions to it are from photons and neutrinos, with are either precisely known using the temperature of the CMB ($T = 2.725\pm 0.001$\,K~\cite{Fixsen:2009ug}, which gives $\Omega_\gamma = 2.47\times10^{-5}\,h^{-2}$) or are precise predictions of the Standard Model (giving that $T_{\nu}\approx(4/11)^{1/3}\,T$ at $T\ll0.5$\,MeV), respectively.\footnote{The Planck collaboration uses the angular acoustic scale $\theta$ as a fundamental parameter instead of $H_{0}$, but this difference is not important for our discussion.} Other quantities, \eg, $\Omega_{\Lambda}$, are derived parameters in the $\Lambda$CDM model. In addition, there are about two dozen nuisance parameters that are introduced to encode instrumental features and observational limitations; we will not discuss these here.

We summarize our current knowledge about the $\Lambda$CDM model -- encoded in the measured values of its parameters -- in \cref{tab:cosmoparams}.

\begin{table}[!h]
\begin{center}
\caption{Cosmological parameters from the 2018 data release by Planck~\cite{Aghanim:2018eyx} (referred to as Planck-18 hereafter). See Tables 1 and 2 in Planck-18. We have quoted the TT+TE+TE+lowE+lensing limits obtained using their default ``Plik'' likelihood. Some of the external or derived parameters are also noted for completeness.}\label{tab:cosmoparams}
\vspace{0.2cm}
  \begin{tabular}{lll}
    \hhline{---}\\[-1.9ex]
    Parameter & Description & Mean and $68\%$ range\\[0.3ex] \hline \\[-0.8ex]
                    $H_0$ & Local Hubble expansion rate& $67.36^{+0.54}_{-0.54}\,{\rm km\,s^{-1}\,Mpc^{-1}}$\\[1.2ex]
    $\Omega_b$ & Baryon fraction & $0.02237^{+0.00015}_{-0.00015}\,h^{-2}$\\[1.2ex]
    $\Omega_c$ & CDM fraction & $0.1200^{+0.0012}_{-0.0012}\,h^{-2}$\\[1.2ex]
    $\log_{10}(10^{10}A_s)$ & Amplitude of perturbations & $3.044^{+0.014}_{-0.014}$\\[1.2ex]
    $n_s$ & Spectral tilt of perturbations & $0.9649^{+0.0042}_{-0.0042}$\\[1.2ex]
    $\tau$ & Optical depth to reionization & $0.0544^{+0.0073}_{-0.0073}$\\[1.6ex] \hline \\
    $\Omega_r$ & Radiation fraction & $9.21^{+0.15}_{-0.15}\times 10^{-5}$\\[1.2ex]
    $\Omega_{m}$ & Matter fraction & $0.3153^{+0.0073}_{-0.0073}$\\[1.2ex]
    $\Omega_\Lambda$ & $\Lambda$ fraction & $0.6847^{+0.0073}_{-0.0073}$\\[1.2ex]
    \hhline{---}
  \end{tabular}
  \end{center}
\end{table}

\subsection{Cosmological History}

The very early Universe presumably began with inflation -- a hypothesized phase of rapid expansion, where spatial patches in the Universe that were already in causal contact grew exponentially with time and became much larger than the Hubble radius. This led to the simple initial conditions assumed in the $\Lambda {\rm CDM}$ model. After inflation, the Universe was essentially empty, but was then reheated through the decay of inflatons (the excitations of the field whose vacuum energy density drives inflation). Inflatons are presumed to have decayed into various particles species that interacted and came into thermal equilibrium. Over the past two decades, the focus of cosmology has been on deciphering the nature of our Universe, assuming these simple initial conditions. Modeling and testing the very early epoch, involving inflation or its alternatives, remains a frontier of modern cosmology. 

For not too extreme values for the parameters $\Omega_{r}$, $\Omega_{b}$, $\Omega_{c}$, and $\Omega_{\Lambda}$, at early times the Universe was radiation-dominated as the radiation energy density scales by the factor $1/a^{4}$. Thereafter, matter composed of dark matter and baryons, which scales as $1/a^{3}$, started to become relatively more important, and at the epoch of matter--radiation equality, $t_{\rm eq}$, the two had equal energy densities: $\rho_{r}=\rho_{m}$. Afterwards, the energy density of the Universe was dominated by matter. From \cref{tab:cosmoparams}, one finds $z_\text{eq} = \Omega_{m}/\Omega_{r} \approx 3402$. At even later times, approximately at $z_{\Lambda} \approx 0.295$, the cosmological constant $\Lambda$, which unlike all other species does not decrease with $a$, began to dominate over everything else. 

At present, we have information about our cosmological history only from a few epochs -- the nucleosynthesis era at $T_{\textrm{BBN}}\approx 0.1$\,MeV and $z_{\textrm{BBN}}\approx 10^{9}$, where light nuclei formed from a soup of protons and neutrons; the recombination era at $T_{\textrm{rec}}\approx 0.3$\,eV and $z_{\textrm{rec}}\approx1100$, when electrons and nuclei formed neutral atoms and the Universe became transparent for the first time; and the relatively recent Universe at $T \approx $~few K and $z \lesssim 1$ with galaxies and other structures. In the future, new observations, \eg, of the 21\,cm line of hydrogen, metal line intensities, and gravitational waves, will provide glimpses into yet unseen epochs of the Universe.

It is a remarkable feat that CMB data alone can determine all parameters of the minimal model to better than 10\% precision~\cite{Aghanim:2018eyx}. The model thus determined by CMB data is internally consistent (the fluctuations are consistent with being statistically isotropic and Gaussian, the $p$-values for the minimal model are reasonable, etc.). Further, the model is externally consistent with the other cosmological datasets that probe subsets of its six cosmological parameters, with only a few notable and widely discussed exceptions. Besides the previously discussed Hubble tension, there are only mild anomalies, e.g., between Lyman-$\alpha$ observations and baryon acoustic oscillations (BAO), between different geometric observables, and between CMB and structure formation measurements of the perturbation amplitude. (In this context, the latter is often parameterized in terms of $\sigma_{8}$, the amplitude of the power spectrum on a scale of \SI{8}{Mpc}, rather than by $A_s$ and $n_s$.) Understanding these tensions usually involves more detailed modeling and they may therefore be considered less robust than the Hubble tension. See Planck-18 for detailed comparisons~\cite{Aghanim:2018eyx}.

The minimal model can be extended in various ways. One way is to simply include known physics that was ignored as a simplifying approximation, for instance active neutrino masses. At present the cosmological data does not distinguish between these more realistic models and the minimal model, but this is likely to change in the future. The other approach is to include new physics to address some of the tensions mentioned above, or to resolve some other theoretical/observational issue. This is where much of the excitement has been. In particular, to address the Hubble tension, modifications to neutrino properties have been hypothesized. Sterile neutrinos, the topic of this review, also fall in this latter category. Below, in \cref{sec:sterile-cosmo}, we will focus on the base $\Lambda$CDM cosmology extended by one or more sterile neutrinos with eV-scale masses and large mixings to the active neutrinos, with or without exotic interactions. However, before we enter a discussion of sterile neutrinos, it is useful to review how various observations constrain cosmology and how various properties of neutrinos affect cosmology.

\subsection{Cosmological Probes}

\subsubsection{Big Bang Nucleosynthesis} 

Big Bang Nucleosynthesis is the era when nucleons in the primordial plasma cooled down sufficiently to combine into light nuclei, in particular $^{2}{\rm H}$, $^{3}{\rm H}$, $^{3}{\rm He}$, $^{4}{\rm He}$, and $^{7}{\rm Li}$. This process depends on the competition of the electromagnetic, weak, and strong interaction rates, and of the Hubble expansion rate in the epoch where the photon temperature was roughly in the range $1\MeV-0.1\MeV$.

At high temperatures, $T\gtrsim\MeV$, neutrons and protons were in chemical
equilibrium owing to their weak interactions. The number densities of protons
and neutrons in this epoch obey the Saha equation,
\begin{align}
  \frac{n_{n}}{n_{p}} = \exp\left(-\frac{\Delta m_{N}}{T}\right)\,,
\end{align}
where $\Delta m_{N}\simeq1.29\,\MeV$ is the neutron--proton mass difference. As
we discussed earlier, the weak interactions of neutrinos decoupled at
$T_{\nu{\rm D}}\sim 1.2\,\MeV$, as determined by the competition of the Hubble
rate $H$, which is proportional to $\sqrt{\rho_{r}} \propto T^2$, and the rate
of weak interactions which scales as $\Gamma\approx n_{\nu}G_{F}^{2}E_{\nu}^{2}
\propto T^5$.

In the same way, conversions between protons and neutrons ceased to be efficient
at $T_D \sim \SI{0.7}{MeV}$.  At this time, the $n/p$ ratio was about
$\exp(-\Delta m_N/T_D) \simeq 1/6$.
Afterwards, neutrons began to decay, and the scaling of the $n/p$ ratio became
\begin{align}
\frac{n_{n}}{n_{p}} = \exp\left(-\frac{\Delta m_{N}}{T_D} \right)
                        \exp\left(-\frac{t}{\tau_{n}}\right)\,.
\end{align}
Here, $t \propto (\MeV/T)^{2}$ during radiation domination, and $\tau_{n} = 880.3 \pm 1.1 \, {\rm sec}$ is the neutron lifetime.

At high temperatures, bound nuclei cannot form. As an example, $^{2}{\rm H}$ nuclei are immediately broken down by photons via the electromagnetic process,
\begin{align}
  \gamma + {^{2}{\rm H}}\to n + p\,.
\end{align}
Up to ${\cal O}(1)$ numerical factors, the equilibrium abundance of $^{2}{\rm H}$ nuclei is 
\begin{align}
  \frac{n_{^{2}{\rm H}}}{n_{\rm H}}=\eta_{b}\left(\frac{T}{m_{N}}\right)^{{3/2}}\exp\left(\frac{B_{^{2}{\rm H}}}{T}\right)\,.
  \label{eq:D-abundance}
\end{align}
The baryon-to-photon ratio $\eta_{ b}=n_{b}/n_{\gamma}$, which can be easily related to $\Omega_b$ and $\Omega_r$ in the minimal model, is the only free parameter. Given the binding energy of the deuteron, $B_{^{2}{\rm H}}\simeq2.2\,\MeV$, the mean nucleon mass $m_{N}$, and the known value of $\eta_{b}\sim10^{-9}$, the equilibrium abundance of $^{2}{\rm H}$ nuclei comes to merely $\sim10^{-12}$ at MeV temperatures. Higher-mass nuclei were even less abundant at high temperature, as they involve additional powers of $\eta_{\rm b}$. \Cref{eq:D-abundance} shows that the deuteron abundance grew significantly only at temperatures much lower than the deuteron binding energy, thanks to the small prefactor $\eta_b$. More quantitatively, one finds that deuterons could only form efficiently at $T_{\rm BBN} \lesssim 0.07\,\MeV$, when their photodissociation became inefficient. This relatively late formation of deuterons is called the ``deuterium bottleneck''.

Once the bottleneck of $^{2}{\rm H}$ formation had been overcome, a chain of nuclear reactions was triggered that created heavier nuclei. As an approximation, it is useful to assume that all neutrons that had not decayed at $T_{\rm BBN}=0.07\,\MeV$ got bound into highly stable $^{4}{\rm He}$ nuclei. This leads to the following result for the helium fraction:
\begin{align}
Y_{^{4}\rm He}=\frac{4 n_{^{4}{\rm He}}}{n_{b}}\sim\frac{2}{1+\exp\left(\frac{\Delta m_{N}}{T_{D}}\right)\exp\left(\frac{t_{\rm BBN}}{\tau_{n}}\right)}\sim0.25\,.
\end{align}

These estimates give an idea of how nuclei formed, but, in practice, the nuclear abundances are calculated by solving a set of coupled kinetic equations that encode a network of nuclear reactions and rates. Fitting to observed abundances allows a determination of the only free parameter, $\eta_{\rm b}\approx6\times 10^{{-10}}$. In a global fit, BBN plays a modest role in determining $\Omega_b$ in the $\Lambda$CDM model. The preceding discussion follows~\cite{Lesgourgues:2018ncw}, which may be consulted for more details and references.

\subsubsection{Cosmic Microwave Background}

After the formation of heavier nuclei by around $T \simeq 0.1\,\MeV$, the Universe remained a charged plasma of electrons and ions until it cooled down sufficiently to form atoms. Naively, one expects this to have occurred at a temperature close to $E_{0}=13.6\,\eV$, when protons and electrons could combine to form neutral hydrogen atoms in their ground state. However, the ratio of the baryon and photon number densities is very small, $\eta_{\rm b}\approx 6\times 10^{-10}$, as independently determined by BBN observations. Therefore, even after the temperature of the Universe had dropped below $E_{0}$, there were initially still enough energetic photons to ionize any hydrogen atoms that may have formed. Assuming each photon with energy larger than $13.6\,{\rm eV}$ is sufficient to ionize an atom, we can determine the temperature $T_\text{rec}$ at which recombination actually happened by solving the equation
\begin{align}
  n_\gamma\int_{E_0}^{\infty}dE\,\exp\left(\frac{-E\,\,~~}{k_BT_\text{rec}}\right) \sim n_{\rm b}\,.
\end{align}
The result is
\begin{align}
  T_\text{rec} \simeq \frac{E_{0}}{-k_B\log\eta_{\rm b}}\sim 0.7\,{\rm eV}\,.
\end{align}
A better estimate is obtained by keeping track of the abundances of neutral, first-excited, and ionized hydrogen in the so-called effective three-level atom model~\cite{Peebles:1968ja, Zeldovich:1969en}, which gives $T_\text{rec} \approx 0.3$\,eV.

Prior to recombination, the photons, baryons, and electrons were all tightly coupled by electromagnetic interactions. The growth of density inhomogeneities via gravitational collapse was counteracted by the radiation pressure. These opposing forces set up density waves that were damped by the expansion of the Universe and, on scales smaller than the photon mean free path, by photon diffusion. Around recombination, however, as protons and electrons formed atoms, the Universe became overall neutral, and the newly formed atoms decoupled from the photons. This stopped the baryon--photon oscillation in its tracks, and the perturbations in the photons became frozen.

Since recombinations, photons have propagated essentially freely, but were gradually redshifted from the optical band into the microwave band. They can thus be observed today as the cosmic microwave background (CMB). The very uniform CMB sky proves that the Universe was in equilibrium at early times. Its observed temperature in the present epoch $T_{0}=2.725\,{\rm K}$~\cite{Fixsen:2009ug}, corresponds to $T_{\rm rec}\sim0.3\,\eV$ with a redshift at recombination $z_{\rm rec}\simeq 1100$~\cite{Aghanim:2018eyx}. This temperature, measured to high precision by the FIRAS instrument on the COBE satellite, gives us a direct handle on the radiation density of photons in the Universe.
 
But probably the most important feature of the CMB are the small fluctuations of up to one part in $10^5$ that are superimposed on the otherwise perfect thermal spectrum. The fluctuation pattern is usually characterized by the average angular correlation, $C_{\ell}^{TT} \propto \langle \delta T(0) \, \delta T(\theta) \rangle$ of the temperature fluctuations at two points on the sky that are separated by an angle $\theta$, which is related to the multipole number $\ell$.  $C_{\ell}^{TT}$ reveals the typical relative phases within the primordial density waves at angular distances of order $\theta$: for wave modes that happen to be at their extremum when recombination happens, $C_{\ell}^{TT}$ is larger, while for modes that encountered a zero crossing at that time, the angular correlation is small.  Peaks in the angular power spectrum, $|C_{\ell}^{TT}|^2$, of CMB anisotropies correspond to points separated by integer and half-integer multiples of the fundamental oscillation wavelength in the baryon--photon fluid. The positions and heights of these peaks, as well as the shape of their overall envelope, encode a wealth of information and allow one to determine the amount of matter, contribution of dark energy, etc. Similar arguments apply to the anisotropies in polarization. Obtaining even an approximate but quantitative understanding of how each cosmological parameter affects the various features of the CMB anisotropy power spectrum takes great care and insight~\cite{Mukhanov:2005sc}.

CMB temperature and polarization observations alone are sufficient to determine all parameters of the minimal $\Lambda {\rm CDM}$ model to better than $10\%$ accuracy, in some cases reaching uncertainties below one percent. Further, most of this is based on easy to understand linear physics. So we will take the attitude that these CMB-derived parameters define our cosmology.

\subsubsection{Large Scale Structure}

While baryon density perturbations could not grow via gravitational collapse before matter--radiation equality at $T_\text{eq} \approx \SI{1}{eV}$, the density perturbations of dark matter began to grow (logarithmically) already much earlier.  In fact, if there were no dark matter and thus no pre-existing gravitational potentials for the baryons to fall into, the huge density perturbations we observe today in the form of galaxy clusters could not have formed within the time elapsed since matter--radiation equality. This is a simple argument demonstrating the need for dark matter.

The growth of the fractional over-density in the $i^{\textrm{th}}$ matter species, $\delta_i = \delta\rho_i/\bar{\rho}$, for the comoving Fourier mode $k$, is given by
\begin{align}
\ddot\delta_{i,k} + 2H\dot\delta_{i,k} = (4\pi G_N\bar{\rho} - c_s^2 k^2)\delta_{i,k} \,,
\end{align}
where $\bar\rho$ denotes the average energy density.  Competition between Hubble expansion and gravitational collapse dictates that perturbations in matter, on scales smaller than the horizon, grew only logarithmically with the scale factor during radiation domination and linearly during matter domination. Ignoring the logarithmic growth, perturbations on small scales that enter the horizon already during radiation domination therefore wait until matter--radiation equality to be able to grow, and suffer a suppression of $(a_{\rm entry}/a_{\rm eq})^2$ relative to the mode that enters only at $T_\text{eq}$. The matter power spectrum $P_m(k) \sim k^{-3}{\cal P}(k)$ therefore has a peak at wavenumber $k_{\rm eq}\sim10^{-2}$\,Mpc$^{-1}$, falls off as $1/k^3$ at $k > k_{\rm eq}$, and grows as $k$ at smaller $k < k_{\rm eq}$. This roughly agrees with the observed shape of the power spectrum of galaxies and other larger objects. 

$P_{m}(k)$ can be measured using a variety of ways. A direct comparison of the predicted linear power spectrum, described above, with the actual observation is quite a challenging task. In no particular order, the typical data on large scale structure come from galaxy and cluster power spectra, the cluster mass function, galaxy weak lensing, CMB lensing, and tomography with Ly-$\alpha$ and 21\,cm lines, etc. A major problem is the fact that galaxies and other baryonic tracers do not exactly follow the underlying dark matter power spectrum. Also, on small scales, roughly $k\gtrsim0.3\,h\,\textrm{Mpc}^{-1}$ the density perturbations are not linear any more. As such, it is often essential to use numerical simulations of structure formation to make contact between theory and data. Nevertheless, one of the most robust measurements of large scale structure is the scale $k_{\rm eq}$, which is directly related to $\Omega_m$. It can be robustly extracted, as can be the position of the baryon acoustic oscillation (BAO) peak, which is roughly the baryonic counterpart of the sound horizon peak in CMB. For these observables, there is excellent agreement with the CMB data.

\section{Neutrinos in Cosmology}
\label{sec:nu-cosmo}

\subsection{Production}
\label{sec:nucosmo:prod}

To understand cosmological neutrino production, including the role of flavor, momentum spectra, chemical potentials, interactions etc., a treatment using quantum kinetic equations is required. These are a matrix generalization of the usual Boltzmann equations, necessitated by the need to track the possibly coherent flavor inter-conversions. Some of the early works on this topic can be found in refs.~\cite{Dolgov:1980cq,Barbieri:1990vx,Enqvist:1990ad,Sigl:1992fn,McKellar:1992ja}. The central entity is the quantum field theoretic expectation value $\langle \hat{a}^{\dagger}_{\beta}(\vec{p}) \, \hat{a}_{\alpha}({\bf p}') \rangle = (2\pi)^{3} \delta^{3}(\vec{p}-\vec{p}') \varrho_{\alpha\beta}$, where $\alpha,\,\beta$ run over the neutrino flavors $e,\,\mu,\,\tau,$ including possibly sterile states denoted by the subscript $s$.  Each momentum mode has its own $\varrho$, but perplexingly $\varrho$ refers to a quantum state at a given position at the time. One may worry that such simultaneous position and momentum specification is not allowed in a quantum treatment. The way out is to think of these $\varrho$ as coarse-grained matrices of densities, i.e., Wigner functions smeared over a scale somewhat larger than the de~Broglie wavelength~\cite{Stirner:2018ojk}. Thus, each diagonal element of $\varrho$ is a non-negative distribution encoding the phase-space density of that neutrino flavor
\begin{align}
\varrho_{\alpha\alpha}=f_{\alpha}({\bf p}, {\bf x},t)\quad\textrm{(no~sum~over~$\alpha$)}\,,
\end{align}
while the off-diagonal terms encode correlations between the flavors. What this construction does is to separate the slow quantum dynamics of the internal state from the faster quantum dynamics of the spacetime motion of the particle itself. The latter is then taken to be in its classical limit. Each $\varrho$ obeys the von~Neumann equation
\begin{align}
  i\frac{d\varrho}{dt} = [{\sf H},\varrho]+{\sf C} \,,
\end{align}
where the Hamiltonian, assuming only Standard Model interactions, is 
\begin{align}
  {\mathsf H}_{\rm SM} = {\sf U}^{\dagger} \frac{{\sf M}_{\nu}^{2}}{2E_{\nu}} \,{\sf U}
                       + \sqrt{2}G_{F} \Bigg[ {\sf L}_{\ell} + {\sf L}_{\nu}
                         -\frac{8 E_{\nu}}{3} \bigg(\frac{\sf E_{\ell}}{M_{W}^{2}}
                                                  + \frac{\sf E_{\nu}}{M_{Z}^{2}} \bigg) \Bigg] \,.
  \label{eq:qke1}
\end{align}
It includes neutrino mass-mixing in the first term and the refractive (i.e., momentum preserving) effects of the medium in the second. Here ${\sf M}_{\nu}^{2}$ is the mass-square matrix for the neutrinos and ${\sf U}$ the mixing matrix. ${\sf E}$ and ${\sf L}$ contain the refractive effects due to thermal and loop corrections to the neutrino self-energy, which add coherently to free propagation. Specifically, ${\mathsf L}_{\nu}=\int d^{3}\mathbf{p}/(2\pi)^{3} \, (\varrho-\bar{\varrho})$ are the particle--anti-particle differences in the neutrino number density, whereas $\mathsf{E}_{\nu} = \int d^{3}\mathbf{p}/(2\pi)^{3} \, E_{\nu}(\varrho+\bar{\varrho})$ are the sums of the corresponding energy densities. Charged lepton energy densities and number densities are similarly defined. They are diagonal matrices with entries that can be interpreted as  potentials. E.g., for the three flavors of Standard Model neutrinos at low temperatures, where only electrons and positrons are present,
\begin{align}
  \sqrt{2}G_{F} \left({\sf L}_{\ell} - \frac{8E_{\nu}}{3M_{W}^{2}}{\sf E}_{\ell}\ \right)
  = \text{diag}(V_\text{eff},0,0) \,,
\end{align}
with $V_\text{eff} = \sqrt{2}G_{F}(n_{e^{-}}-n_{e^{+}})-8\sqrt{2}G_{F}\,E_{\nu}/(3M_{W}^{2})$~\cite{Weldon:1982bn,Notzold:1987ik}. When there is no particle--anti-particle asymmetry, the ${\sf L}$ terms are zero, whereas the ${\sf E}$ terms remain nonzero. This is why the latter are usually more important at high temperatures, unless one considers lepton asymmetries that are significantly larger than the observed baryon asymmetry of the Universe.

The matrix ${\sf C}$ in \cref{eq:qke1} encodes the collision integral, which includes the effect of all momentum changing processes and is given by
\begin{align}
  {\sf C} = -\frac{i}{2} G_{F}^{2} \big[ \{ {\sf S}^{2}, \varrho - {\sf I}\}
                                       - 2 {\sf S} (\varrho-{\sf I}) {\sf S}
                                       + \{ {\sf A}^{2},\varrho-{\sf I} \}
                                       - 2 {\sf A}(\varrho-{\sf I}) {\sf A} \big]\,,
\end{align}
where ${\sf S}$ and ${\sf A}$ are diagonal matrices in flavor basis that encode various scattering and annihilation rates for each flavor. For illustration, in a 2+1 flavor approximation with 2 active flavors $e$ and $\mu$ and 1 non-interacting sterile flavor, one has ${\sf S}={\rm diag}(g_{s}^{e},g_{s}^{\mu},0)$ and ${\sf A}={\rm diag}(g_{a}^{e},g_{a}^{\mu},0)$, with $(g_{s}^{e})^{2}\approx 3.06$, $(g_{s}^{\mu})^{2}\approx 2.22$, $(g_{a}^{e})^{2}\approx 0.5$ and $(g_{a}^{\mu})^{2}\approx 0.22$ encoding the strengths of scattering and annihilation processes affecting $e$ and $\mu$ flavored neutrinos. A systematic and detailed discussion of these coupling-matrices can be found in ref.~\cite{Sigl:1992fn}. Numerical values of the various $g_{i}$ are available in ref.~\cite{Gariazzo:2019gyi}. If the sterile neutrinos have additional interactions they will add refractive potentials in ${\sf L}$ and ${\sf E}$, as well as modify the ${\sf S}$ and ${\sf A}$ matrices.

Note that in the above formalism anti-neutrinos of energy $p$ are represented by $\bar\varrho_{\alpha\beta} \sim \langle \hat{b}^{\dagger}_{\alpha} \hat{b}_{\beta} \rangle$, and they are treated as if they were neutrinos but with a negative $p$. This means in particular that the $\mathsf{M}_{\nu}^{2}$-term and the $\mathsf{E}$-terms in \cref{eq:qke1} enter with an overall minus sign.

The matrices of densities obey the same rules as usual density matrices, i.e., $\textrm{Tr}\varrho$ is conserved as long as there are no number changing processes, whereas $\textrm{Tr}\varrho^{2}$ is only conserved if there is no decoherence. Typically, the state $\varrho$ decoheres, and the off-diagonal entries eventually vanish. This can happen, e.g., owing to reduced spatial overlap of originally coherent components that get separated out by being scattered away due to collisions (or by traveling at slightly different speeds, an effect that we will ignore). In that limit, it is sufficient to work with the diagonal components individually, which simply obey the usual Boltzmann equations.

These quantum kinetic equations are challenging to solve. In the context of sterile neutrinos, the problem is further complicated by not just the increase in the dimension of the Hilbert space, but also chemical potentials, CP violation, and new scales introduced by the flavor-universal neutral current interactions that can be dropped when dealing with active neutrinos only, but not when also sterile neutrinos are present. Numerically, these challenges have prevented a full solution. However, over the past two decades increasingly more complete treatments have been attempted~\cite{Dolgov:2003sg, Hannestad:2012ky, Hannestad:2015tea, Gariazzo:2019gyi}.

\subsubsection{1+1 Flavors}

To obtain some more insight, it is very useful to focus on a two-flavor approximation, with one active neutrino flavor $\nu_{a}$ and one sterile flavor $\nu_{s}$. In this case, one can express all density matrices in terms of Bloch vectors by writing $\varrho = \frac{1}{2}[P_{0} + {\sigma} \cdot \vec{P}]$, with the Pauli matrices $\sigma_i$. Here $P_{0}$ encodes the occupation number of neutrinos with momentum $\vec{p}$, i.e., $f(\vec{p})$, and $\vec{P}$ carries information on the flavor composition including coherence. One commonly uses the basis $\{ \hat{\vec{e}}_{1}, \hat{\vec{e}}_{2}, \hat{\vec{e}}_{3} \}$ in flavor space, where $\hat{\vec{e}}_{3}$ points in the direction of a pure $\nu_{a}$ state.  In Bloch vector notation, the quantum kinetic equations take the following form:
\begin{align}
  (\partial_{t} - H p \, \partial_{p}) \vec{P}
    &= \vec{H} \times \vec{P}
    - D (P_{1} \hat{\vec{e}}_{1} + P_{2} \hat{\vec{e}}_{2}) + R \hat{\vec{e}}_{3} \,,
                                          \label{eq:pol1} \\
  (\partial_{t} - H p \, \partial_{p}) {P}_{0} &= R \,,
                                          \label{eq:pol2}
\end{align}
where $\vec{H}$ is the Bloch vector corresponding to the Hamiltonian in \cref{eq:qke1}, the decoherence function is
\begin{align}
  D = \frac{\Gamma_\text{tot}}{2} \,,
\end{align}
and the repopulation function is
\begin{align}
  R = \Gamma_\text{tot} \left[ f_\text{eq} - \frac{1}{2} (P_{0} + P_{3}) \right] \,.
\end{align}
It is worth noting that the decoherence term depends on the total collision rate $\Gamma_\text{tot} = (\Gamma_{a} + \Gamma_{s})$, which means that, even if one of the flavors does not have any interactions, it can still undergo collisional decoherence, owing to the collisions experienced by the other flavor.

In standard cosmology, the active neutrinos decouple from weak interactions roughly at $T_{\nu\textrm{D}} = 1.2\MeV$, as dictated by a competition of the neutrino interaction rate $n \langle\sigma v\rangle \simeq G_{\text{F}}^{2} T^{5}$ with the Hubble expansion rate $H$, where we ignore factors of order one. By examining the steady state solution of \cref{eq:pol2}, one can see that before neutrino decoupling, the large interaction rate drives $f$ to $f_\text{eq}$, that is to a Fermi-Dirac distribution with temperature $T$ and chemical potential $\mu$. The chemical potential is nonzero when there is a neutrino--anti-neutrino asymmetry. It was shown that the stringent BBN bounds on an electron neutrino asymmetry apply to all flavors, since neutrino oscillations led to approximate flavor equilibrium before BBN~\cite{Dolgov:2002ab, Wong:2002fa, Abazajian:2002qx}. Therefore, in standard cosmology, it is a good approximation to ignore any initial asymmetry and one can set $\mu \to 0$. After active neutrinos decouple, they initially have a temperature that is the same as that of the photon bath. However, around $T \approx m_{e}$ the electrons and positrons annihilate and dump energy mainly into the photon bath, and far less into neutrinos. As a result, the photon bath cools more slowly than $1/a^3$, while the neutrinos continue to be diluted unhindered. By convention, the temperature of the photon bath is referred to as the temperature of the Universe, so in relative terms the neutrino temperature becomes lower due to $e^+ e^-$ annihilation. Assuming that the entropy transfer from electron and positrons to photons does not affect the neutrinos at all, one finds the neutrino temperature is lower by the ratio $T_\nu^\text{std}/T = (4/11)^{1/3}$ afterwards.
%

Sterile neutrinos may initially not have been in contact with the rest of the thermodynamic bath. Confining our discussion to thermal scenarios, the simplest possibility is that they were produced purely by oscillations. If the active neutrinos mix ever so slightly with sterile neutrinos, they create a small sterile component through oscillations, i.e., the precession of $\vec{P}$ around $\vec{H}$ in \cref{eq:pol2} allows the component of $\vec{P}$ along $\hat{\vec{e}}_{3}$ to reduce. If the collision rate $\Gamma_\text{tot}$ is smaller than the Hubble rate $H$, neutrinos can be taken to be oscillating without scattering.  Then if $\Delta m^2/(2E_{\nu}) \gg H$ the final sterile neutrino number density is $n_{\nu_s} \simeq \tfrac{1}{2} \sin^2 2\theta_\text{m}\,n_{\nu_a}$, where
the mixing angle in matter $\theta_m$ is given by~\cite{Akhmedov:1999uz}
\begin{align}
  \sin^2 2\theta_\text{m}
    = \frac{\sin^2 2\theta}
           {\Big(\cos 2\theta + \tfrac{2 E_{\nu}}{\Delta m^2} V_\text{eff} \Big)^2
           + \sin^2 2\theta} \,,
  \label{eq:s22thm}
\end{align}
Refractive effects are contained in the potential $V_\text{eff}$. The final population of sterile neutrinos thus remains small due to the matter-suppressed mixing angle. The small $\nu_s$ population that is produced inherits the Fermi-Dirac momentum spectrum of the bath. 

If $\Gamma_\text{tot}$ exceeds the Hubble rate $H$, then sterile neutrinos
cannot be treated as non-collisional~\cite{Stodolsky:1986dx}. Instead, the coherent
evolution of the flavor ensemble is constantly interrupted by collisions, which
can be viewed as ``measurements'' in the quantum mechanical sense, projecting the neutrino
state onto either $\nu_a$ or $\nu_s$.  The $\nu_a$ start oscillating again, once
again developing a small $\nu_s$ component.  Thus, after many collisions, a
sizable abundance of $\nu_s$ can be produced.  The $\nu_a \to
\nu_s$ production rate is $\Gamma_\text{prod} \simeq \frac{1}{2} \sin^2 2\theta_m
\, \Gamma_\text{tot}$, where we can interpret
the first factor as the average probability that an initially active
neutrino is in its sterile state at the time of collision.  The second factor
gives the scattering rate that keeps ``measuring'' it in a flavor state. The
fraction of $\nu_a$ converted to sterile neutrinos is not limited by the mixing
angle, and all neutrino flavors can end up with equal phase space densities. In the Bloch vector language, $\vec{P}$ precesses around $\vec{H}$ with a small angle, but constantly gets projected onto the $\vec{e}_{3}$ direction due to collisions. As a result, the vector shrinks in length and can eventually vanish, i.e., both flavor states can be equally occupied. If the collision rate is finely tuned, so as to not fully thermalize the sterile neutrinos, sub-thermal abundances can also be produced. The momentum distribution of the sterile neutrinos is approximately the same as that of the active neutrinos in this case. We will revisit this later when we discuss keV-mass sterile neutrinos as dark matter. This is the Dodelson--Widrow mechanism~\cite{Dodelson:1993je}.

The above picture can change drastically if there is a Mikheyev--Smirnov--Wolfenstein (MSW) resonance~\cite{Wolfenstein:1977ue, Mikheev:1986gs}. Oscillations can be enhanced due to matter effects (${\sf L}$ and ${\sf E}$ contributions from $V_\text{eff}$) that make the $\vec{e}_{3}$ component of $\vec{H}$ small, so that the precession of $\vec{P}$ around $\vec{H}$ in \cref{eq:pol1} happens at a large angle, i.e., $\theta_\text{m}$ is large due to a cancellation in the denominator of \cref{eq:s22thm}. This happens either for neutrinos or anti-neutrinos (but not both) and thus leads to different abundances of sterile neutrinos and anti-neutrinos. The neutrino momentum spectra can then depart significantly from a Fermi-Dirac spectrum. In the cosmological context, the MSW potential is dominated by the self-energies of the particles and usually there are no resonances. However, if large lepton asymmetries are present, there can be a large contribution from them and resonances can occur. This leads to resonant production and, because the resonance is sensitive to the neutrino momentum, the resulting neutrino spectrum can be significantly different from a thermal spectrum. This is referred to as the Shi--Fuller production mechanism~\cite{Shi:1998km}.

One issue that has attracted attention here is that the final sign, but not the absolute value, of the neutrino asymmetry was said to be unpredictable. In a sense it was being argued that it depended in some chaotic manner on initial conditions. Recent numerical studies with higher resolution have demonstrated that the sign oscillation is a numerical artifact, and in fact the sign is completely deterministic~\cite{Hannestad:2013pha}.

\subsection{Observables}
\label{sec:cosmo-observables}

The effects of neutrinos on cosmology have been studied and elaborated upon previously in great detail, focusing especially on their number and mass. However, for sterile neutrinos, not only their number and mass can be probed, but also their interaction strength, which can be quite different from that of active neutrinos. Concretely, the mostly-sterile mass eigenstates experience weak interactions suppressed by a factor of the active--sterile  mixing angle. Moreover, neutrinos, whether active or sterile, may also have new interactions beyond the Standard Model, \eg, so-called ``secret interactions'' among themselves, or perhaps with a component of dark matter. The interactions among sterile states could be much stronger than SM weak interactions. Thus it is useful to delineate the effects of neutrinos on cosmology -- as a function of their number, their mass, and their interaction strength. Of course, cosmology can be sensitive to more detailed properties, and we outline them in brief at the end.

\subsubsection{Number of Neutrinos}

Relativistic neutrinos contribute to the energy density of the Universe as radiation. Their contribution is parameterized in terms of an effective number of neutrino flavors $N_\text{eff}$, defined by
\begin{align}
&\frac{\rho_r-\rho_\gamma}{\rho_\nu^{\rm std}} = \Neff\,,
\end{align}
where $\rho_r$ is the total radiation energy density, $\rho_\gamma$ is the photon contribution to it, and $\rho_\nu^{\rm std} = 2 \times \frac{7}{8} \, \frac{\pi^{2}}{30} \, \left( \frac{4}{11} \right)^{4/3} T^4$.
Obviously, any new relativistic species could lead to an increase in $N_\text{eff}$, but one might naively think that $N_\text{eff}$ cannot be larger than 3 if no exotic particles exist. In reality, it is slightly larger even without any new physics. The earlier precision calculations predicted that $\Neff = 3.046$~\cite{Mangano:2001iu}, but recent detailed studies suggest $N_{\rm{eff}} \approx 3.044\pm0.0005$ in the Standard Model~\cite{Akita:2020szl, Froustey:2020mcq}. The excess over 3 encodes the effective heating of neutrinos during $e^+e^-$ annihilation. The remaining theoretical uncertainty comes from approximations made in the treatment of finite temperature and/or higher-loop QED effects~\cite{Bennett:2019ewm}, non-perturbative effects, etc. The deviations of $N_\text{eff}$ from 3 reflects the fact that the active neutrinos are not described by an exact Fermi--Dirac distribution with $T_\nu^\text{std} = (4/11)^{1/3}T$.

Neutrinos with a mass smaller than $\sim\eV$ contribute as radiation during recombination and earlier. Therefore, both BBN and CMB data are sensitive to them via $N_{\rm eff}$. BBN is treated within homogeneous cosmology. A larger $\Neff$ leads to a larger expansion rate. This affects the synthesis of light nuclei because the weak interaction will freeze out earlier, increasing the $n/p$ ratio and thus resulting in a larger $^{4}\textrm{He}$ fraction. On the other hand, if the extra neutrinos are $\nu_{e}$, they decrease the $n/p$ ratio. Overall, taking both of these effects together, one finds~\cite{Dolgov:2003sg}
\begin{align}
  \Delta N_\text{eff,\,BBN} = \frac{4}{7} \left[ \frac{4\times10.75 + 7\Delta N_{\nu}}
                                                      {(1+n_{\nu_{e}})^{2}} - 10.75 \right]\,,
\end{align}
for the change in $\Neff$ at BBN due to $\Delta N_\nu$ additional neutrino flavors and a possible change to $n_{\nu_{e}}$. 

\begin{figure}
  \centering
  \includegraphics[width=0.48\textwidth]{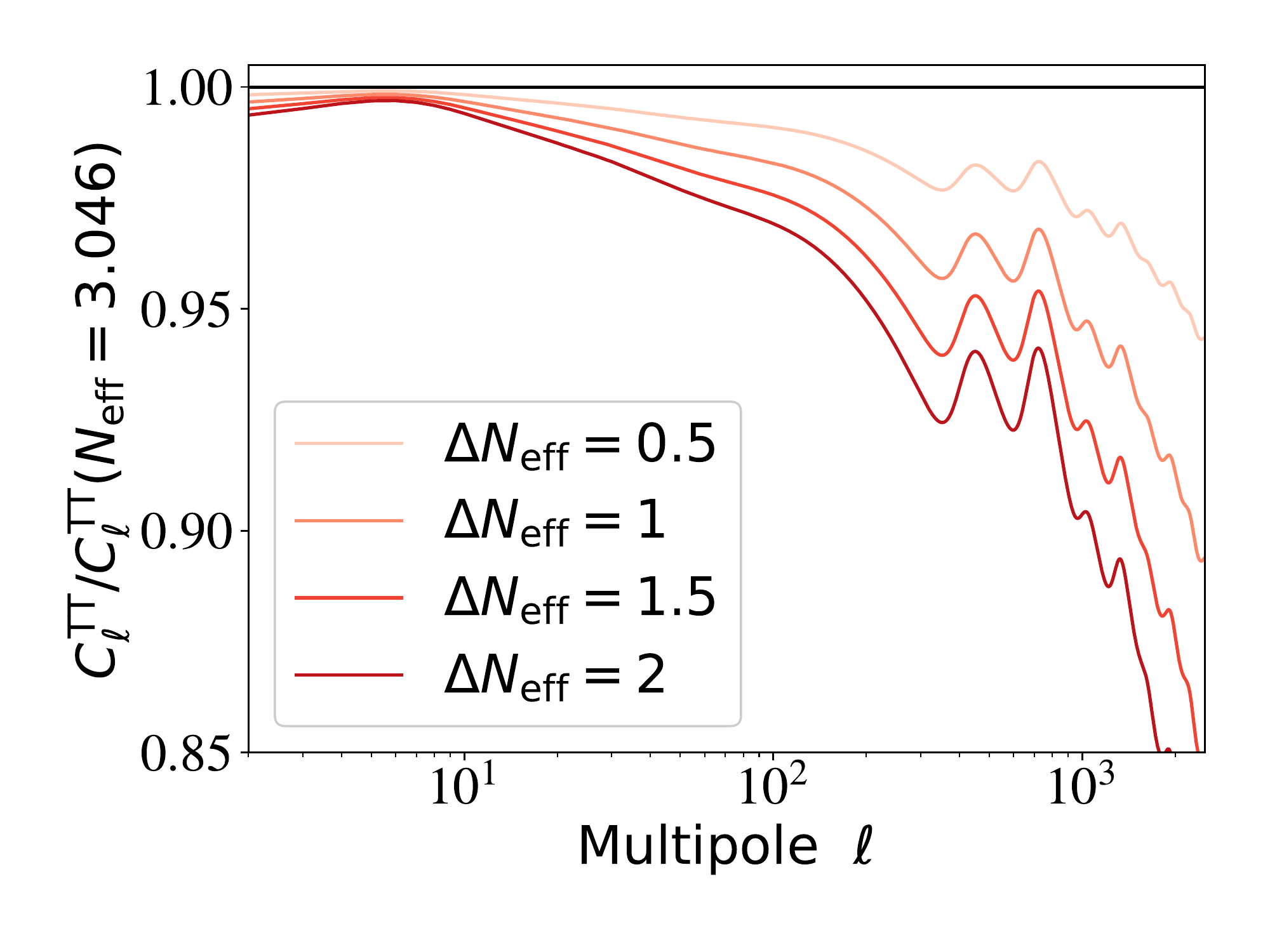}\quad
  \includegraphics[width=0.48\textwidth]{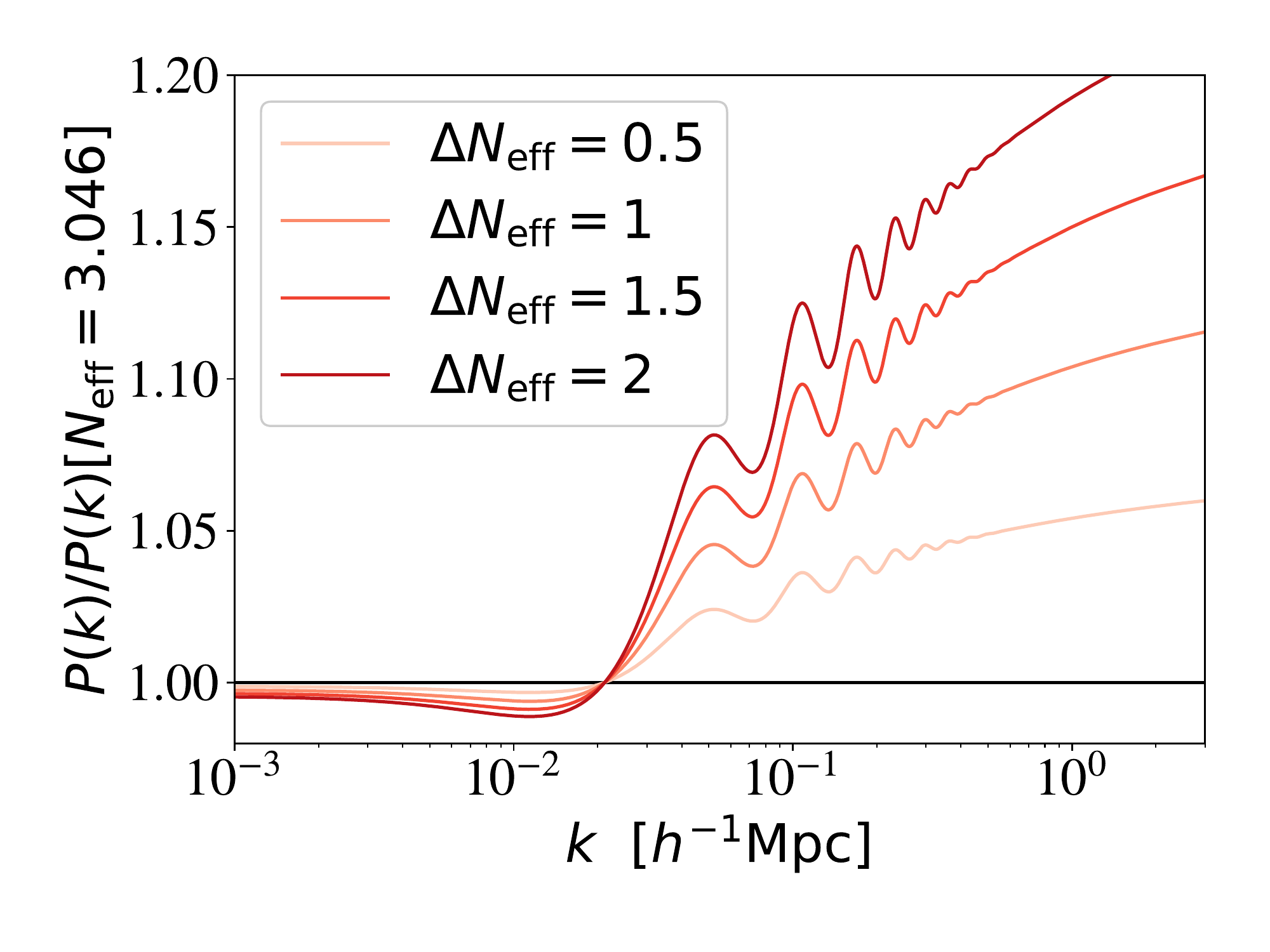}
  \caption{\emph{Left panel:} the relative change in the angular correlation $C_{\ell}^{TT}$ of the CMB power spectrum with and without extra neutrinos as function of the multipole number $\ell$. (Larger $\ell$ corresponds to smaller distance scales.)  The energy density carried by the extra neutrinos is parameterized as $\Delta N_\text{eff} = \Neff - 3.046$.  \emph{Right panel:} the relative change in the power spectrum of matter fluctuations, $P(k)$, as a function of wave number $k$ and for different values of $\Delta N_\text{eff}$. (Once again, larger $k$ corresponds to smaller distance scales.). The cosmological parameters $z_{\text{eq}},\,z_{\Lambda},\,\omega_{b},$ and $\tau$ are kept unchanged. Figures taken from the PDG review by Lesgourgues and Verde~\cite{Tanabashi:2018oca}.}
  \label{fig:Neff}
\end{figure}

Unlike BBN, the CMB is sensitive not only to the homogeneous component of the Universe, but also to the perturbations on top of it.  This allows the CMB to probe radiation-like neutrinos in a more detailed manner. The effect of increasing $N_\text{eff}$ on the CMB depends on which other parameters are allowed to vary simultaneously. Increasing $N_\text{eff}$, and changing nothing else, delays matter--radiation equality, which would drastically change many cosmological observables. But this is not a specific signature of extra neutrino energy density, because other changes, \eg, decreasing the energy density of dark matter, can lead to a similar effect. In the context of CMB observations, it is useful to realize that the ratio of the length-scale corresponding to the first acoustic peak in the CMB power spectrum,
\begin{align}
  r_{\rm a}=\int_{0}^{a_{\rm rec}} da\frac{c_{s}}{a^{2}H}\,,
\end{align}
and the photon diffusion scale
\begin{align}
  r_{\rm d}^{2}=\int_{0}^{a_{\rm rec}} da\frac{1}{a^{3}Hn_{e}\sigma_{T}}\,,
\end{align}
depends essentially only on the Hubble rate and the electron density, \ie,
\begin{align}
  \frac{r_{\rm d}}{r_{\rm a}}=\frac{\theta_{\rm d}}{\theta_{\rm a}}\propto\sqrt{\frac{H}{n_{e}}}\,.
  \label{eq:rd-over-ra}
\end{align}
All other cosmological parameters mainly affect the angular diameter scale, which is eliminated in \cref{eq:rd-over-ra} by taking the ratio.  In the above expression, $c_s$ is the speed of sound in the electron--photon plasma and $a_\text{rec}$ is the scale factor at recombination. The electron density $n_e$ entering in \cref{eq:rd-over-ra} is independently determined from BBN, so any putative increase in $N_\text{eff}$ directly increases $H^{2}$. Given that the angular scale of the first acoustic peak, $\theta_{a}$, is directly measured to 0.03\% precision by Planck, the diffusion scale $\theta_{d}$ increases if $\Neff$ is increased. In other words, CMB photons react to the gravitational effects of neutrinos. Increasing the neutrino energy density increases the overall energy density, but reduces density perturbations at small scales where neutrinos do not cluster. Thus the primary effect of increased neutrino density is a relatively stronger damping at large multipole number $\ell$. The reduction in $C_{\ell}^{TT} \sim \exp(-\ell^{2} \theta_\text{d}^{2})$ is roughly proportional to $\Delta\Neff$, and quantitatively one finds~\cite{Hu:1995en}
\begin{align}\label{CelNeff}
  \frac{\Delta C_{\ell}}{C_{\ell}} = 
    \left[ 1 + \frac{4}{15} \left(\frac{0.2271\Neff}{1+0.2271\Neff}\right) \right]^{-2}
  \approx - 0.072 {\Delta\Neff}
\end{align}
This roughly agrees with what is shown in the left panel of \cref{fig:Neff}.

The matter power spectrum is sensitive to $\Neff$ due to its effect on the background geometry. The basic idea is that one must not change the epoch of matter--radiation equality or $\Omega_{b}$, which are both very well measured by the CMB. Thus, a putative increase in $N_{\rm eff}$ must be accompanied with a related increase in $\Omega_{c}$, which in turn reduces the ratio $\Omega_{b}/\Omega_{c}$. This effective reduction of the baryon-to-dark-matter ratio, owing to increased $N_{\rm eff}$, allows more clustering at small scales where perturbations in baryonic matter are otherwise allowed to damp. The reduction in damping of the matter power spectrum at small scales due to increased $N_{\rm eff}$, leads to more small-scale power as seen in the right panel of \cref{fig:Neff}.

\subsubsection{Neutrino Mass}

Non-relativistic neutrinos behave like dark matter. In fact, given the laboratory limits on the mass and interactions of active neutrinos, they have decoupled and at least two out of the three species have become non-relativistic sometime between the epoch of recombination and now. Early Universe constraints on neutrinos depend quite crucially on whether neutrinos act like matter or radiation at the relevant epoch. eV-scale sterile neutrinos will certainly have become non-relativistic by the era of recombination.

The signature of neutrino mass in cosmology is the smearing of structure on distances smaller than their maximal free-streaming scale. The comoving free-streaming horizon at a time $t$ is
\begin{align}
  r_\text{fs}=\int_{0}^{t}dt\frac{\langle v_{\rm th}\rangle}{a}\,,
  \label{eq:r-fs}
\end{align}
where $\langle v_\text{th}\rangle$ is the average thermal velocity of the neutrino. This is an integrated quantity that legitimately decides the largest distances up to which neutrino free-streaming erases density perturbations therein. A related instantaneous quantity is the free-streaming length-scale $\lambda_\text{fs} = 2\pi / k_\text{fs} \simeq 2\pi\sqrt{\frac{2}{3}}\frac{\langle v_{\rm th}\rangle}{H}$, which is essentially the ratio of the average neutrino speed to the Hubble rate. The wide usage of this instantaneous quantity is that it allows one to display the dependence on various parameters more transparently than \cref{eq:r-fs}; initially it increases in the radiation dominated era, reaching its maximum value at the time when neutrinos transition from being relativistic to non-relativistic, and then decreases as the thermal velocities become smaller.  For neutrinos that turn nonrelativistic in the matter-dominated regime $r_\text{fs}$ and $2\pi/k_\text{fs}$ are close. This is the case with eV-scale neutrinos. For heavier neutrinos, such as the keV-scale neutrinos that we will encounter in \cref{sec:keVchapt}, the transition occurs already in the radiation-dominated era. Then, the correct free-streaming horizon, $r_\text{fs}$, is significantly larger than $2\pi/k_\text{fs}$, making the latter a poor approximation to it. The difference comes from the logarithmic growth of the comoving free-streaming horizon in the epoch between when the neutrinos turn nonrelativistic and when the Universe turns matter-dominated. 

The value of the maximal free-streaming scale, $k_\text{nr}$, encodes the scale corresponding to the Hubble horizon at the time neutrinos became non-relativistic. As we noted, it depends on whether the neutrino became non-relativistic in the radiation era or the matter era. For neutrinos with mass $\lesssim1.5$\,eV, this transition occurs in the matter dominated epoch and it has been shown that the maximal free-streaming scale in this case corresponds to~\cite{Lesgourgues:2018ncw}
\begin{align}
  k_\text{nr} \simeq 0.0178 \, \Omega_{m}^{1/2} 
                     \bigg( \frac{T_{\nu}^{\text{std}}}{T_{\nu}} \bigg)
                     \bigg( \frac{m_{\nu}}{1\,\text{eV}} \bigg)^{1/2} \,
                     h \, \text{Mpc}^{-1} \,.
\label{eq:knreV}
\end{align}

\begin{figure}
\centering
\includegraphics[width=0.48\textwidth]{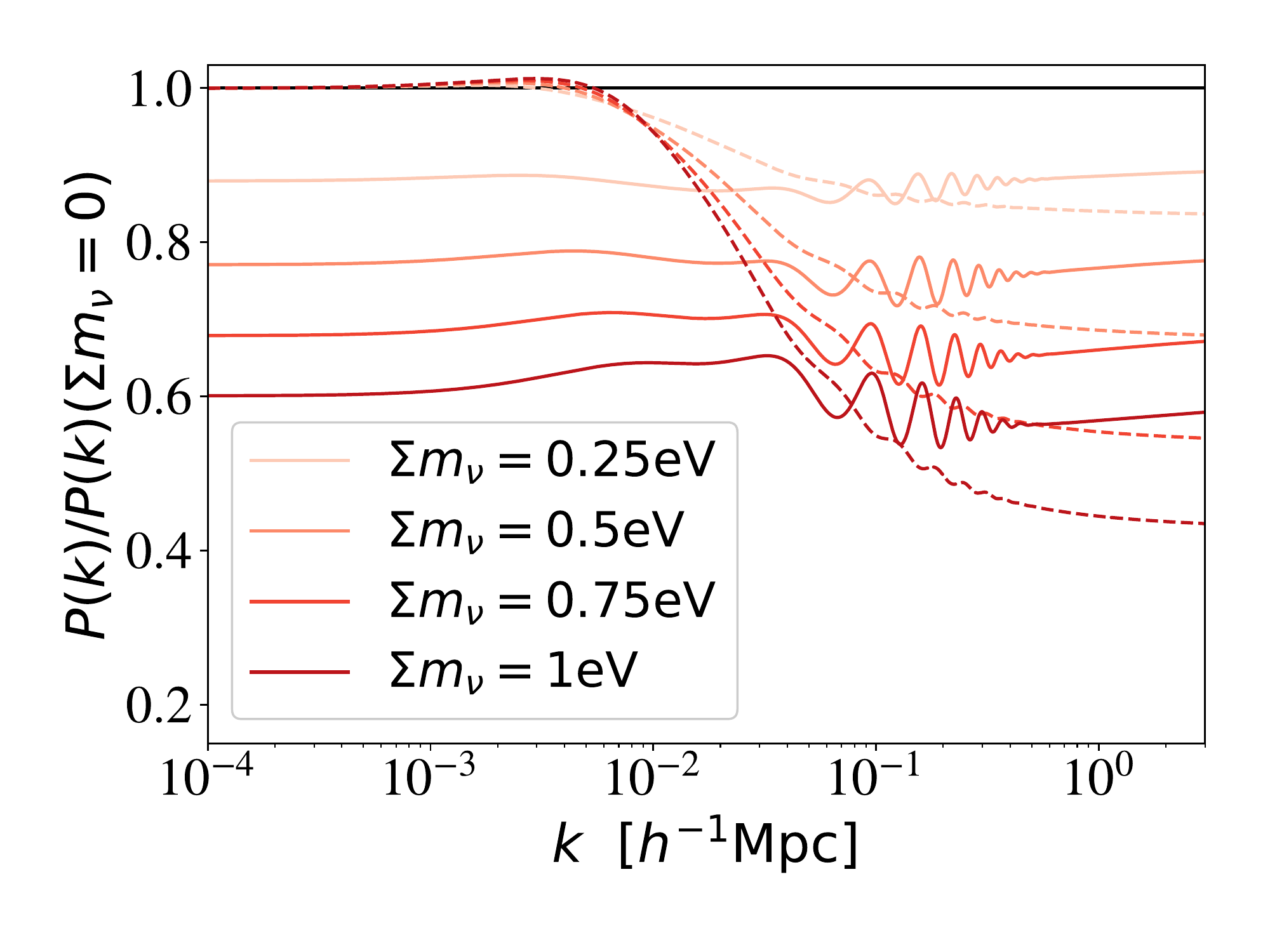}\quad
\includegraphics[width=0.48\textwidth]{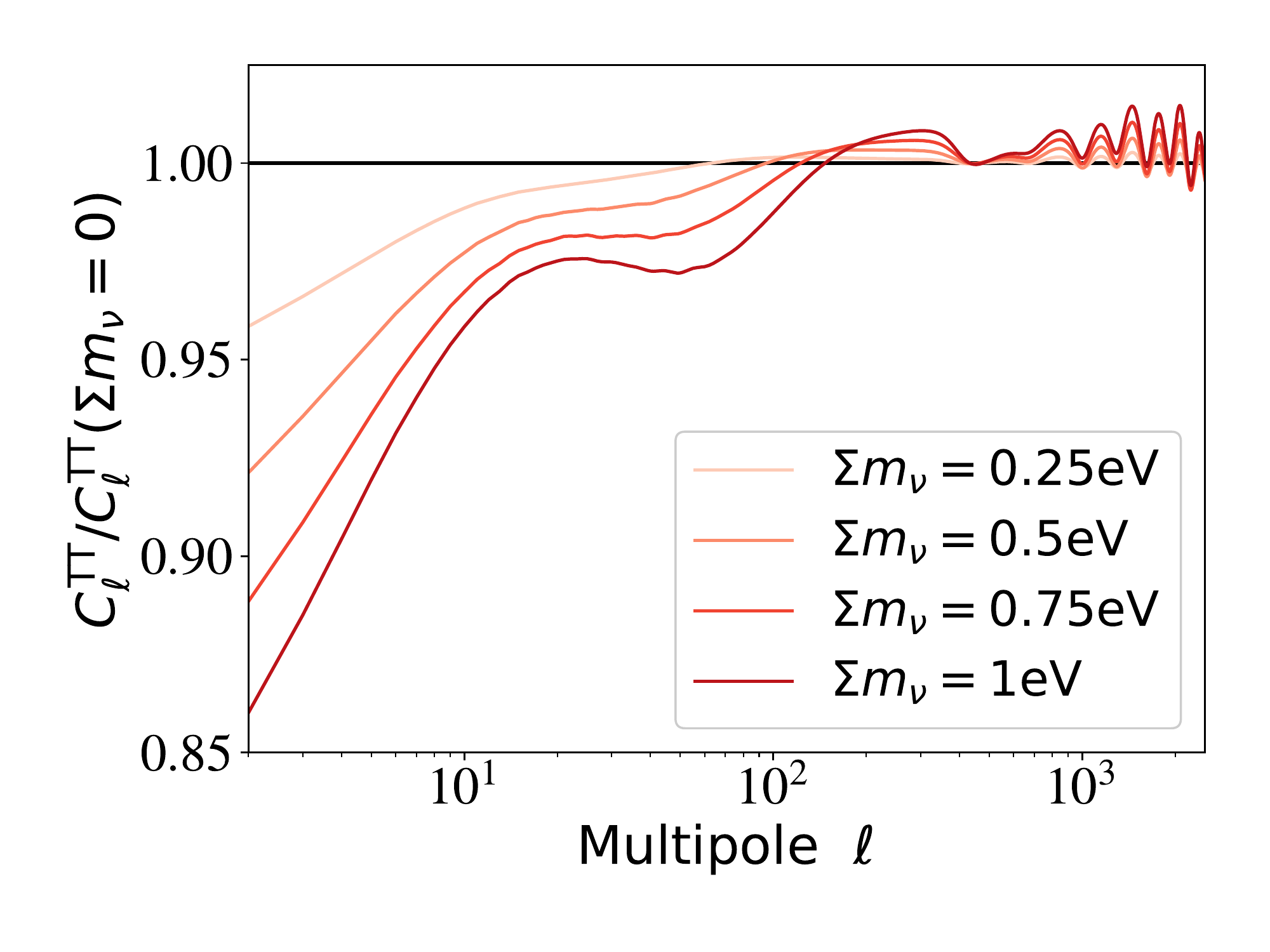}
\caption{Plot of the ratio of the matter power spectrum $P(k)$ to its value without any neutrinos vs.\ wave number $k$ (left panel) and of the CMB angular correlation $C_{\ell}^{TT}$ vs.\ multipole order $\ell$ (right panel) for varying neutrino mass $m_{\nu}$. The baryon density $\omega_{b}$, the cold dark matter density $\omega_{c}$, the optical depth to reionization $\tau$, the acoustic scale $\theta_{a}$, and the parameters $A_s$ and $n_s$ are kept fixed; for the dashed lines in the left panel, the total matter density $\omega_{m}$, the baryon density $\omega_{b}$, and the dark energy density $\Omega_{\Lambda}$ are held fixed to distill the effect of neutrino free-streaming. Figures taken from the PDG review by Lesgourgues and Verde~\cite{Tanabashi:2018oca}.}
\label{fig:mnu}
\end{figure}

Neutrinos act as radiation as long as they remain relativistic and therefore do not take part in gravitational clustering. This means that neutrino density perturbations get erased on length-scales smaller than the largest neutrino free-streaming length. For these scales $k \gg k_\text{nr}$, the power spectrum reduces by a factor proportional to the neutrino energy density. The ratio of the matter power spectrum, including effects of neutrino mass, to the putative neutrino power spectrum is~\cite{Hu:1997mj}
\begin{align}
\frac{P_{m}}{P_{m}^{0}}=\left(1-\frac{\Omega_{\nu}}{\Omega_{m}}\right)^{2}\left[\frac{\delta_\text{c}}{\delta_\text{c}^{0}}\right]^{2}\approx1-8\frac{\Omega_{\nu}}{\Omega_{\rm m}}\,.
\label{eq:PmeV}
\end{align}
In the deeply nonlinear regime, the prefactor of 8 in the above equation increases to almost 10 at $k\approx1\,h\,\text{Mpc}^{-1}$~\cite{Brandbyge:2008rv}.
This fractional reduction of the matter power spectrum upon varying $f_{\nu}$ and keeping all other parameters fixed, i.e., with the CDM density adjusted to keep a fixed total dark matter density, can be seen in \cref{fig:mnu} (dashed lines in left panel). However, as pointed out in the review by Lesgourgues and Verde~\cite{Tanabashi:2018oca}, this way of looking at the problem does not leave the redshift of matter--radiation equality invariant, which itself has very large effects on the CMB spectra, as we noted in our discussion of $N_{\rm eff}$. Following the logic of minimizing the impact on the CMB and fixing $z_{\rm eq}$, as we did in the case of $N_{\rm eff}$, the increase in the total matter energy density $\rho_{m}$ brought on by a larger $m_{\nu}$ must be accompanied by a related increase in $\rho_{\rm crit}$ by increasing $H_{0}$, so as to keep $z_{\rm eq}$ unchanged. This tends to suppress the large-scale power spectrum, by approximately the same amount as the neutrino free-streaming effect. The impact of neutrino masses on the matter power spectrum then appears as an overall amplitude suppression, which can be seen in \cref{fig:mnu} (solid lines in left panel). This effect is not degenerate with reduced primordial power, because the power in the CMB spectra is unaffected by this procedure. Of course, $A_{s}$ enters the equations through $A_{s}e^{-2\tau}$, and thus a measurement of $\tau$ is necessary in order to fix $A_{s}$ from CMB data, and avoid a parameter degeneracy between $m_{\nu}$, $A_{s}$, and $\tau$.

The CMB data are sensitive to eV-mass neutrinos mainly through reduced gravitational lensing of the CMB owing to the small-scale suppression of the matter power spectrum and through the modification of the distance to the last scattering surface. Roughly one has~\cite{Lesgourgues:2018ncw}
\begin{align}
  \frac{\Delta C_{\ell}^{TT}}{C_{\ell}^{TT}}=-\frac{m_{\nu}}{10\,\text{eV}}\,.
\end{align}
This agrees well with the dependence shown in the right panel of \cref{fig:mnu}.

As a historical note, a quantitative treatment of the effect of massive neutrinos on cosmology was first provided by Bond, Efstathiou and Silk~\cite{Bond:1980ha}. Thereafter, this was a subject of intense theoretical investigation~\cite{Bond:1983hb}, in particular with the development of $N$-body simulations for studying cosmological structure formation~\cite{White:1984yj}. During the 1980s, neutrinos were considered a plausible dark matter candidate, and these studies provided concrete predictions that have now been used to rule out the standard neutrinos as a major component of dark matter. This was perhaps the second time that a fundamental fact about our Universe was derived out of a simulation~\cite{White:2018kji}.\footnote{The first, to our knowledge, being the remarkable discovery by Fermi, Pasta, Ulam and Tsingou that Hamiltonian evolutions of even seemingly complicated systems, with many degrees of freedom, do not often lead to ergodicity.}

\subsubsection{Neutrino Interactions}

Neutrino interactions can hinder the free-streaming of neutrinos and affect both the CMB and the large scale structure observables. For definiteness, let us consider that neutrinos interact with each other only via a contact interaction with a cross section
\begin{align}
  \sigma_{X}=G_{X}^{2} E^{2}\,,
\end{align}
where $G_{X}$ has the same dimensions as the Fermi constant $G_{F}$ and encapsulates the strength of these so-called secret neutrino interactions. 

Consider that the neutrinos were in kinetic equilibrium until some time $t_{\rm kd}$, until which time they only propagated via diffusion, and free-stream only thereafter. In contrast to always free-streaming neutrinos, interacting neutrinos have an effectively lower free-streaming length:
\begin{align}
  r_\textrm{fs} \approx \sqrt{r_\textrm{dif}^{2} + r_\textrm{bal}^{2}}\,,
  \label{eq:rfsint}
\end{align}
where
\begin{align}
r_\textrm{dif}^{2} = \int_{0}^{t_{\rm kd}} \! dt \,
                    \frac{\langle v \rangle^{2}}{a^{2}}
                    \frac{1}{n_{\nu} \langle \sigma_{X} v \rangle}
\end{align}
encodes the typical comoving distance a neutrino can wander off following a diffusive Brownian motion, until it can free-streaming at $t_{\rm kd}$. Thereafter, it ballistically travels a comoving distance
\begin{align}
  r_\textrm{bal} = \int_{t_{\rm kd}}^{t} \! dt \, \frac{\langle v\rangle}{a} \,.
\end{align}
So all the smearing that occurs must be limited to scales $k \gg k_\text{nr}$, where the largest free-streaming scale $k_\text{nr}$ should now be derived from $r_\textrm{fs}$ in \cref{eq:rfsint}, instead of the $r_\textrm{fs}$ in \cref{eq:r-fs}, where we had considered the neutrinos to be always free-streaming. It is also possible that neutrino interactions are initially absent and become strong at later times. This is possible, e.g., in a pseudoscalar-mediated interaction. We will discuss this later in \cref{sec:non-standard-cosmo}.

\begin{figure*}
\centering
\includegraphics[width=0.55\textwidth]{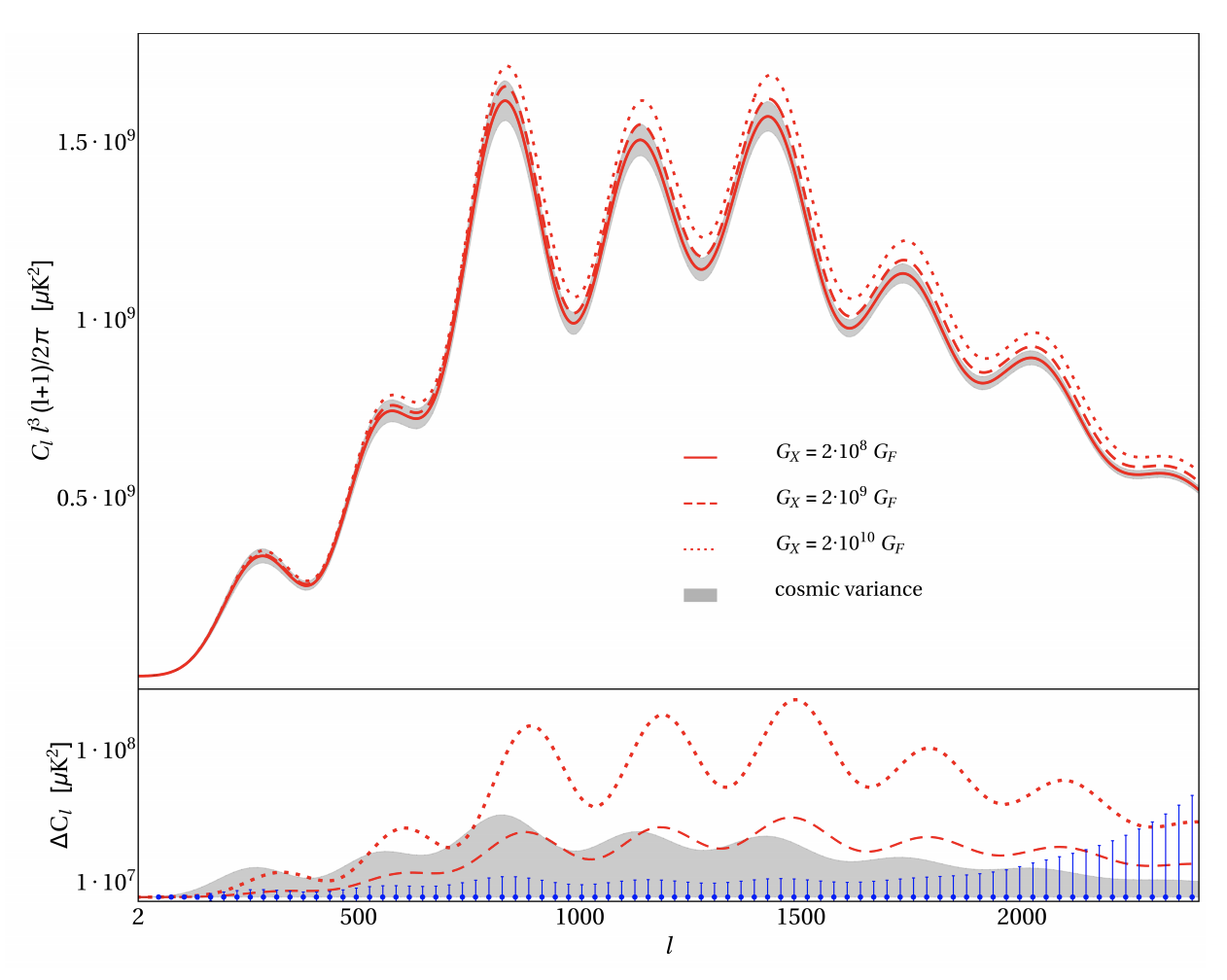}\quad
\includegraphics[width=0.37\textwidth]{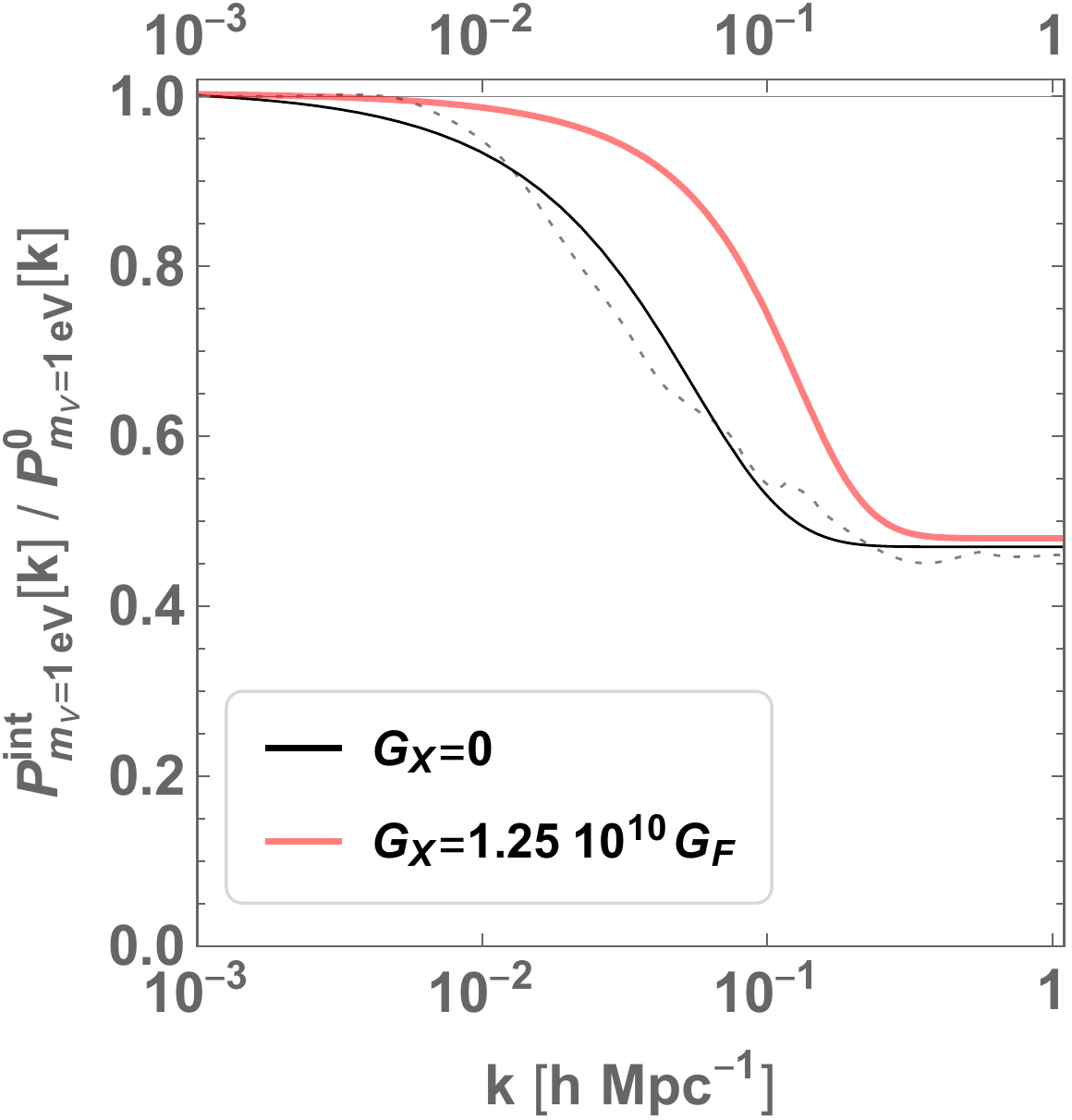}
\caption{Left: Increase in $C_{\ell}$ with larger $G_{X}$; Figure taken from~\cite{Forastieri:2017oma}. Right: Increase of $k_\text{fs}$ for sterile neutrinos with self-interactions. For this illustration we fit simulated data (gray dashed) with a tanh-transition at $k_\text{fs}=0.018\,h/\text{Mpc}$ for 1\,eV sterile neutrinos with no self-interactions (black), and at $0.085\,h/\text{Mpc}$ for $G_{X}=1.25\times10^{10}G_{F}$ (red)~\cite{Chu:2015ipa}, to show the approximate effect on the cut-off in $P_{m}(k)$.}
\label{fig:cdkLSS}
\end{figure*}

Reduced free-streaming of neutrinos can be relevant in two ways. Firstly, the active neutrinos do not free-stream if they develop a collisional nature through mixing with a strongly collisional sterile neutrino. However, said free-streaming of active neutrinos is required by the CMB data, where it has the effect of reducing the heights of the acoustic peaks. If neutrino interactions inhibit neutrinos free-streaming, then the CMB's acoustic peaks are enhanced at $k \gg k_\text{nr}$.  The fractional increase of power due to reduced free-streaming is roughly similar to \cref{CelNeff}, but the $\ell$-range where the increase of power is seen is given by the decreased free-streaming length in \cref{eq:rfsint}. The second aspect is that the energy density contained in a collisional sterile neutrino species smears structure and reduces power on scales $k \gg k_\textrm{nr}$. This is similar to the effect of non-interacting sterile neutrinos, except now $k_\textrm{nr}$ is increased due to diffusion. As the interaction strength becomes large, the smearing of the matter power spectrum occurs only at very small length-scales where data is scarce and where nonlinear effects may be important, making reliable predictions more difficult. These features are shown in \cref{fig:cdkLSS}.

\subsubsection{Other Neutrino Properties}

Cosmology can be sensitive to various other neutrino properties, even those that distinguish between different neutrino flavors or masses, etc. It is difficult to give an encyclopedic discussion of all possibilities.  As an example, BBN is sensitive to neutrinos in more detail than can be described by $\Neff$ alone. For example, presence of a significant chemical potential, or changes in the momentum distribution of neutrinos (e.g., via resonant oscillations) without a change to $\Neff$ can affect BBN by changing momentum-dependent reaction rates. Similarly, adding only  $\nu_{e}$, and not $\bar\nu_{e}$ can have an even more pronounced impact on BBN by changing the reaction kinetics that control nuclear abundances. As such changes cannot be parameterized in terms of just the number of neutrino species, they require a more complete Boltzmann treatment coupled to a BBN code. Neutrino coupling to other species, e.g., dark matter, can also be constrained.

\section{Cosmological Viability of eV Sterile Neutrinos}
\label{sec:sterile-cosmo}

\subsection{Standard Sterile Neutrinos}

Sterile neutrinos with only mass-mixing with active neutrinos, but no other non-gravitational interactions, are referred as standard sterile neutrinos, as opposed to more exotic variants that we will discuss later. Bounds on standard sterile neutrinos come essentially from two directions.  Additional neutrinos that had reached approximately thermal abundances at any point in cosmological history contribute to the energy density of the Universe and can be constrained through the $\Neff$ parameter. Robust constraints on $\Neff$ are obtained from BBN, CMB, as well as BAO. In addition, and in mild contrast to the lighter active neutrinos, eV-mass sterile neutrinos become semi-relativistic already around the epoch of matter--radiation equality where structure formation begins. They act essentially as warm dark matter and smear structures on scales smaller than their maximal free-streaming length. The degree of smearing is proportional to the energy density they carry. Thus, knowing their abundance, CMB and BAO put a constraint on their mass.

\subsubsection{BBN Constraints}

The latest estimates of the primordial helium abundance come from the data compilations of Aver et al.~\cite{Aver:2015iza} and Peimbert et al.~\cite{Peimbert:2016bdg}, which are consistent with each other. Izotov et al.~\cite{Izotov:2014fga} find a higher value in moderate tension with the previous two. Aver et al.\ discuss the differences between their results and Izotov et al.\ and attribute them to modeling differences. The primordial deuterium abundance was reported in Cooke et al.~\cite{Cooke:2017cwo}. Note that for the purpose of comparing with BBN data, eV-scale neutrinos can be taken as effectively massless, and a $\Lambda$CDM+$\Neff$ model is sufficient to describe their impact on BBN. The BBN constraint on $\Neff$ is $\Neff = 3.45\pm0.51$ at $95\%$\,CL~\cite{Cooke:2017cwo}, with similar results obtained by other groups.

\begin{figure}
  \centering
  \includegraphics[width=0.48\textwidth]{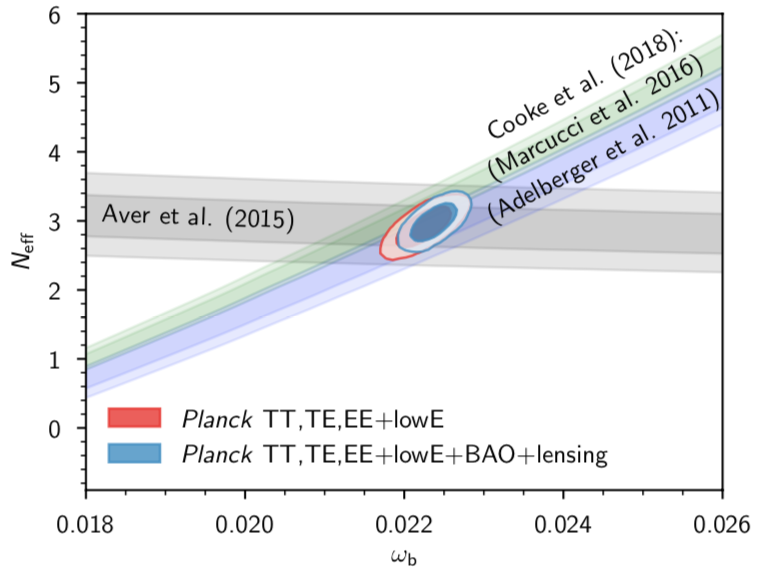}
  \caption{BBN constraints on standard eV-scale sterile neutrinos. Shaded regions are allowed at the 68\% and 95\% confidence levels, respectively. Figure taken from Planck-18~\cite{Aghanim:2018eyx}.}
  \label{fig:NeffBBN}
\end{figure}

Planck-18~\cite{Aghanim:2018eyx} take the helium abundance from Aver et al.~\cite{Aver:2015iza} and the deuterium abundance from Cooke et al.~\cite{Cooke:2017cwo}, employing a variety of assumptions (related to using different BBN codes and determinations of the $d\,(p,\gamma)^{\,3}\textrm{He}$ reaction rate) to relate the abundances to cosmological parameters and obtain bounds on a $\Lambda$CDM+$\Neff$ model. They take $\Neff$ as a free parameter, but the electron fraction $Y_{e}$ is calculated knowing $\omega_{b}$ and $\Neff$. Combining with CMB and BAO data, they find $\Neff = 3.0\pm0.2$ at $95\%$\,CL, the last significant digit depending mildly on the different assumptions. This result strongly disfavors one fully thermalized sterile neutrino. The bounds are shown in \cref{fig:NeffBBN}.

\subsubsection{CMB and Structure Constraints}
\label{CMBstructconst}

As discussed before, cosmology is more sensitive to the aggregate of all the neutrino species, encoded in a few numbers like $\Neff$ and $\sum m_{\nu}$, than it is to the individual properties of each species. In practice, however, the data has now improved to an extent that some choices for the ``substructure'' of the neutrino sector, such as having eV-scale sterile neutrinos be lighter than the active neutrinos, are strongly ruled out. Other well-motivated particle physics models for the sterile sector are indistinguishable as their predictions for $\Neff$ and $\sum m_{\nu}$ do not differ much.

\begin{figure}
\centering
\includegraphics[width=0.48\textwidth]{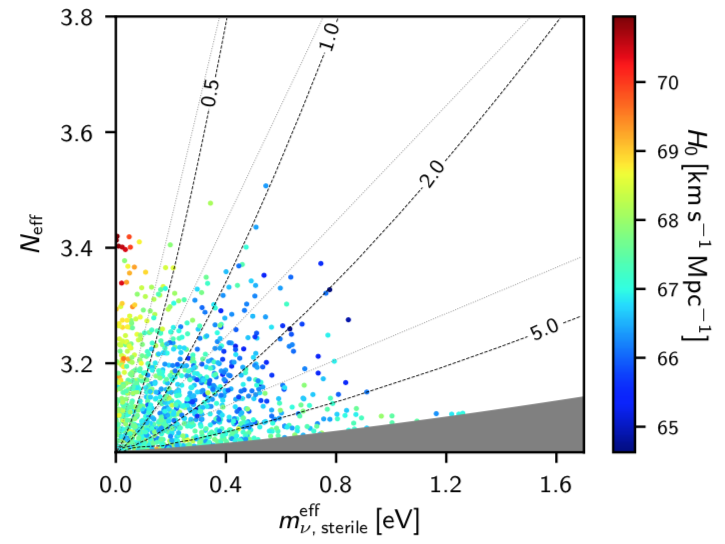}
\caption{CMB constraints on standard (non-interacting) eV-scale sterile neutrinos. Colored dots show samples from Planck's Markov Chain Monte Carlo determination of the likelihood for a model with one additional sterile neutrino. The mass and abundance of this sterile neutrino is encoded in $m_{\nu,\,\textrm{sterile}}^\textrm{eff}$, see text for details. The physical mass is constant along the diagonal dotted lines (thick for thermally produced sterile neutrinos; thin for production via the Dodelson--Widrow mechanism). The gray region is excluded by the prior, and the color code indicates the value of the Hubble constant for each point. Figure taken from Planck-18~\cite{Aghanim:2018eyx}.}
\label{fig:cmbbao}
\end{figure}

\begin{figure}
\centering
\includegraphics[width=0.95\textwidth]{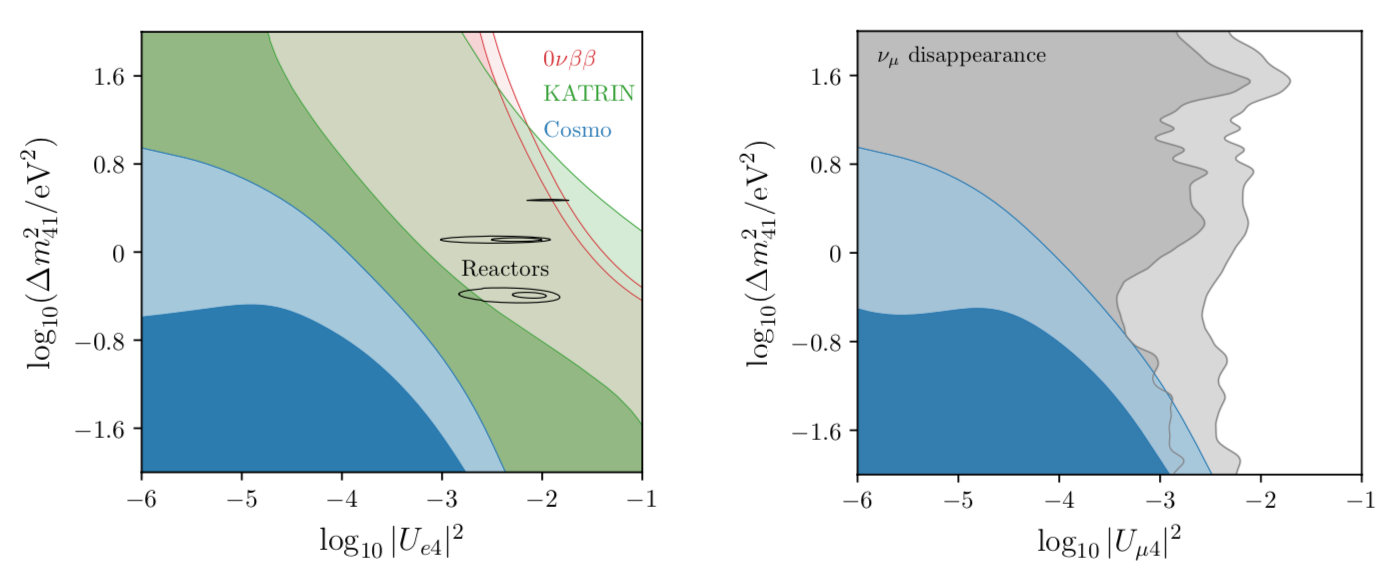}
\caption{Incompatibility of standard sterile neutrinos, relevant for the short baseline anomalies, with other data from laboratory experiments and from cosmology. Red regions show constraints from neutrinoless double beta decay (valid only if neutrinos are Majorana particles), green regions correspond to direct kinematic constraints from the beta-endpoint spectrum measured in KATRIN, and blue regions are derived from a fit to CMB+BAO data. Darker colors correspond to 68\%~CL constraints, while lighter colors are for a CL of 95\%.  The black contours in the left panel show a fit to reactor data, while the gray regions in the right panel are based on IceCube and MINOS+ data on $\nu_\mu$ disappearance.  In the left (right) panel it has been assumed that sterile neutrinos mix predominantly with $\nu_{e}$ ($\nu_{\mu}$). Figure taken from ~\cite{Hagstotz:2020ukm}.}
\label{fig:snuconflict}
\end{figure}

The base model assumed by the Planck collaboration in their latest study \cite{Aghanim:2018eyx}, takes $\sum m_{\nu} = 0.06$\,eV as sum of neutrino mass. This is the minimum possible value, which is realized when choosing the lightest neutrino mass eigenstate state as exactly massless, taking into account the mass-squared differences measured in oscillation experiments, and setting $\Neff=3.046$.  This choice leads to the most conservative limits on any deviations from the base model, in particular on sterile neutrinos whose presence introduces changes to both $\Neff$ and $\sum m_\nu$.  Allowing both $\Neff$ and $\sum m_{\nu}$ to vary, the data sets upper limits on $\sum m_{\nu} < 0.65~(\textrm{resp.}~0.23)$\,eV and $\Neff<3.29~(\textrm{resp.}~3.34)$ at 95\%~CL with the conservative (resp.\ aggressive) prior $m_{4} < 10\,\textrm{eV}~(\textrm{resp.}~2\,\textrm{eV})$ on the fourth mass eigenvalue. This constraint is shown in \cref{fig:cmbbao}, where the density of the different colored dots indicates the likelihood of different parameter regions. Note that the quantity on the abscissa is not $\sum m_{\nu}$, but rather $m_{\nu,\,\textrm{sterile}}^\textrm{eff}$ which maps onto the mass of the sterile state, e.g., as $m_{4}=(\Delta\Neff)^{-3/4}m_{\nu,\,\textrm{sterile}}^\textrm{eff}$ for a thermal sterile neutrino and as $m_{4}=(\Delta\Neff)^{-1}m_{\nu,\,\textrm{sterile}}^\textrm{eff}$ for production via the Dodelson--Widrow mechanism. The darker (resp.\ lighter) diagonal lines correspond to loci of constant $m_{4}$ for sterile neutrinos produced thermally (resp.\ via the Dodelson--Widrow mechanism). The gray shaded region is excluded from the analysis by the prior. See Sec.~7.5.3 of Planck-18~\cite{Aghanim:2018eyx} for details. Joint constraints using terrestrial data and cosmology show that eV-scale sterile neutrinos in the parameter range motivated by the short baseline experiments discussed in \cref{sec:sbl-anomalies}, are strongly disfavored~\cite{Berryman:2019nvr,Adams:2020nue,Hagstotz:2020ukm}. This tension is explicitly highlighted in \cref{fig:snuconflict}.

\subsection{Non-Standard Sterile Neutrinos}
\label{sec:non-standard-cosmo}

Several ideas have been proposed to modify sterile neutrinos in some way that makes them viable from a cosmological standpoint. The key idea is that cosmology is only sensitive to particles that have been produced in sufficient (\ie, close to thermal) abundance. If, for some reason, a particle is not produced abundantly before the cosmologically observable epochs, such a particle is very difficult to probe using cosmology. Accordingly, all of the proposed models that aim to reconcile oscillation-motivated eV-scale sterile neutrinos with cosmology are based on a reduction of the effective sterile neutrino abundance.

One idea along these lines is a non-negligible chemical potential for the active neutrinos, which would make sterile neutrinos less abundant in the Universe~\cite{Foot:1995bm}. Another possibility is that eV-scale sterile neutrinos with large mixing angles may not be cosmologically abundant because the Universe did not reheat sufficiently. This would require a reheating temperature after inflation of $\ll100$\,MeV~\cite{Gelmini:2004ah}. Yet another idea is reheating of the photon bath after neutrinos have decoupled.  This can happen if there are additional exotic particles, which decay sufficiently late. The result would be an effective reduction of the sterile neutrino abundance~\cite{Fuller:2011qy, Ho:2012br}.  Given that these mechanisms still predict $\Neff > 3$ at BBN, they are likely ruled out by now.

\subsubsection{Interacting Sterile Neutrinos}

This brings us, in a full circle, back to the Hubble tension. The discrepancy between the local and cosmological measurements of $H_{0}$ can be reduced if a sterile-neutrino-like hot dark matter component were to be added to the minimal $\Lambda$CDM model.  This extended model is variously dubbed $\nu\Lambda$CDM (for addition of neutrinos) or $\Lambda$MDM (where M stands for ``mixture'' of cold and hot dark matter)~\cite{Wyman:2013lza,Hamann:2013iba}. The typical increase in $\Neff$ required to ameliorate the $H_0$ tension is $\Delta\Neff=0.61\pm0.30$, with an effective hot dark matter mass $m_{\nu}^\textrm{eff} = 0.41\pm0.13$\,eV. Although naively this motivates sterile neutrinos, the parameter region motivated by the short baseline anomalies is in tension with the above constraint~\cite{Wyman:2013lza, Hamann:2013iba}. Two independent studies~\cite{Hannestad:2013ana, Dasgupta:2013zpn} showed that an exciting possibility is to endow the sterile neutrinos with new, so-called, hidden or secret interactions. Such interactions could be mediated, e.g., by a new massive gauge boson or a pseudoscalar.

The first variant involves charging the sterile flavor eigenstate $\nu_s$
under a new $U(1)_s$ gauge group, with a gauge boson $X$  at the 100\,MeV-scale
or somewhat below, but not so light that long-range forces become important.  
The relevant interaction is
\begin{align}
  \mathcal{L}_\text{int} = g_{X} \bar\nu_s \gamma^\mu P_L \nu_s X_\mu \,,
  \label{eq:Lintsec}
\end{align}
where $g_X$ is the $U(1)_X$ coupling constant, and \mbox{$P_L = \tfrac{1}{2} (1
- \gamma^5)$} is the projection operator onto left-chiral fermion states.  Most
studies on this topic are agnostic about the mechanism that breaks $U(1)_s$ and
endows the $X$ boson with a mass. In particular, possible additional degrees of
freedom introduced to break $U(1)_s$, such as a sterile sector Higgs boson $H_s$,
are often neglected.  A sterile Higgs boson is typically needed, however,
to generate $\nu_a$--$\nu_s$ mixing, for instance via an effective operator of
the form $(LH)(\nu_{s}H_{s})$. ($H_s$ is not strictly required to generate
a mass $M_X$ for $X$, as the latter can also arise from the St\"uckelberg mechanism.)
At energy scales smaller than the mass of $X$, the gauge boson can be integrated
out and sterile neutrino self-interactions can be described by four-$\nu_s$
contact operator of dimension 6 with an effective coupling $G_{X} = g_{X}^{2} /
(2\sqrt{2} M_{X}^{2})$. If $M_{X} \gtrsim 10$\,MeV, this effective field theory
description is valid at all relevant cosmological epochs from BBN onwards.
(Note that, in some papers, the notation $e_{s}$ is used instead of $g_{X}$ for the
sterile sector gauge coupling, and the mediator is denoted by $A'$ instead of
$X$, corresponding to a hidden photon with mass denoted as $M$ instead of $M_{X}$.
Some of the figures in this section also use this alternate notation.)

\begin{figure}
  \centering
  \includegraphics[width=0.5\textwidth]{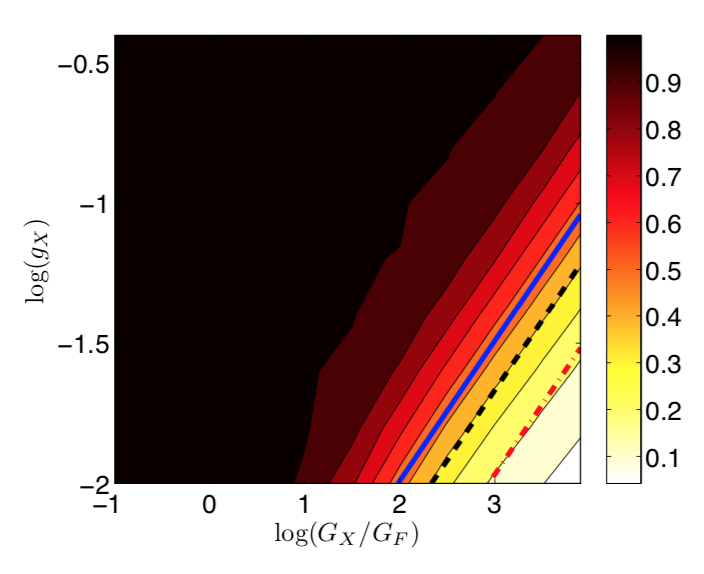}
  \caption{The cosmological abundance of self-interacting sterile neutrinos,
    parameterized in terms of $\Delta N_\text{eff}$ (color code).  The interaction
    is parameterized by the $U(1)_s$ gauge coupling $g_X$ (vertical axis) 
    and the effective four-fermion coupling $G_X$ (horizontal axis).  We see that,
    for parameters in the bottom right-hand corner of the plot, full
    thermalization is avoided.  The solid blue, dashed black, and dot-dashed
    red lines correspond to mediator masses of $M_X = 300$, 200, and \SI{100}{MeV},
    respectively.  Figure taken from ref.~\cite{Hannestad:2013ana}.}
  \label{fig:dk14}
\end{figure}

When a sterile
neutrino with energy $E$ propagates through a thermalized background of sterile
neutrinos and $X$ bosons at temperature $T_s$, it experiences a
matter potential~\cite{Dasgupta:2013zpn}
\begin{align}
  V_\text{eff,s} \simeq
  \left\lbrace 
    \begin{array}{lcl}
      -\dfrac{7 \pi^2 g_X^2 E T_s^4}{45 M_{X}^4}  &\quad& \text{for $T_s \ll M_{X}$} \\[3ex]
      +\dfrac{g_X^2 T_s^2}{8 E}               &\quad& \text{for $T_s \gg M_{X}$}
    \end{array}
  \right.\,.
  \label{eq:Veff}
\end{align}
Like the conventional Mikheyev--Smirnov--Wolfenstein (MSW)
potential~\cite{Wolfenstein:1977ue, Mikheev:1986gs},
$V_\text{eff,s}$ changes the neutrino mixing angle. At the high temperatures
prevalent in the early Universe, the effective mixing angle $\theta_m$ can be
much smaller than the vacuum mixing angle $\theta_0$ because $|V_\text{eff,s}|
\gg \Delta m^2 / (2E_{\nu})$.  Experiments today, on the other hand, will observe
$\theta_m = \theta_0$ to a very good approximation. This model is therefore
capable of significantly suppressing non-collisional production of sterile neutrinos,
as shown more quantitatively in \cref{fig:dk14}.

What about collisional production? Initial estimates for the effect of collisional production~\cite{Hannestad:2013ana} suggested that the contribution to $\Neff$ can be made negligible for $M_{X}\sim100$\,MeV and $G_{X}\sim10^{3}G_{F}$. Thus, it would appear that the model successfully reconciled eV-scale sterile neutrinos with large mixing with the stringent cosmological bounds. However this dream turns out to be short-lived. Although $\Neff$ is not affected, active--sterile oscillations cannot be suppressed for too long after neutrino decoupling, and they occur before the end of BBN. As a result, although $\Neff$ remains close to 3, the $\nu_{e}$ abundance and spectrum are altered and this has observable consequences: the $^{2}\textrm{H}/\textrm{H}$ abundance ratio increases beyond what is allowed by data~\cite{Saviano:2014esa}.

\begin{figure}
  \centering
  \includegraphics[width=0.5\textwidth]{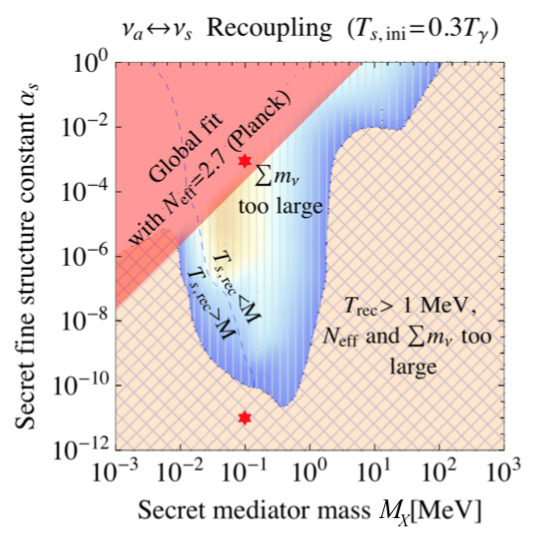}
  \caption{Constraints on the parameter space of self-interacting sterile
  neutrinos.  We see that in the simplest secret interaction model there is no
  parameter space allowed where the model remains safe from both $N_\text{eff}$
  and $\sum m_{\nu}$ constraints as well as the bound from required
  free-streaming of active neutrinos.  At small secret gauge
  coupling $g_X$, the model runs afoul of both $N_{\text{eff}}$ and $\sum
  m_{\nu}$ constraints, while at large coupling, active neutrinos
  are prevented from free-streaming. In the intermediate range the
  thermalization of secretly interacting sterile neutrinos is less efficient,
  but their final abundance is still too large to avoid the bounds on $\sum
  m_{\nu}$ in particular. Figure taken from \cite{Chu:2018gxk}.}
  \label{fig:cdk18}
\end{figure}

The situation with lighter mediators, $M_{X} \lesssim \si{MeV}$, is more interesting as it can postpone the onset of collisional $\nu_s$ production to an epoch well beyond BBN. Moreover, it has been shown that, if the new gauge boson $X$ is coupled not only to sterile neutrinos, but also to dark matter, it can mitigate some of the problems associated with structure formation in models of purely cold, non-interacting dark matter (see \cref{sec:keVchapt}) and make testable predictions~\cite{Dasgupta:2013zpn, Bringmann:2013vra}.  A coupling of $X$ to dark matter could also be motivated by the possible cancellation of the $[U(1)_{X}]^3$ anomaly that would be present if $X$ only coupled to $\nu_{s}$.  However, this scenario, too, has issues. The $\nu_{s}$ production rate $\Gamma_s$ is proportional to $n_{\nu_s}$, and thus rapidly approaches its final value
\begin{equation}
  \Gamma_s \simeq {1\over2}\sin^2 2\theta_m \times
    \frac{3}{4} n^\text{SM}_{\nu_a} \times
    \left\lbrace 
      \begin{array}{lcl}
        g_{X}^4\frac{E^2}{M_{X}^4}
                             &\quad& \text{for $T_s \ll M_{X}$} \\[1.5ex]
        g_{X}^4\frac{1}{M_{X}^2}
                             &\quad& \text{for $T_s \gg M_{X}$}
      \end{array}
    \right. \,,
  \label{eq:Gamma-s}
\end{equation}
where $T_s$ is the temperature of the sterile neutrinos, which may be different from the temperature of the active neutrinos.  Of crucial importance here is the proportionality of $\Gamma_s$ to $1/M_X^2$ in the regime $T_s \gg M_{X}$. Originally, it was mis-estimated that $\Gamma_s$ would be proportional to $1/E^{2}$ in this regime, but the authors of~\cite{Cherry:2016jol} have pointed out that the cross section in fact is larger, namely $\propto 1/M_{X}^{2}$, and that because of this thermalization of $\nu_s$ with the $\nu_a$ is inevitable. There is another issue with the light mediator scenario. As shown by the authors of~\cite{Mirizzi:2014ama}, collisional production of $\nu_s$ at the expense of the $\nu_a$ after neutrino decoupling at $T_{\nu D}$, while not changing $\Neff$ appreciably, still leads to a substantial $\nu_{s}$ population at late times. Thus the bound on $\mnusum$ applies and could in principle rule out the model. It was then counter-argued in~\cite{Chu:2015ipa} that sterile neutrinos with sufficiently strong interactions would not free-stream, but rather diffuse, and thus the bound from constraints on $\mnusum$ would have to be appropriately recalculated and weakened. The downside is that such strongly interacting sterile neutrinos would lead to lowered free-streaming of at least one species of active neutrinos (through $\nu_a$--$\nu_s$ mixing), in slight tension with CMB data.  Moreover, as shown in \cref{fig:cdk18}, one cannot avoid overproducing sterile neutrinos at late times.

\begin{figure}
  \centering
  \includegraphics[width=0.5\textwidth]{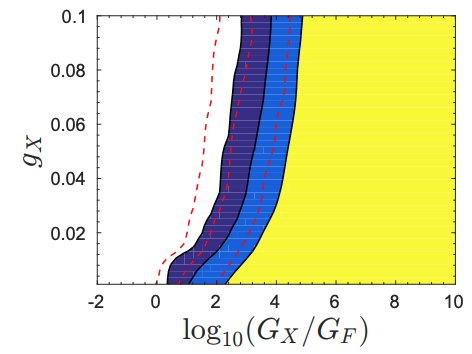}
  \caption{BBN constraint on secret interactions of sterile neutrinos at the
    $2\sigma$ (dark blue), $3\sigma$ (light blue), and $4\sigma$ (yellow)
    confidence levels. Shaded regions are ruled out. The dashed lines are the
    same constraints obtained by assuming that the entropy of the decoupled
    active and sterile neutrino population is conserved. This may not be a good
    approximation when there are no number changing interactions that establish
    chemical equilibrium. Figure taken from~\cite{Song:2018zyl}.}
\label{fig:bbnnust}
\end{figure}

\begin{figure}
  \centering
  \includegraphics[width=0.8\textwidth]{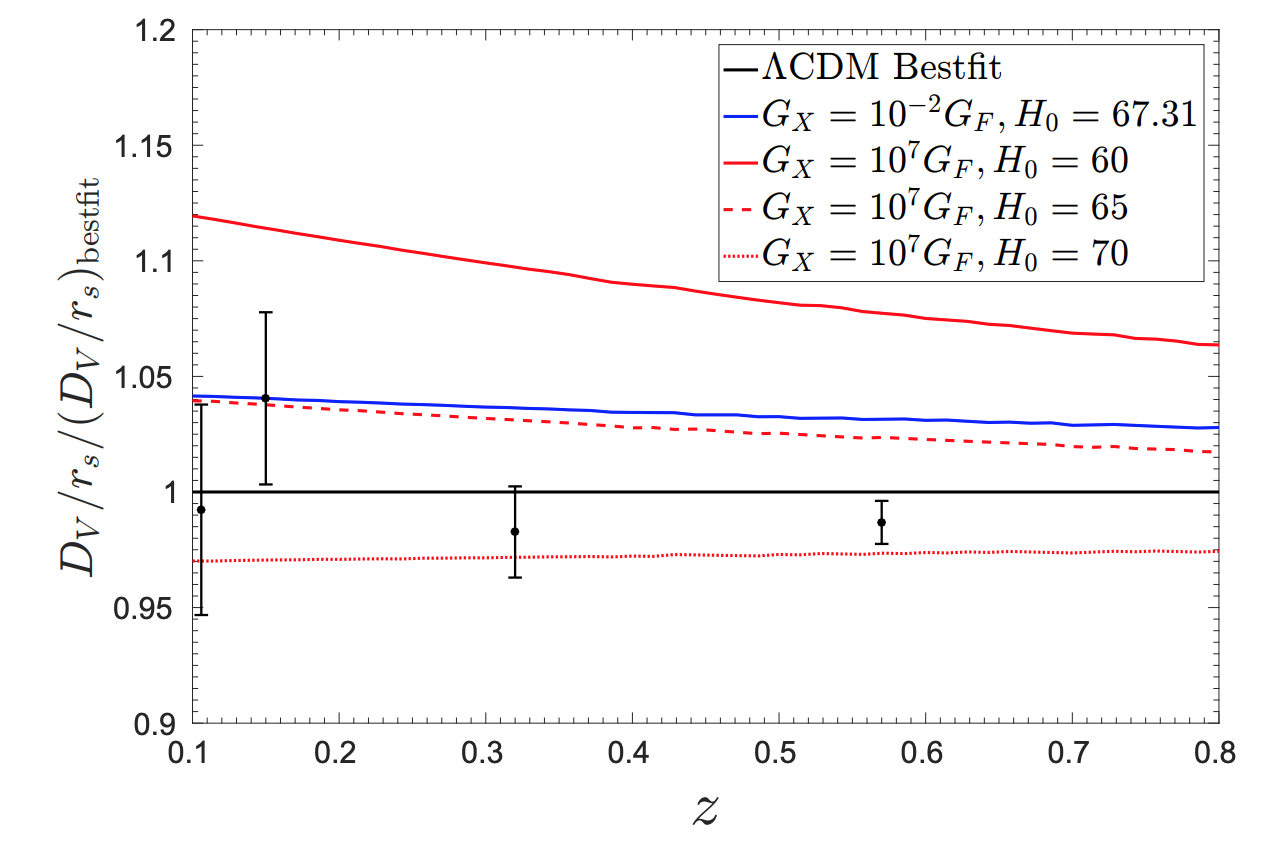}\\
  \includegraphics[width=0.85\textwidth]{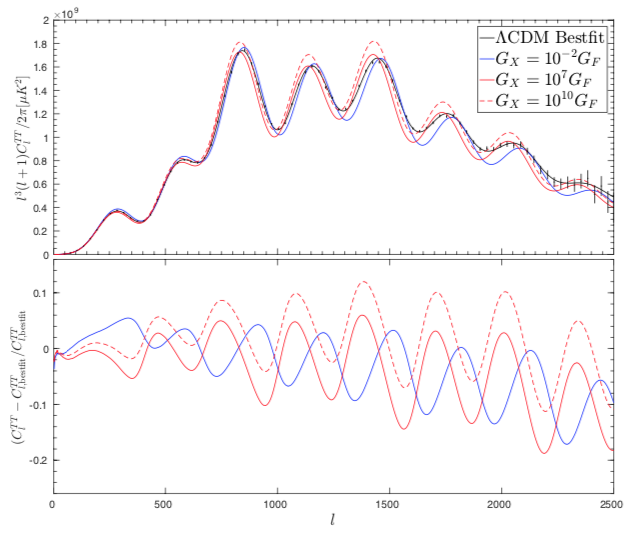}
  \caption{Comparison of BAO (top panel) and CMB (bottom panel) data  to the corresponding theoretical predictions by including secret interaction of sterile neutrinos. Figures taken
    from~\cite{Song:2018zyl}.}
  \label{fig:cmbbaonust}
\end{figure}

While the arguments presented so far are rather qualitative, sterile neutrino scenarios with secret interactions have since also been studied in a more quantitative way by computing BBN yields and by implementing the model within cosmological perturbation theory to obtain reliable predictions for the anisotropy power spectrum of the CMB. BBN is sensitive to both the coupling $g_{X}$ and the mass $M_X$ of the secret mediator, as they enter \cref{eq:Veff,eq:Gamma-s} separately and not in a fixed combination like $G_{X}$. In \cref{fig:bbnnust} we show the BBN constraints on the two-dimensional parameter space (parameterized there in terms of $g_X$ and $G_X$) from ref.~\cite{Song:2018zyl}. CMB and structure formation are mainly sensitive to the combination $G_{X}$ if $\Neff$ is kept unchanged. \Cref{fig:cmbbaonust} shows the CMB and large scale structure constraints on $G_{X}$. These results clearly rule out the vanilla version of the secret interactions scenario~\cite{Forastieri:2017oma, Song:2018zyl} for the sterile neutrino mass ($\sim \si{eV}$) and mixing angles ($\sim 0.1$) suggested by the short baseline anomalies, for both heavy and light mediators~\cite{Chu:2018gxk}.

Before we end the discussion of secretly interacting sterile neutrinos in cosmology, we clarify a conceptual issue. When sterile neutrinos are produced from active neutrinos purely due to collisional decoherence in their flavor oscillations, number density and energy density are conserved, but not entropy. This somewhat unfamiliar result has caused some confusion in the literature. Strong sterile neutrino scattering and efficient oscillations establish a common kinetic equilibrium, and the final momentum distribution is a Fermi--Dirac spectrum with a common chemical potential $\mu_s$ and temperature $T_s$. As no number-changing interactions are occurring, however, there cannot be any change in the total number of neutrinos. The energies of all species are redshifted as $1/a^{4}$ as long as they are relativistic. Conservation of comoving number density and energy then allow one to compute $\mu_s$ and $T_s$~\cite{Chu:2015ipa}. Entropy, though, increases due to the irreversible process corresponding to collisional decoherent production of sterile neutrinos. This is akin to the irreversible mixing of two gases, wherein number and energy is constant, but entropy increases.

\vspace{1em}
Besides the secret vector interactions discussed so far, the other avatar of secretly interacting sterile neutrinos is a scenario in which the interaction is instead mediated by a pseudoscalar $X$~\cite{Babu:1991at, Bento:2001xi, Archidiacono:2014nda}:
\begin{align}
  {\cal L}_\textrm{int}=g_{X}\overline{\nu}_{s}\gamma_{5}\nu_{s}X\,.
  \label{selfintpseudo}
\end{align}
This kind of interaction, first proposed in the context of the somewhat heavier and ill-fated \SI{17}{keV} neutrino interacting with a Majoron~\cite{Babu:1991at}, was revisited recently in the context of eV-scale sterile neutrinos~\cite{Archidiacono:2014nda}. A key feature of this model is that the pseudoscalar-mediated neutrino self-interactions are negligible at early times, but become very strong at late times.  Two other features also set this model apart: first, the pseudoscalar can be very light, thus sterile neutrinos can annihilate or decay into it; second, the self-interaction in \cref{selfintpseudo} can be diagonal in the mass basis, unlike what is possible for vector-mediated models. Such an alignment of the couplings ensures that the mostly sterile mass eigenstate $\nu_4$ does not free-stream at late times. Thus, the $\mnusum$ constraint is avoided, while free-streaming of active neutrinos is not affected. The constraint on $\Neff$ is avoided in the same way as for the vector-mediated scenario due to the increased thermal mass of the sterile neutrino that arises from the secret interaction. More recent studies of this model show that, for a sterile neutrino mass of \SI{1}{eV}, a small part of parameter space is allowed by both short baseline data and cosmology~\cite{Archidiacono:2020yey}.

\subsubsection{Sterile Neutrinos Interacting with Ultralight Dark Matter}

\begin{figure}
  \centering
  \includegraphics[width=0.55\textwidth]{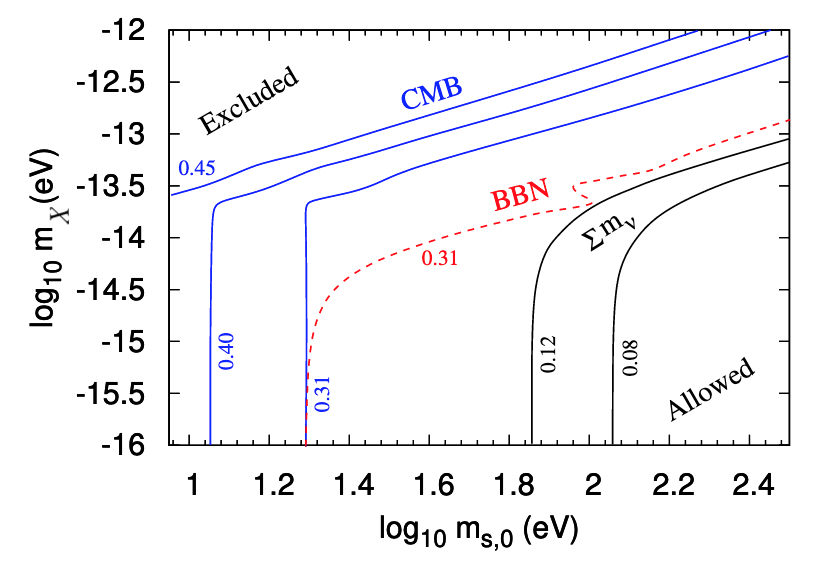}
  \caption{Constraints on the parameter space of sterile neutrinos with a coupling
    of the form $g_{X} X \overline\nu_{s} \nu_{s}$ to a dark, ultralight scalar field $X$
    with $g_{X} \sim 10^{-23}$. The horizontal axis shows the sterile neutrino mass
    at the present time, while the vertical axis shows the mass of $X$.  The labels
    on the contours indicate the change in $\Neff$ at the CMB (blue) and BBN (red)
    epochs, as well as the change in $\sum m_\nu$ (black).
    Figure taken from~\cite{Cline:2019seo}.}\label{fig:cline}
\end{figure}

As we have shown, the secret interaction scenarios discussed above have limited potential to reconcile sterile neutrinos with cosmology, especially if the idea was to significantly mitigate the Hubble tension~\cite{Archidiacono:2020yey}. However, a new class of models seems promising and attractive. In these models secret interactions exist between sterile neutrinos and dark matter. This was already attempted in~\cite{Dasgupta:2013zpn}, however as we showed above the model parameter space discussed in that paper is no longer viable. In newer models, however, the dynamics of the dark sector are such that sterile neutrinos are not produced appreciably in the early Universe, rendering cosmology insensitive~\cite{Bezrukov:2017ike,Farzan:2019yvo,Cline:2019seo}. Consider a coupling of the form
\begin{align}
  {\cal L}_\textrm{int} = g_{X}{\nu}_{s}^{T}C\nu_{s}X\,.
\end{align}
between the sterile neutrino and a dark scalar field $X$.  If $X$ has a non-zero vacuum expectation value, this coupling obviously leads to a Majorana-type mass term for $\nu_{s}$. One could also consider couplings of the form
\begin{align}
  {\cal L}_\textrm{int} = g_{X} \overline{\nu}_{s} \nu_{s} X \,,
\end{align}
but only at the expense of adding additional degrees of freedom. The novel idea here is that $X$ is an ultralight boson that could be the dark matter. As is well known, bosonic dark matter can be as light as $\sim 10^{-21}$\,eV as long as it is extremely weakly interacting and never comes into equilibrium. Thus, despite its tiny mass it can be ultra cold because of its non-thermal production akin to axions. As long as it is not too dense, i.e., $n_{X} < 1/\lambda_\textrm{th}^3$, where $\lambda_\textrm{th}$ is the thermal de-Broglie wavelength, such ``fuzzy'' dark matter can be treated as a classical field. The time evolution of the field value is the given by
\begin{align}
  X = \frac{\sqrt{2\rho_{X}}}{M_{X}}\cos(M_{X}t-{\bf p}.{\bf x})\,,
\end{align}
where $p \ll M_{X}$, as required for the particle to be non-relativistic. In the early Universe, the $\nu_{s}$--$X$ coupling gives a time-dependent mass to $\nu_{s}$, which is $m_{s} \approx (\sqrt{2\rho_{X}}) / {M_{X}} \cos(M_{X}t)$. Choosing $M_{X}<5\times10^{-17}$\,eV makes the sterile neutrino mass large enough for active--sterile mixing to be strongly suppressed until $T \approx \SI{0.01}{MeV}$, and the model is safe from both $N_\text{eff}$ and $\sum m_{\nu}$ constraints. Below $T\approx0.01$\,MeV the $\nu_{s}$ acts like a standard sterile neutrino. The parameter space for these models is viable, as seen in \cref{fig:cline}, and perhaps testable in KATRIN~\cite{Farzan:2019yvo}.

Another approach is to make the sterile neutrinos exactly massless at early times. This requires a symmetry that is unbroken at high temperatures and may break at lower temperatures~\cite{Bezrukov:2017ike, Chu:2018gxk}.

\subsubsection{Other Related Developments}

One outcome of all these studies is a multitude of new phenomenological studies they have initiated. As we have argued above, active neutrinos can also feel the secret interactions, suppressed by the mixing angle. Therefore, high energy neutrinos from faraway astrophysical sources can get absorbed by the cosmic neutrino background through resonant scattering at a center-of-mass energy corresponding to the mass of the hidden mediator~\cite{Cherry:2014xra}. Such an absorption feature would show up as a dip in the neutrino flux now measured at neutrino telescopes such as IceCube. The data do not show any statistically significant dips, so the observed flux can be used to set constraints on hidden mediator models. It turns out that the diffuse supernova neutrino background is also sensitive to these secret interactions of the sterile neutrinos~\cite{Jeong:2018yts, Farzan:2018pnk}.

Another field of activity has been in discovering new production mechanisms for sterile neutrinos~\cite{Hansen:2017rxr,deGouvea:2019phk}. Self-scattering of sterile neutrinos allows a modification of the Dodelson--Widrow mechanism and can expand the parameter space where some desired quantity of sterile neutrinos can be produced. We will discuss this possibility in \cref{sec:keVchapt} in the context of sterile neutrino dark matter.

Further, secret interactions of active neutrinos in cosmology~\cite{Cyr-Racine:2013jua, Archidiacono:2013dua}, that were considered contemporaneously with the sterile neutrino proposals, have attracted interest. From the cosmological standpoint, self-interactions of neutrinos lead to lowering of shear stress. This is simply because of the isotropizing action of the scatterings. However, CMB data prefers most of the neutrino energy density to free-stream. Careful implementation of self-interactions in Boltzmann codes has recently been achieved both for secret interactions mediated by heavy and by light mediators~\cite{Cyr-Racine:2013jua, Archidiacono:2013dua, Lancaster:2017ksf, Oldengott:2017fhy, Kreisch:2019yzn, Oldengott:2014qra}. Notwithstanding contrary claims~\cite{Kreisch:2019yzn}, it was already known that self-scattering through heavy mediators does not offer a better fit to the cosmological data, especially when CMB polarization measurements are included. It also does not reduce the Hubble tension~\cite{Aghanim:2018eyx}. The light mediator scenario does better until BAO data is included~\cite{Archidiacono:2020yey}. Many developments in active neutrinos with secret interactions mirror the issues discussed here for sterile neutrinos~\cite{Ioka:2014kca, Ng:2014pca, Farzan:2018gtr, Denton:2018xmq, Bakhti:2018avv, Blinov:2019gcj, Berlin:2017ftj,Ghosh:2019tab}.

\chapter{keV Sterile Neutrinos as Dark Matter}
\label{sec:keVchapt}

So far, in this review, we have mostly focused on light (eV-scale) sterile neutrinos. We now extend this discussion to sterile neutrinos with larger masses. We have already elaborated in \cref{sec:theory-motivation} on the role which heavy sterile neutrinos may play in generating the active neutrino masses.  Let us in this chapter focus on sterile neutrinos with keV-scale masses, which are a promising dark matter candidate, before moving on in \cref{sec:leptogenesis} to the tantalizing possibility that sterile neutrinos play a role in explaining the excess of matter over anti-matter in the Universe.  Both possibilities show that sterile neutrinos are theoretically well-motivated from an Occam-like perspective: one gets to explain the nature of dark matter and the observed matter--antimatter asymmetry at little extra cost, having already invoked sterile neutrinos to explain the lightness of active neutrino masses, albeit at possibly different mass-scales. Both sterile neutrino dark matter and the generation of the particle--anti-particle asymmetry are vast fields of research in themselves, therefore we here limit ourselves to outlining the key underlying ideas and to briefly summarizing recent developments.

There is overwhelming evidence that the Universe contains a nonbaryonic form of matter, called dark matter, but the nature of this substance is not known. Active neutrinos, given their small masses, are still relativistic at the onset of structure formation and can therefore not be a major contributor to the dark matter as they would render structure formation far less efficient than what is observed.  While they still constitute a small, hot, component of the dark matter, it is believed that one or several new species of cold particles form the bulk of the dark matter. There are hundreds of proposed dark matter candidates, if not more, but Weakly Interacting Massive Particles (WIMPs), axions, and sterile neutrinos are perhaps the three best-motivated ones.

Sterile neutrinos with keV-scale masses and $\nu_s$--$\nu_a$ mixing angles $\sin^{2} 2\theta \sim 10^{-10}$, act like dark matter. They can be produced with the correct abundance to make up a large fraction of dark matter, and are not yet ruled out by any data.  However, there are several subtle aspects to sterile neutrino dark matter. In the following, we first outline how sterile neutrino dark matter can be produced with the correct abundance, what the key features are that distinguish it from other candidates, and how one can detect it. We then summarize some recent developments. The interested reader can dig deeper into the subject by following our references, especially those to dedicated review articles by Kusenko~\cite{Kusenko:2009up} and by Abazajian~\cite{Abazajian:2017tcc}, which have previously appeared in this series. Most of the material we discuss here is covered in greater detail in a community white paper dedicated to keV sterile neutrinos, ref.~\cite{Adhikari:2016bei}.

\section{Production of keV Sterile Neutrino Dark Matter}
\label{sec:keV-production}

The Dodelson--Widrow mechanism introduced in \cref{sec:nu-cosmo} can produce the correct abundance of keV-mass sterile neutrinos as dark matter, requiring only mixing of active and sterile neutrinos, and no additional ingredients~\cite{Dodelson:1993je}. Flavor mixing between active and heavier sterile neutrinos is a generic and almost unavoidable prediction of neutrino mass models, so this production mechanism is extremely well-motivated. Only two parameters, viz., the sterile neutrino mass, $m_{s}$,\footnote{Or, more precisely, the mass of the mostly sterile mass eigenstate.}  and the dominant mixing angle with the active neutrinos, $\theta_s$, control the flavor oscillations that lead to dark matter production.  The next-simplest mechanism, the Shi--Fuller mechanism~\cite{Shi:1998km}, includes the possibility of large lepton asymmetries which can lead to resonantly enhanced production and is thus viable at smaller mixing angles.  We have already discussed the basic idea of both of these mechanisms in \cref{sec:nucosmo:prod}, so we do not repeat them here. Only the aspects specific to a keV-scale sterile neutrino are noted below.

The main point is that at high temperatures $\Gamma_\text{prod}/H \propto T^{-9}$, due to the finite temperature propagators and collisional damping (ignoring lepton asymmetries), whereas at low temperatures $\Gamma_\text{prod}/H \propto T^{3}$. The production rate peaks at~\cite{Dolgov:2003sg,Gelmini:2019wfp} 
\begin{align}
  T_\text{peak} \approx \SI{108}{MeV} \,
                        \bigg( \frac{3T} {\langle E_{s} \rangle} \bigg)^{1/3}
                        \big( \cos2\theta_{s} \big)^{1/6}
                        \bigg( \frac{m_{s}}{\SI{1}{keV}} \bigg)^{1/3} \,.
\end{align}
For small mixing angles, $\sin^{2} 2\theta_{s} \lesssim10^{-5} \, (1\,\text{keV}/m_{s})$, the additional suppression at high temperatures ensures that the sterile neutrinos never reach thermal equilibrium. Their final abundance is decided largely by the production at $T_\text{peak}$. An accurate fitting formula for the final sterile neutrino abundance is~\cite{Abazajian:2005gj}
\begin{align}
  {\Omega_{s}} h^{2} \approx
    0.12 \, \bigg(\frac{m_{s}}{\SI{1}{keV}} \bigg)^{2}
            \bigg(\frac{\sin^{2}2\theta_{s}}{7.3\times10^{-8}} \bigg)^{1.23}\,,
\end{align}
where we have assumed $T_\text{QCD}=\SI{170}{MeV}$ for the temperature of the QCD phase transition.  Finetuning the production rate by adjusting $\theta_{s}$ for each $m_{s}$ allows one to dial-in exactly the observed amount of dark matter, thus leading to a very tight correlation between the two parameters. This mechanism in a more general avatar has gained currency as  ``freeze in'' production of dark matter. Including lepton asymmetries, as in the Shi--Fuller mechanism, can lead to MSW-like level crossing(s), which may result in resonant production of the sterile neutrinos. As the resonance occurs only for a narrow range of sterile neutrino energies, it may preferentially produce sterile neutrinos that are either hotter or colder than those produced by the Dodelson--Widrow mechanism. This extra freedom provided by the asymmetry allows the Shi--Fuller mechanism to access a larger part of the $m_{s}$--$\theta_{s}$ parameter space.

\begin{figure}
  \centering
  \includegraphics[width=0.55\textwidth]{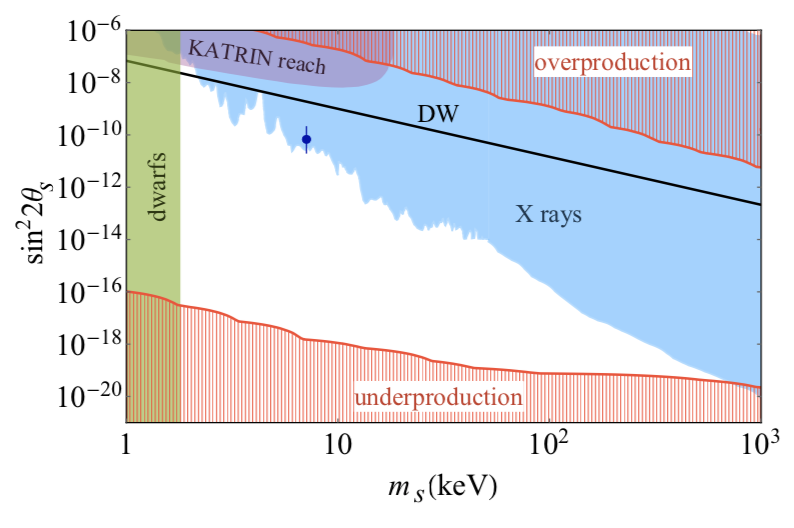}
  \caption{Viable regions of parameter space for sterile neutrino dark matter. Shaded/hatched regions are ruled out.  The shaded region marked ``dwarfs'' on the left is ruled out by observations of small-scale structure, in particular dwarf galaxies,  whereas ``X-rays'' denotes the constraints from non-observation of photons from radiative sterile neutrino decay. Along the black line, Dodelson--Widrow production yields the correct dark matter abundance. Including a lepton asymmetry (Shi--Fuller production) distorts and shifts the black line towards smaller mixing angles. An even larger region of parameter space can be reached in extended models, such as the ones discussed in~\cite{Hansen:2017rxr,deGouvea:2019phk,Merle:2013wta,Brdar:2017wgy}. The black dot with an error bar corresponds to the claimed detection of a possible X-ray signal \cite{Bulbul:2014sua, Boyarsky:2014jta}, which we will discuss in \cref{sec:keV-searches}. The red hatched regions labels ``overproduction'' and ``underproduction'' delimit the parameter ranges accessible in the specific model of ref.~\cite{deGouvea:2019phk}, from which this figure is taken.}
  \label{fig:DWSI}
\end{figure}

Recent developments in this area include more formal treatments of thermal and radiative effects, and including Pauli blocking~\cite{Asaka:2006rw, Ghiglieri:2015jua, Venumadhav:2015pla}. A thorough discussion, including open questions, can be found in ref.~\cite{Adhikari:2016bei}. The same reference and the references therein also explore alternative production mechanisms, often involving decays of additional particles. Similar to the secret interactions scenario invoked for lighter sterile neutrinos in \cref{sec:non-standard-cosmo}, hidden interactions of sterile neutrinos also extend the reach of thermal production for keV sterile neutrinos~\cite{Hansen:2017rxr, deGouvea:2019phk}. \Cref{fig:DWSI} shows regions of parameter space where sterile neutrino dark matter can be produced with acceptable abundance using the Dodelson--Widrow mechanism or the extended mechanisms.

\section{Features and Signatures}
\label{sec:keV-signatures}

Sterile neutrinos with keV masses can be cosmologically stable, sufficiently noninteracting, and have small enough velocity dispersion to qualify as cold, collisionless dark matter. They also have a number of unique signatures.

\subsection{Stable and Collisionless}
\label{sec:keV-stability}

A keV-mass sterile neutrino dominantly decays to 3 active neutrinos, with a rate~\cite{Pal:1981rm}
\begin{align}
  \Gamma_\text{decay}
    \approx \frac{G_F^2 \sin^2 2\theta_s m_s^5}{384 \pi^3}
    \approx \frac{1}{\SI{6e27}{sec}} \,
            \bigg( \frac{\sin^2 2\theta_s}{\num{e-8}} \bigg) \,
            \bigg( \frac{m_s}{\SI{1}{keV}} \bigg)^5 \,. 
\end{align}
We have encountered this expression already in \cref{eq:nu-3nu}, and the corresponding Feynman diagram in \cref{fig:nu-decay}~(a) in \cref{sec:nu-s-decay}.  The sterile neutrino lifetime is thus sufficiently large to render it essentially stable over timescales comparable to the age of the Universe. It is also sufficiently non-interacting, as its interactions are the same as those of the active neutrinos (which are already inert enough), but with cross sections that are further suppressed by a factor of $\sin^2 2\theta_{s}$.  We note that $\theta_{s}$ is the effective mixing angle in a 1+1 flavor scenario; in general one should replace $\sin^{2}2\theta_{s}$ by the net mixing of the sterile to active flavors.


\subsection{Free-streaming}

Sterile neutrinos had either reached thermal equilibrium at some point in cosmological evolution or they had not. As we discussed above, it turns out that the latter is in fact the minimal scenario, at least when the mixing of sterile neutrinos to active neutrinos is small. As a result, keV sterile neutrinos may be significantly colder than a particle of the same mass in equilibrium. If produced through a resonance, they can be colder still. This makes their free-streaming length smaller than would be naively expected from a thermally produced sterile neutrino of the same mass.  The free-streaming cut-off for keV-scale sterile neutrino dark matter depends on the momentum distribution, and for the Dodelson--Widrow case is approximately~\cite{Lesgourgues:2018ncw}
\begin{align}
  r_\text{fs} = \SI{70}{Mpc} \,
                         \Bigg( \frac{\SI{1}{keV}}{m_{s}} \Bigg)
                         \Bigg( \frac{T_{\nu_{s}}}{T_{\nu_{a}}} \Bigg)\Bigg[ 1+\frac{1}{2}\ln\Bigg(\bigg({\frac{1.7\times10^{2}}{\Omega_{s}+\Omega_{b}}\bigg)\bigg(\frac{T_{\nu_{a}}}{T_{\nu_{s}}}\bigg)\bigg(\frac{m_{s}}{\SI{1}{keV}}\bigg)}\Bigg)\Bigg]\,.
\end{align}
Compare this expression with $2\pi/k_\text{nr}$ from \cref{eq:knreV} in \cref{sec:cosmology}.
The term inside the logarithm encodes the effect of keV neutrinos becoming nonrelativistic in the radiation-dominated era and raises the free-streaming scale by roughly a factor of 5. For very distorted momentum spectra no such simple estimate is possible. Typically, $r_\text{fs} \gtrsim \,\SI{1}{Mpc}$ is required for a dark matter candidate to qualify as cold. Sterile neutrinos with mass of a few keV just fail to meet the requirement, and are therefore dubbed as ``warm'', instead of either hot or cold. Of course, if $\langle p_{s}\rangle/T_{\nu_{a}}$ is much smaller than is typical, they can also be cold.

The relatively large free-streaming distance for sterile neutrinos leads to constraints from observations of clustering of matter. The standard lore is that baryons fall into the gravitational potential created by dark matter halos and form structures. If this were not the case, there would not be enough time for the at most ${\cal O}(10^{-5})$ fluctuations inherited from the epoch of recombination to become ${\cal O}(1)$ by now. If dark matter clustering is hindered, as it will be for scales smaller than the free-streaming scale for sterile neutrinos, there is less clustering of matter on the relevant scales. We have already discussed this issue in the context of eV-scale neutrinos, where observations of clustering at the galaxy-scale and slightly below give strong constraints on $\sum m_{\nu}$, see \cref{sec:cosmo-observables}. 

For keV-mass sterile neutrino warm dark matter the matter power spectrum is reduced, relative to a cosmology with only cold dark matter, by a factor~\cite{Boyarsky:2008xj}
\begin{align}
\frac{P_{m}^\text{warm}}{P_{m}^\text{cold}}\approx(1-f_\text{warm})^{2}\bigg(\frac{a_{0}g_{\star}(a_{0})}{a_\text{eq}}\bigg)^{-\frac{3}{2}f_\text{warm}}\,,
\end{align}
on scales $k\gg k_\text{fs}$, where $f_\text{warm}={\Omega_{s}}/{\Omega_{m}}$. Contrast this expression with \cref{eq:PmeV} in \cref{sec:cosmology}. Constraints arise mainly from measurements of the two-point correlation function of the distribution of galaxies and from observations of the Lyman-$\alpha$ forests. See section 4.2 in~\cite{Adhikari:2016bei} for an expert review of this subject.

\subsection{Galaxy-sized Halo Abundance and Density Profiles}

We have discussed the difficulties associated with the mildly non-relativistic nature of keV sterile neutrino dark matter and the relevant constraints from large scale structure. However, the warm nature of sterile neutrinos may actually be desirable in a certain sense.

Although the paradigm of cold collisionless dark matter is quite successful on cosmological scales, its efficacy at galactic scales is less easy to establish. On such small scales the impact of baryons can be quite important. This impact is very hard to calculate. While cold collisionless dark matter can be simulated using N-body simulations, treating baryons requires including hydrodynamics at the very least, and to be entirely realistic one must include stellar evolution and the feedback of supernovae on the gravitational potential. This is obviously extremely challenging. Not only is the dynamic range, from the scale of a supernova explosion to cosmological scales, very large. But also the simulation of individual aspects, such as the evolution of stars, their emission, their impact on the gas and through gravity on the dark matter, is very computationally intensive. Nevertheless, such simulations are being done with increasing levels of sophistication, see for instance \cite{Springel:2020plp} for a publicly available state-of-the-art structure formation code. The best simulations include baryons and include a variety of feedback processes through ``sub-grid'' prescriptions. The parameters of these prescriptions are tuned to match a number of astrophysical observations, and once this has been achieved, it is possible to make predictions for how dark matter and baryons ought to be distributed in a galaxy and around it.

It appears that there are some difficulties in trying to match the predictions of cold dark matter to observations. Cold dark matter predicts a universal Navarro--Frenk--White density profile for dark matter halos. This density profile has a cusp that grows as $1/r$ in the inner regions of the halo. Determinations of the dark matter density profile of dwarf galaxies seem to indicate that halos are not always cuspy. This is known as the cusp--core problem. Another prediction of cold dark matter is that in a halo, say the size of the Milky Way's dark matter halo, there ought to be about $10^{3}$ smaller dark matter halos that should host observable satellite galaxies. We certainly do not see so many satellites. This is known as the missing satellites problem. These issues have been known since the 1980s, and it had already been appreciated that the resolution to these problems, especially the latter, might lie in baryonic feedback and lack of adequately sensitive observations. However, in the late 2000s a refined version of these problems was pointed out. If one were to restrict one's attention to the most massive subhalos of a Milky-Way-like galaxy in an N-body simulation, they are too dense in comparison to the corresponding satellite galaxies in real life. Usually this comparison is made by using the maximal circular velocity of stars in the satellite galaxy as a proxy for its mass. Somehow, the real satellites of the Milky Way are less massive than they should be according to simulations. This is called the too-big-to-fail problem, as in the simulation, these halos are too big to have failed to form stars, and there is no reason why they are not seen. Subsequently, rapid developments in hydrodynamic simulations have shown that the mismatch is not as large as originally thought. However some mismatch still persists. For more details on small scale problems of dark matter, see the reviews~\cite{Tulin:2017ara, deMartino:2020gfi}.

One might wonder if the problem is perhaps with assuming that dark matter is completely cold? Warm dark matter is implementable within N-body codes, essentially requiring only some change in the initial conditions. This has been studied  quite extensively, and it was shown that warm dark matter, e.g., comprising of sterile neutrinos, leads to smearing of the inner density profiles of galaxies and a cut-off in the linear matter power spectrum that restricts the number of satellites. While qualitatively these changes go in the correct direction, the most recent comparisons show that at a quantitative level warm dark matter does not fully address the small-scale issues highlighted above~\cite{Garzilli:2015iwa, Irsic:2017ixq}. The main issue is that the required particle mass to  address the issues is quite low, $m_{s}<2$\,keV for a thermal particle, which is forbidden by Lyman-$\alpha$ data. There are, however, escape routes; see~\cite{Lovell:2016nkp} for example.

\subsection{Phase Space Restrictions}

One of the most remarkable features of sterile neutrino dark matter is that it finds itself under attack from an unexpected but fundamental restriction -- the Pauli exclusion principle. Consider a self-gravitating halo of mass $M$ and radius $R$, composed of non-relativistic fermionic dark matter particles of mass $m_{s}$. Obviously, one has
\begin{align}
  M = m_{s}\int \frac{d^{3}{\bf x}\, d^{3}{\bf p}}{(2\pi)^{3}}\,f({\bf p},{\bf x})\,,
\end{align}
where $f$ is the phase space distribution. The Pauli principle says that no elementary volume of phase space can contain more than $1$ fermion, where we ignore a factor of $g_{s}$ that counts the internal degrees of freedom of the particle. If we assume  $p=mv$ and that $v$ extends to at most the escape velocity of the halo $v_\text{esc}\approx \sqrt{2G_N M/R}$ (ignoring order one factors from the density distribution), then demanding that $f\leq 1$ gives a conservative lower bound $m_{s}^{4}\gtrsim (9\pi/2)\,M / (R^{3} v_\text{esc}^{3})$ in the degenerate limit. Putting in numbers for $R$ and $M$ corresponding to a dwarf galaxy halo, leads to a constraint $m_{s}\gtrsim0.05$\,keV. This is the degeneracy bound, which can be further improved by considering properties of the phase space distribution.

Astronomical observations do not directly see the fine-grained phase space distribution, but rather the coarse grained phase space distribution today. However, the \emph{maximal} fine-grained true distribution $f$, predicted by a dark matter model, must always exceed the observed coarse grained $f$. This is simply because coarse-graining can only lead to reduction in height of the maximal peaks. Enforcing this requirement gives the Tremaine--Gunn bound on $m_{s}$~\cite{Tremaine:1979we}. Depending on various details, bounds in the range $0.2$\,keV to $6$\,keV have been obtained by various groups~\cite{Boyarsky:2008ju, Gorbunov:2008ka, Domcke:2014kla}. A recent conservative study shows that~\cite{DiPaolo:2017geq}
\begin{equation}  
m_{s} \gtrsim \SI{0.1}{keV} \, \bigg(\frac{0.2}{\Omega_{s}}\bigg)^{1/3}\,,
\end{equation}
by using astronomical data from dwarf galaxies to estimate the coarse-grained $f$. The main uncertainty that leads to this slightly weaker bound, is the lack of knowledge of the velocity dispersion anisotropy.  

These exact bounds are sensitive to the momentum distribution assumed for the sterile neutrinos, i.e., the production mechanism, in addition to what astronomical data is used. In particular, the lower bound weakens slightly for resonantly produced neutrinos with slightly colder spectra. Ref.~\cite{Boyarsky:2008ju} provides a good understanding of how the lower bound on $m_s$ depends on what is assumed about the momentum distribution. For example, the Tremaine--Gunn bound changes from $2.79$\,keV for non-resonantly produced neutrinos to $2$\,keV for resonantly produced neutrinos~\cite{Boyarsky:2008ju}. These bounds are stronger than what we quote in the equation above; the difference stems mainly from the more conservative assumptions in ref.~\cite{DiPaolo:2017geq}.

\subsection{Searches}
\label{sec:keV-searches}

So far we have discussed effects of sterile neutrinos on the large scale structure, focusing on the smaller scales therein. Now we return to the issue of possible decays of sterile neutrinos. We have already argued in \cref{sec:keV-stability} that the dominant sterile neutrino decay mode, into three active neutrino, is unfortunately invisible. The most easily observable decay mode is instead the radioactive decay $\nu_{s} \to \nu_{a}+\gamma$, via a one loop process involving a $W$ boson and a charged lepton (see \cref{fig:nu-decay} in \cref{sec:nu-s-decay}). This two-body decay gives an essentially monochromatic photon of energy $\approx m_s/2$, as $m_a \ll m_s$. For $m_{s} \approx \SI{10}{keV}$, these photons lie in the X-ray band. Of course, there are many atomic lines in this energy range so there is great scope for confusion.

For Majorana neutrinos, the decay rate is given by~\cite{Pal:1981rm}
\begin{align}
  \Gamma_{\nu_{s} \to \nu_{a}+\gamma}
    \approx \SI{1.35e-29}{sec^{-1}} \,
            \bigg( \frac{\sin^2 2\theta_s}{\num{e-8}} \bigg)
            \bigg( \frac{m_s}{\SI{1}{keV}} \bigg)^{5} \,,
\end{align}
see also \cref{eq:nu-nu-gamma-Majorana} in \cref{sec:nu-s-decay}. This rate is suppressed by a factor of $27\alpha/(8\pi)$ with respect to the decay to 3 active neutrinos. As we previously noted, this is very small and the lifetime exceeds the age of the Universe, so that the sterile neutrino is after all a good dark matter candidate. Nevertheless,  given that modern X-ray telescopes see about $10^{78}$ dark matter particles in their line of sight to a nearby galaxy cluster of mass $\SI{e15}{M_{\odot}}$, a signal may be detectable. As one is looking at a decay, as opposed to an annihilation process, the dependence on the dark matter density (and its uncertainties) is weaker than for dark matter searches targeting annihilation, especially if the field of view is somewhat larger than the scale height of the halo in question.

\begin{figure}
  \centering
  \includegraphics[width=0.55\textwidth]{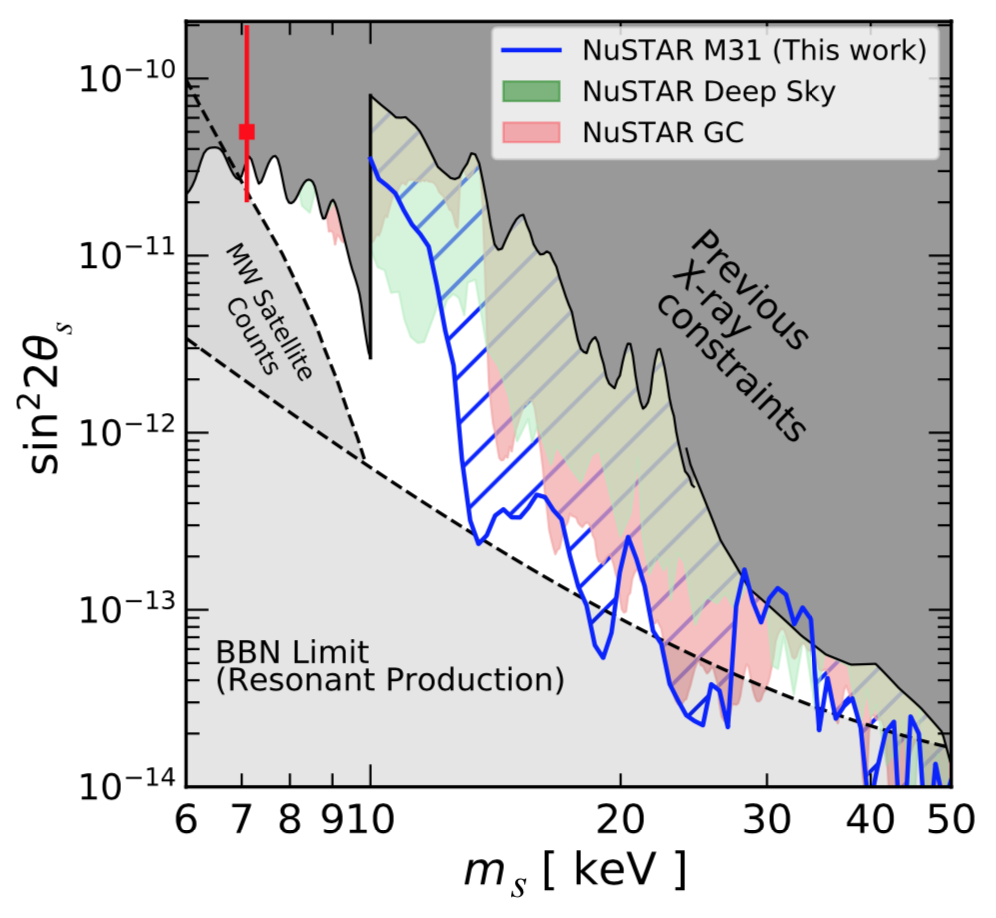}
  \caption{Constraints on the parameter space of keV-scale sterile neutrino dark matter from X-ray observations (colored and dark gray regions and contours) and from structure formation arguments (medium gray region labeled ``MW Satellite Counts''~\cite{Cherry:2017dwu}).  The light gray region labeled ``BBN Limit (Resonant Production)'' is disfavored by BBN constraints on the lepton asymmetry of the Universe if production via the Shi--Fuller mechanism is assumed. The red point denotes the claimed detection of a \SI{3.5}{keV} line~\cite{Bulbul:2014sua, Boyarsky:2014jta}. Figure taken from~\cite{Ng:2019gch}.}
  \label{fig:kevcon}
\end{figure}

Given a smorgasbord of astrophysical objects, one is limited only by money and imagination when choosing where to look for potential signals of radiatively decaying sterile neutrinos. The first proposals suggested targeting the diffuse signal~\cite{Drees:2000wi} and the signals from massive dark matter halos like galaxy clusters~\cite{Abazajian:2001nj, Abazajian:2001vt} and galaxies. Since then constraints using a variety of sources have been derived; see for example~\cite{Boyarsky:2005us, Watson:2006qb, RiemerSorensen:2006fh,Loewenstein:2009cm, Ng:2019gch,Roach:2019ctw,Bhargava:2020fxr}. A more complete list of references can be found in~\cite{Boyarsky:2018tvu}. A representative compilation of constraints is shown in \cref{fig:kevcon}. As is obvious, at least for the simplest model the parameter space is squeezed tight from all directions. 

In 2014, stacked observations of galaxy clusters using XMM-Newton data has led to the detection of an unidentified line near 3.55\,keV with high significance~\cite{Bulbul:2014sua}. The possible interpretation of this signal as a hint for a sterile neutrino of mass $m_{s}\approx7.1$\,keV has sparked  frenetic activity and heated discussion about the reality and origin of the line. This has been adequately reviewed in reviews focused on keV sterile neutrino dark matter, such as refs.~\cite{Abazajian:2017tcc, Boyarsky:2018tvu}. The main development since the appearance of these reviews is that the constraints proposed in~\cite{Dessert:2018qih} appear to rule out a dark matter interpretation of the 3.55\,keV line. Other groups appear to not be convinced yet because of non-conservative choice made for the dark matter density and supposedly improper treatment of nearby lines~\cite{Boyarsky:2020hqb, Abazajian:2020unr}.

Among other constraints, it has been suggested that core-collapse supernovae and the natal velocities of the resultant pulsar can be affected by keV sterile neutrinos~\cite{Shi:1993ee,Kusenko:1996sr}. These have been recently studied by several groups~\cite{Warren:2016slz, Arguelles:2016uwb, Suliga:2020vpz}.

Sterile neutrinos can be produced in the laboratory through their mixing with active neutrinos. Observation of anomalous Tritium beta decay in KATRIN is touted as a possible technique to look for this possibility, even though reaching cosmologically interesting mixing angles seems challenging using this method. We have already discussed this type of searches in \cref{sec:kinematics}.

\chapter{Heavy Sterile Neutrinos and the Cosmic Baryon Excess}
\label{sec:leptogenesis}

An outstanding problem of theoretical physics is to explain why there is more matter than antimatter in the Universe. This observation is encoded in the baryon-to-photon ratio $\eta_{b}=n_{b}/n_{\gamma}\approx 6\times 10^{-10}$ and in the observational fact that there are essentially no anti-baryons in the Universe. Evidently, the baryon asymmetry $\eta_{b}-\eta_{\bar{b}}$ is the same as $\eta_{b}$ itself, which is tiny but nonzero. One might think that explaining this small asymmetry ought not to be too difficult. If matter and antimatter are almost but not exactly identical to each other, there would be a small excess of one over another and that would be the end of the story. It turns out, though, that it is very challenging to implement this idea in practice. We will only provide a pedagogical introduction to this rich subject, with the goal of establishing the connection of sterile neutrinos to this important problem.  For more details and references, see the dedicated review by Davidson, Nardi, and Nir~\cite{Davidson:2008bu} that has previously appeared in this series. We comment on some recent developments at the end.

It was pointed out by Sakharov already in 1967 that a baryon asymmetry could arise in a Universe with completely symmetric initial conditions only if three conditions are met:
\begin{itemize}
  \item baryon number $B$ is not strictly conserved (otherwise, there can be no $B$ asymmetry)
  \item $C$ and $CP$ are not exact symmetries of nature (otherwise, baryons and anti-baryons, or the left-handed and right-handed sectors, respectively, would contain equal but opposite asymmetries, preventing the generation of a net asymmetry)
  \item there is departure from thermal equilibrium (otherwise, any asymmetry in a non-conserved quantum number such as $B$ gets erased in chemical equilibrium).
\end{itemize}
When Sakharov proposed these condition, baryon number was thought to be a strictly conserved quantum number. Only in the late 1970s, after the advent of grand unified theories, the idea of baryon number violation began to be taken seriously. The other two conditions are already met within the Standard Models of particle physics and cosmology -- there is $C$ and $CP$ violation in the weak interactions, and the expansion of the Universe allows out-of-equilibrium conditions. This allowed the development of a theoretical framework to explain baryogenesis. Unfortunately, it is not possible to obtain successful  baryogenesis within the Standard Model. Interestingly, the road block is not baryon number violation -- in fact, $B$ is violated in the Standard Model by non-perturbative processes.  The key obstacle is instead the somewhat heavy Higgs boson: at a Higgs mass of 125\,GeV, the electroweak phase transition is not strongly first order but rather a smooth crossover, and the departure from thermal equilibrium is not strong enough. Further, the extent of $CP$ violation in hadronic interactions, encoded in the Jarlskog invariant of the CKM matrix, is not sufficient either.

The preceding discussion might appear to imply no connection to neutrinos at all. All of this changed in the 1980s, following some observations by 't~Hooft. In a seminal paper Kuzmin, Rubakov, and Shaposhnikov pointed out that nonperturbative solutions of the Standard Model, called sphalerons, do not independently conserve $B$ and $L$, but only their linear combination $B-L$. The other combination, $B+L$, is in fact violated~\cite{Kuzmin:1985mm}. This led to the realization that there is a remarkable connection between baryogenesis and neutrinos. Fukugita and Yanagida showed that if heavy right-handed Majorana neutrinos exist, as strongly motivated by the seesaw mechanism and several grand unified theories, they can decay in a lepton number-violating way to Standard Model leptons and the Higgs boson.  The decay rate can be smaller than the Hubble expansion rate~\cite{Fukugita:1986hr}, leading to out-of-equilibrium decays. If these decays violate $C$ and $CP$, all three Sakharov conditions (now substituting $L$ for $B$) are satisfied, and there can be an asymmetry between leptons and anti-leptons. Further, the lepton asymmetry can be converted to a baryon asymmetry through sphalerons, thereby explaining the matter--antimatter asymmetry.

\section{Standard Leptogenesis}

The text-book variety of a leptogenesis model involves extending the Standard Model by three heavy right-hand neutrinos $N_{iR}$. This is the same model that we described in \cref{sec:typeI-seesaw} as the simplest implementation of the type-I seesaw mechanism as a way to explain the smallness of active neutrino masses. In complete generality, but using a slightly different notation, the relevant part of the Lagrangian density is
\begin{align}
  -{\cal L}_{R} = \sum_{i} \frac{m_{\ell}}{v} \bar{\psi}_{iL} H \ell_{iR}
                + \sum_{i,j} h_{ij} \bar{\psi}_{iL} (i \sigma^2 H^*) N_{jR}
                + \sum_{i} M_{i} N_{iR}^{T} C^{-1} N_{iR}\,,
  \label{eq:L-leptogenesis}
\end{align}
where $i,\,j$ are generation indices, $\psi_{iL}$ are the left-handed Standard Model lepton doublet, $\ell_{iR}$ are the right-handed charged lepton fields, and $H$ is the Higgs doublet. Here the basis has been chosen such that the mass-matrices for the charged leptons and right-handed neutrinos are diagonal, and the second term causes mixing between left-handed and right-handed neutrinos. The last term implies Majorana masses, and there is lepton number violation in the model through these $\Delta L=2$ operators. If the dimensionless Yukawa couplings $h_{ij}$ are complex, there is $CP$ violation; $C$ is violated already by the weak interaction. If the decay rate of the $N_{iR}$ fields can be made slower than the Hubble rate, all the Sakharov conditions will be satisfied.

The relevant decay rate of a right-handed neutrinos into a charged lepton and a Higgs boson is
\begin{align}
  \Gamma_{i} \simeq \frac{M_{i}}{8\pi}\sum_{j} \lvert h_{ij}\rvert^{2} \,,
\end{align}
assuming the three $N_{iR}$s have hierarchical masses, $M_{1} \ll M_{2} \ll M_{3}$. The out-of-equilibrium condition requires $\Gamma_{i} \lesssim H$, which during radiation-domination translates to
\begin{align}
  M_{i} \gtrsim \frac{M_\text{Pl}}{8\pi\sqrt{g_{\star}}}
                \sum_{j} \lvert h_{ij}\rvert^{2} \,.
\end{align}
As such, we have no information on the Yukawa couplings $h_{ij}$. However, if we assume $h_{ij}\sim m_{u_{i}}/v$, i.e., related to the up-quark masses as motivated by some grand unified theories, one finds $M_{1}\gtrsim10^{7}$\,GeV, $M_{2}\gtrsim10^{12}$\,GeV, and $M_{3}\gtrsim10^{16}$\,GeV. These are clearly very large masses, but independently well-motivated by the small masses of the active neutrinos that would arise through the seesaw mechanism. It is easy to see that $2\to2$ processes involving $N_{iR}$ are strongly Boltzmann suppressed at any temperature below their masses.

Once the Universe cools to a temperature below $M_{1}$, most of the $N_{iR}$ decay away. The asymmetry in this decay is parameterized as
\begin{align}
  \epsilon_{1} = \frac{\Gamma_{N_{1} \to \ell H} - \Gamma_{N_{1} \to \bar\ell H}}
                      {\Gamma_{N_{1} \to \ell H} + \Gamma_{N_{1} \to \bar\ell H}} \,.
\end{align}
Note that this asymmetry would be predicted to be zero if only the tree-level diagram generated by the second term in \cref{eq:L-leptogenesis} were considered. At the one-loop level, however, it can be non-zero thanks to interference between tree and loop diagrams. It should also be noted that at the very high temperatures under consideration, above the electroweak symmetry breaking, all four degrees of freedom of the Higgs doublet $H$ are physical because they have not yet been eaten up by the weak bosons.  One finds~\cite{Davidson:2008bu}
\begin{align}
  \epsilon_{1} = -\frac{3}{16 \pi (hh^{\dagger})_{11}}
                  \sum_{j\neq1} \text{Im}[(hh^{\dagger})^{2}_{1j}]
                                \frac{M_{1}}{M_{j}}\,.
\end{align}
This lepton asymmetry converts to a baryon asymmetry through the sphaleron processes. Roughly, $B = r (B-L)$, where $r$ is a model-dependent conversion factor determined by the conserved chemical potentials. It is $-28/79$ for the Standard Model. Further, there is partial washout, encoded in an efficiency factor $\kappa$, due to scattering and inverse decays.  Finally, $n_{b}/n_{\gamma}$ is further reduced due to the reduction in the number of degrees of freedom from its high-temperature value $g_{\star} \approx 106.75$ to the value $g_{\star}^{0}=3.75$ in the present-day Universe. All in all, one finds the baryon-to-photon ratio
\begin{align}
  \eta_{b} = r \, \kappa \, \frac{g_{\star}^0}{g_{\star}} \,
             \epsilon_{1}
             \simeq 10^{-3}\,\epsilon_{1} \,,
\end{align}
where in the last equality we have plugged in the typical efficiency $\kappa = 0.1$. To obtain the observed baryon asymmetry, one thus requires $\epsilon_{1} \sim 10^{-6}$ or larger.

Is this a good prediction? It naively seems to be. There is a clear constraint on $\epsilon_{1}$, and surely it puts constraints on the parameter space of the neutrino mass Lagrangian? Unfortunately, this turns out to be a mirage. The problem is the $h$-matrix, which, even though connected to neutrino masses, is sufficiently underdetermined that essentially no constraints on low-energy neutrino parameters are possible. This is easy to see because, even if we ignore $CP$ violation, $h$ has nine parameters and the $M_{i}$ amount to 3 more, whereas at low energy we only have the 3 masses and 3 mixing angles (again assuming no $CP$ violation). Even including $CP$ violation, there are far too many parameters at high energy to make a robust connection with observations at low energy.  This disconnect is most transparently encoded in the Casas--Ibarra parameterization, which explicitly factorizes $h$ in such a way that low-energy and high-energy parameters are separated~\cite{Casas:2001sr}.  On the other hand, if one assumes that nature has drawn neutrino masses and mixing parameters from some random distribution, as in ``neutrino anarchy'' scenarios~\cite{Hall:1999sn, Haba:2000be, deGouvea:2003xe, Espinosa:2003qz, deGouvea:2012ac, Brdar:2015jwo}, it is more likely that $CP$ violation at high scales implies observable $CP$ violation at low scales, and vice-versa.

We have thus hit upon the fly in the leptogenesis ointment: is this even a testable theory? One has enough freedom of parameters to get the correct baryon asymmetry, essentially no matter what experimental constraints one imposes on the values of the low-energy parameters. In other words, even the simplest model works far too well. The other phenomenological challenge in this paradigm is that the high-energy sector is at energies that are too high to be accessible at laboratory experiments.

The response to this quandary has been two-fold. On one hand the above mechanism has been analyzed and computed in more detail, with many subtle effects taken into account. This is partly motivated by the hope that such detailed calculations would reduce the viable parameter space for the model, enhancing its predictive power. The other approach has been to look at other models of leptogenesis, often aimed at lowering the scale of the high energy sector so that collider experiments can put additional constraints. We now discuss these more recent developments.

\section{Modern Developments}

Since the 1990s, a great deal of effort has gone into making the treatment of basic thermal leptogenesis, i.e., the model described above, increasingly more sophisticated. These developments include improved treatments using semi-classical Boltzmann equations and the closed time path formalism in thermal quantum field theory in order to consistently account for thermal masses, decay widths, etc.  Modern calculations also account for the full flavor structure of the neutrino sector, i.e., they do not assume any more that the asymmetries in each generation of light leptons is identical, but instead track each of them individually, including correlations. Other developments include better treatments of so-called spectator processes, which do not change lepton number themselves but affect the washout efficiency, rates of sphaleron processes, etc. See the recent review in~\cite{LeptoReview} for a detailed discussion of these issues.

Another development has been the advent of alternative leptogenesis models. The hierarchical masses in the thermal leptogenesis scenario introduced above make the lightest $N_{iR}$ still too heavy -- the reheating temperature must exceed its mass, and in supersymmetric models leads to production of too many gravitinos whose decays spoil BBN~\cite{Ellis:1984eq}. Choosing quasi-degenerate masses for the right-handed neutrinos, on the other hand, allows stronger $CP$ violation in $N_{iR}$--$N_{jR}$ mixing and helps lower the reheating temperature. Of course, there might not be supersymmetry at all in nature, but given that typical leptogenesis models are not really testable, this conditional constraint was hoped to be the guiding light to the correct model. After all, supersymmetry was to be insisted upon for other reasons that do not bear repeating here.

Among the alternative models, we find the model proposed by Akhmedov, Rubakov, and Smirnov~\cite{Akhmedov:1998qx} to be a particularly attractive alternative to standard leptogenesis. Loosely speaking, the starting point here is the neutrino mass sector of the $\nu$MSM model described in~\cref{sec:nuMSM}. One of the right handed states is at the keV-scale, and the other two are at the GeV-scale, but strongly degenerate. In this model, the right-handed neutrinos do not start out in equilibrium. Instead, like for the Dodelson--Widrow production mechanism described before, their abundance grows with time until it freezes in to a desired level. Their decays then transfer the asymmetry to the light leptons. A key feature is that in this model the $CP$ violation occurs in oscillations between the two right-handed sterile neutrinos, so there is no lepton number violation as such. Instead, part of the lepton number is hidden in the sterile states, where the sphalerons do not see it. This leads to the generation of an effective lepton asymmetry in the active sector. $CP$ violation in oscillation can be strongly enhanced if the two heavy neutrinos are degenerate. This also brings the heavy neutrino mass scale below the electroweak scale, and in fact the model can be probed at colliders and fixed-target experiments. One can also obtain leptogenesis in type-II and type-III seesaw models, as well as in various non-thermal mechanisms that can tolerate a lower reheating temperature.

The decays of the right-handed neutrinos may have signatures in collider and fixed target experiments if their masses are at the few hundred GeV-scale or below. They either decay promptly or after traveling a small but macroscopic length, and can be detected in displaced vertex searches at LHC, or SHiP, MATHUSLA, etc. Observation of lepton number violation at the LHC could set a decisive constraint on thermal leptogenesis, possibly falsifying it because of the strong washout effect that is predicted in this case. However, there are ways of evading this constraint, e.g., flavor-dependence or conserved quantum numbers that protect the baryon asymmetry from washout. A very promising idea in this context is the low-scale seesaw, which has been studied in great detail in recent times, especially in the context of left--right symmetric models.  The discovery of a right-handed gauge boson at the LHC would pose a strong constraint on leptogenesis in such scenarios. We point the reader to a recent review~\cite{10.1007/978-3-030-29622-3_40}, where more references can be found.

Our brief review of leptogenesis does not do justice to the rich and complex subject. Nevertheless, we hope to have highlighted that sterile neutrinos that are as ``sterile'' as they can possibly get, i.e., even when they are experimentally almost inaccessible, can still play an important and interesting role in cosmology, and may in fact be the key to understanding one of the grandest mysteries of the Universe, namely the dominance of matter over antimatter.

\chapter{Summary and Outlook}
\label{sec:summary}

In summary, we have outlined the manifold applications of sterile neutrinos in
particle physics and cosmology, beginning in \cref{sec:sbl-anomalies} with an
overview over several statistically significant experimental anomalies that
have spurred interest in particular in sterile neutrinos at the eV scale.  It
is interesting that, when interpreted as hints for sterile neutrinos, all
anomalies point towards the same parameter region.  Explaining them in terms of
more mundane effects -- which may be possible because each of them suffers from
large and hard-to-quantify systematic uncertainties -- on the other hand
requires several independent effects.  Moreover, no convincing Standard Model
explanation exists for any of the anomalies so far.

We have seen in \cref{sec:theory-motivation} how eV-scale sterile neutrinos
could be embedded into well-motivated theoretical models of the neutrino
sector, and we have discussed their phenomenology in \cref{sec:pheno}.  There,
we have in particular highlighted the tension between the simplest sterile
neutrino scenario and non-anomalous data, in particular from $\nu_\mu$
disappearance searches.

Given the strong and fairly consistent hints on the one side, and the exclusion
of the minimal sterile neutrino models on the other side, it is clear that
resolving the anomalies should be a top priority for neutrino physics in the
coming years: either, a fundamental discovery with far-reaching consequences
will be made, or we will learn important lessons about systematic effects in
neutrino experiments, for instance about the subtleties of neutrino--nucleus
interactions (relevant to MiniBooNE) or about precision predictions for nuclear
beta decay spectra (relevant to reactor experiments).  Such lessons would be
indispensable for the successful operation of future billion-euro scale
facilities like DUNE and Hyper-Kamiokande.  Fortunately, a large number of new
experiments that are already running or are in advanced R\&D stages will
hopefully achieve this goal -- provided they are accompanied by an equally
vigorous theory effort.

Besides constraints from other oscillation experiments, eV-scale sterile
neutrinos also have to face cosmology.  We have offered an introduction to
this topic in \cref{sec:cosmology}, highlighting in particular the main
cosmological observables relevant here: the effective number of neutrino
species, $N_\text{eff}$, the sum of neutrino masses, $\sum m_\nu$, and
observables related to possible self-interactions of sterile neutrinos.  Once
again, minimal models are already in gross conflict with cosmological
observations, while the extended scenarios reviewed in \cref{sec:sterile-cosmo}
can avoid this conflict, typically by preventing sterile neutrinos from being
produced abundantly in the early Universe.  As cosmology continues to
progress at breakneck speed, sterile neutrino cosmology is sure to flourish
for many years to come.  Future large scale structure surveys will reach
sensitivity to even the active neutrino mass scale, making it more and more
difficult to hide extra neutrino states from them.  The precision at which
$N_\text{eff}$ at recombination can be measured will take another leap
forward with the next-generation CMB-S4 experiment.  And, finally, we
have shown that sterile neutrinos might play a role in relaxing the
tension between local and cosmological measurements of the Hubble constant.
As this tension is one of the biggest unsolved problems in modern
cosmology, massive observational and theoretical efforts are directed
towards its resolution, resulting in a thorough reappraisal of sterile neutrino cosmology at every step.

In the final part of the review, we have extended our focus to sterile
neutrinos with masses significantly heavier than \SI{1}{eV}.  We have
in \cref{sec:keVchapt} briefly reviewed the possibility that keV-scale
sterile neutrinos constitute the dark matter in the Universe.  And while
this possibility is getting more and more constrained observationally,
it remains quite intriguing.  We refer the reader to the dedicated reviews
on this topic, cited in \cref{sec:keVchapt}, for a more detailed discussion
of sterile neutrino dark matter.

In \cref{sec:leptogenesis}, we have outlined the basics of leptogenesis,
wherein the decays of superheavy sterile neutrinos generate the observed
particle--anti-particle asymmetry in the Universe, which is another major unsolved problem in cosmology.  Once again, readers
interested in a more thorough discussion of this topic are referred
to the literature cited in \cref{sec:leptogenesis}.

We see that sterile neutrino appear in manifold forms in theories
beyond the Standard Models of particle physics and cosmology, and
could be related to many of the most pressing unsolved problems in
both fields.  We therefore hope that this review will be a useful reference for
researchers working on this active field. Simultaneously, we expect that many of the results highlighted here as ``state-of-the-art'',
will be rendered obsolete by newer and more impressive results by the next generation of experiments and observations.

Finally, let us reiterate that a sterile neutrino is simply a SM-singlet
fermion, so having one in nature is really not asking for much. If the authors
had to make a bet, one of us would bet that there is a sterile neutrino in
nature -- though with no idea at what mass scale it may be hiding. The other
would bet the same, but has no idea as to what its ``mixing'' is with known
particles.  Both of us dream of a day when sterile neutrinos will transcend
their status as a hypothesized explanation for a potpourri of anomalies, become
a reality, and open a new window into the nature of matter and the Universe.

\chapter*{Acknowledgments}

We would like to thank all those colleagues and friends with whom we have
collaborated on sterile neutrinos over the years, in particular Carlos A.\
Arg\"{u}elles, Vedran Brdar, Xiaoyong Chu, Mona Dentler, Ivan Esteban, Roni
Harnik, \'{A}lvaro Hern\'{a}ndez-Cabezudo, Pedro Machado, Michele Maltoni, Ivan
Martinez-Soler, Ninetta Saviano, Thomas Schwetz, Xiao-Ping Wang, and Johannes
Welter.

BD's work is partly supported by the Dept.\ of Atomic Energy (Govt.\ of India)
research project 12-R\&D-TFR5.02-0200 and the Max Planck-Gesellschaft through a
Max Planck Partner Group.  JK's work has been partially supported by the European
Research Council (ERC) under the European Union's Horizon 2020 research and
innovation program (grant agreement No.\ 637506, ``$\nu$Directions''). He has
also received funding from the German Research Foundation (DFG) under grant
No.\ KO~4820/4-1.

\bibliographystyle{JHEP}

\providecommand{\href}[2]{#2}\begingroup\raggedright\endgroup

\end{document}